\documentclass{aa}

\usepackage{graphicx}
\usepackage{txfonts}
\usepackage{graphicx}
\usepackage{txfonts}
\usepackage[utf8]{inputenc}
\usepackage{dcolumn}
\usepackage{natbib}
\usepackage{enumerate}
\usepackage{color}
\usepackage{grffile}
\usepackage{array,float}
\usepackage{multirow}
\usepackage{lscape}
\usepackage{hyperref}
\newcolumntype{d}{D{.}{.}{-1} } 

\newcommand{\msun}{M$_\odot$}

\newcommand{\mum}{$\mu$m}

\newcommand{\hi}{H\,{\sc i}}

\newcommand{\hii}{H\,{\sc ii}}

\newcommand{\Scu}{Scutum-Centaurus}
\newcommand{\Sag}{Sagittarius}

\newcommand{\cat}{catalogue}

\begin{document}

   \title{ATLASGAL-selected massive clumps in the inner Galaxy}
   \titlerunning{ATLASGAL-selected massive clumps in the inner Galaxy III.}

   \subtitle{III. Dust Continuum Characterization of an Evolutionary Sample}

   \author{C. K\"onig\inst{1}
			\and
			J. S. Urquhart\inst{1,2}
			\and
			T. Csengeri\inst{1}
			\and
			S. Leurini\inst{1}
			\and
			F. Wyrowski\inst{1}
			\and
			A. Giannetti\inst{1}
			\and
			M. Wienen\inst{1}
			\and
			T. Pillai\inst{1}
			\and\\
			J. Kauffmann\inst{1}
			\and
			K. M. Menten\inst{1}
			\and
			F. Schuller\inst{1}
          }
   \authorrunning{C. K\"onig et al.}

   \institute{Max-Planck-Institut f\"ur Radioastronomie (MPIfR), 
              Auf dem H\"ugel 69, 53121 Bonn, Germany\\
              \email{koenig@mpifr-bonn.mpg.de}
         \and
         	  School of Physical Sciences, University of Kent, 
              Ingram Building, Canterbury, Kent CT2 7NH, UK
             }

   \date{Received MMMM dd, YYYY; accepted MMMM dd, YYYY}

 
  \abstract
   {Massive star formation and the processes involved are still poorly understood. The ATLASGAL survey provides an ideal basis for detailed studies of large numbers of massive star forming clumps covering the whole range of evolutionary stages. The ATLASGAL Top100 is a sample of clumps selected from their infrared and radio properties to be representative for the whole range of evolutionary stages. }
   {The ATLASGAL Top100 sources are the focus of a number of detailed follow-up studies that will be presented in a series of papers. In the present work we use the dust continuum emission to constrain the physical properties of this sample and identify trends as a function of source evolution.}
   {We determine flux densities from mid-infrared to submm wavelength (8--870$\,\mu$m) images and use these values to fit their spectral energy distributions (SEDs) and determine their dust temperature and flux. Combining these with recent distances from the literature including maser parallax measurements we determine clump masses, luminosities and column densities.}
   {We define four distinct source classes from the available continuum data and arrange these into an evolutionary sequence. This begins with sources found to be dark at 70\,$\mu$m, followed by 24\,$\mu$m weak sources with an embedded 70\,$\mu$m source, continues through mid-infrared bright sources and ends with infrared bright sources associated with radio emission (i.e., \hii\,regions). We find trends for increasing temperature, luminosity and column density with the proposed evolution sequence, confirming that this sample is representative of different evolutionary stages of massive star formation. Our sources span temperatures from approximately 11 to 41\,K, with bolometric luminosities in the range $57$\,L$_\odot$-$3.8\times10^6$\,L$_\odot$. The highest masses reach $4.3\times10^4$\,M$_\odot$ and peak column densities up to $1.1\times10^{24}$\,cm$^{-1}$, and therefore have the potential to form the most massive O-type stars. We show that at least 93 sources (85\%) of this sample have the ability to form massive stars and that most are gravitationally unstable and hence likely to be collapsing.}
   {The highest column density ATLASGAL sources cover the whole range of evolutionary stages from the youngest to the most evolved high-mass star forming clumps. Their study provides a unique starting point for more in-depth research on massive star formation in four distinct evolutionary stages whose well defined physical parameters afford more detailed studies. As most of the sample is closer than 5\,kpc, these sources are also ideal for follow-up observations with high spatial resolution.}


   \maketitle
%

\section{Introduction}
Massive stars ($>$8\,\msun) play an important role in the evolution of their host galaxies \citep{kennicutt2005}. During the early stages of their formation they drive powerful molecular outflows that inject momentum into the surrounding environment. At their later stages they drive strong stellar winds and emit copious amounts of ionizing radiation that shape their local environments and regulate further star formation. Massive stars are also primarily responsible for the production of nearly all heavy elements, which are returned to the interstellar medium (ISM) through stellar winds and supernovae leading to an enrichment of the local and global environment and changes to the chemistry. Yet the formation and early evolutionary stages of massive star formation are still not well understood (see \citealt{Zinnecker2007} for a review).

The main hurdles to an improved understanding are that high mass stars are rare and are therefore typically found at large distances from the Sun ($>$ 2\,kpc), and that the earliest stages of their formation take place while they are still deeply embedded in their natal molecular clouds. Consequently, the earliest stages are hidden from more traditional observations at optical and near-infrared wavelengths and therefore observations in the far-infrared and submillimeter regimes are required to probe these extremely dense environments. Furthermore, massive stars are known to form almost exclusively in clusters \citep{de-wit2004} and therefore high-resolution is required to separate individual (proto-)cluster members.

In recent years, a number of Galactic plane surveys have been undertaken that probe large volumes of the Galaxy and provide a straightforward way to identify a large sample of embedded massive stars and clusters that will include examples of sources in all of the important evolutionary stages. These surveys provide almost complete coverage of near-infrared to radio wavelengths (e.g., the UKIRT Infrared Deep Sky Survey Galactic Plane Survey (UKIDSS GPS), \citealt{lucas2008};  Galactic Legacy Infrared Mid-Plane Survey Extraordinaire (GLIMPSE), \citealt{benjamin2003}; Midcourse Space Experiment (MSX), \citealt{Price2001}; Wide-Field Infrared Survey Explorer (WISE) \citealt{Wright2010}; Multiband Infrared Photometer for Spitzer survey of the inner Galactic Plane (MIPSGAL), \citealt{Carey2009}; \textit{Herschel} infrared Galactic Plane Survey (Hi-GAL), \citealt{molinari2010}; APEX Telescope Large Area Survey of the Galaxy (ATLASGAL), \citealt{schuller2009}; Bolocam Galactic Plane Survey (BGPS), \citealt{aguirre2011} and the Co-ordinated Radio and Infrared Survey for High-Mass Star Formation (CORNISH), \citealt{purcell2013} and  \citealt{hoare2012}).

Dust emission is generally optically thin at submillimetre wavelengths and therefore surveys at these wavelengths are an excellent tracer of column density and total mass. The ATLASGAL survey covers a total area of 420 square degrees, tracing dust throughout the inner Galaxy ($300\degr < \ell < 60\degr$ with $|b|\leq1.5\degr$  and was subsequently extended to $280\degr < \ell < 300\degr$ with $-2\degr < b < 1\degr$ \citep{schuller2009}. The survey was conducted with the Large APEX Bolometer Camera \citep[LABOCA,][]{Siringo2009} using the Atacama Pathfinder EXperiment 12\,m telescope \citep[APEX;][]{Gusten2006}, which is located at a height of $\sim$5100\,m on the Chajnantor plateau in the Atacama desert in Chile. APEX has an angular resolution of 19.2\arcsec\ at 870\,\mum. 

Subsequently, the ATLASGAL compact source catalog \citep[CSC;][]{Contreras2013,urquhart2014c} and the ATLASGAL Gaussclumps source catalog \citep[GCSC;][]{Csengeri2014} were extracted, identifying $\sim$10,000 dust clumps located throughout the inner Galaxy. These catalogues include large numbers of potential high mass star forming regions in different evolutionary stages from massive starless regions to clumps associated with (ultra-) compact \hii\ regions on the verge of destroying their natal environment. In the past, studies into massive star formation have focused on a single well defined evolutionary stage (e.g., UCH\textsc{ii} regions (\citealt{wood1989}), high mass protostellar objects (\citealt{sridharan2002}) or Class II methanol masers, which exclusively pinpoint the locations of high mass protostars (\citealt{walsh1997,walsh2003})). Although many of these have been successful in parameterising these stages they tell us little about how these various phases are connected, how their properties change as the embedded star evolves or the relative lifetimes of each stage. However, covering the full mid-infrared to submm wavelength regime with unprecedented sensitivity and resolution, the above-mentioned unbiased Galactic plane surveys and in particular the ATLASGAL catalogues provide an excellent starting point to study the complete evolutionary sequence of massive stars in a robust statistical manner (\citealt{urquhart2014c}).

We have used the ATLASGAL survey to select a sample of $\sim$100 massive clumps that likely represent different evolutionary stages \citep{Giannetti2014}. We have also correlated this sample with methanol masers (\citealt{Urquhart2013}), YSOs and \hii\ regions (\citealt{urquhart2013b,Urquhart2014a}) to constrain the evolutionary state of the sample's sources. This sample has also been the subject of molecular line follow-up studies to fully characterize the evolutionary sequence and derive the physical properties of different stages. The results of these studies have been  discussed in a series of papers. For example, CO depletion and isotopic ratios have been investigated by \citet{Giannetti2014}, SiO emission for the northern sources to trace shocked gas was studied by \citet{Csengeri2016} and NH$_3$ has been used to investigate infall towards selected sources by \citet{Wyrowski2016} with a number of papers in preparation focusing on their associated outflows (Navarete et al in prep.) and modeling of their chemistry (Giannetti et al. in prep.).  Using these spectroscopic and continuum surveys will provide the most detailed view of the evolutionary sequence of massive stars and robust constraints on the physical properties, chemical conditions and kinematics of this unique sample of high-mass star forming regions selected from the whole inner Galactic plane.

In this paper we use multi-wavelength dust continuum emission to characterize this sample of candidate massive star forming regions, in terms of dust temperatures, bolometric luminosities and clump masses. Together with the latest distances these quantities are key to the further analysis of the sample. To derive the sources' physical properties from dust continuum spectral energy distributions, we complement the 870~$\mu$m ATLASGAL data with publicly available Herschel/Hi-GAL \citep{molinari2010}, MSX \citep{Egan2003} and WISE \citep{Wright2010} data in order to cover a wavelength range from 8~$\mu$m to 870~$\mu$m. Furthermore, we show that our sample comprises a representative set of sources covering all of the important embedded evolutionary stages of massive star formation.

This paper is organized as follows: in Section\,\ref{sect:source_selection} we describe how the initial selection was made and briefly discuss how the source classification has evolved as new survey data have become available. In Section\,\ref{data.atlasgal} we explain how the photometry and spectral energy distributions (SEDs) have been obtained. In Section\,\ref{results} we derive physical parameters from the results of the SED modeling, while in Section\,\ref{discussion} we discuss the assignment of our sources to different stages of development and an evolutionary sequence. In Section\,\ref{conclusion} we summarize our findings and present a brief outlook on future work.



\section{Sample Selection and Classification}\label{sect:source_selection}

We selected 110 sources from the ATLASGAL compact source \cat that are likely in different evolutionary stages, using ancillary data to trace their star formation activity. We originally selected 102 sources as being the brightest sources at submillimeter wavelengths in 4 distinct groups as described by \citet{Giannetti2014}, using ancillary mid-infrared and radio data. During various ongoing follow-up projects 8 sources were added, and here we investigate the physical properties of this sample of 110 sources as a necessary reference for future studies. In the following sections we refer to this sample as the ATLASGAL ``Top100''.

\subsection{Classification}\label{sect.classification}

Since the initial classification of the sources by \citet{Giannetti2014}, new catalogs have become available, allowing for a refined classification that better reflects the physical properties of the sources in different evolutionary stages. In this paper, we reclassify our sample using four distinct phases of massive star formation. Three of these phases are drawn from the scheme originally outlined by \citet{Giannetti2014} and \citet{Csengeri2016} (i.e., mid-infrared dark/weak, mid-infrared bright and H\textsc{ii} regions). Here we refine this classification and extend it to include the youngest starless/pre-stellar phase based on the physical properties of the sample.

\begin{figure}[htp!]
\begin{center}
\includegraphics[width=0.98\linewidth]{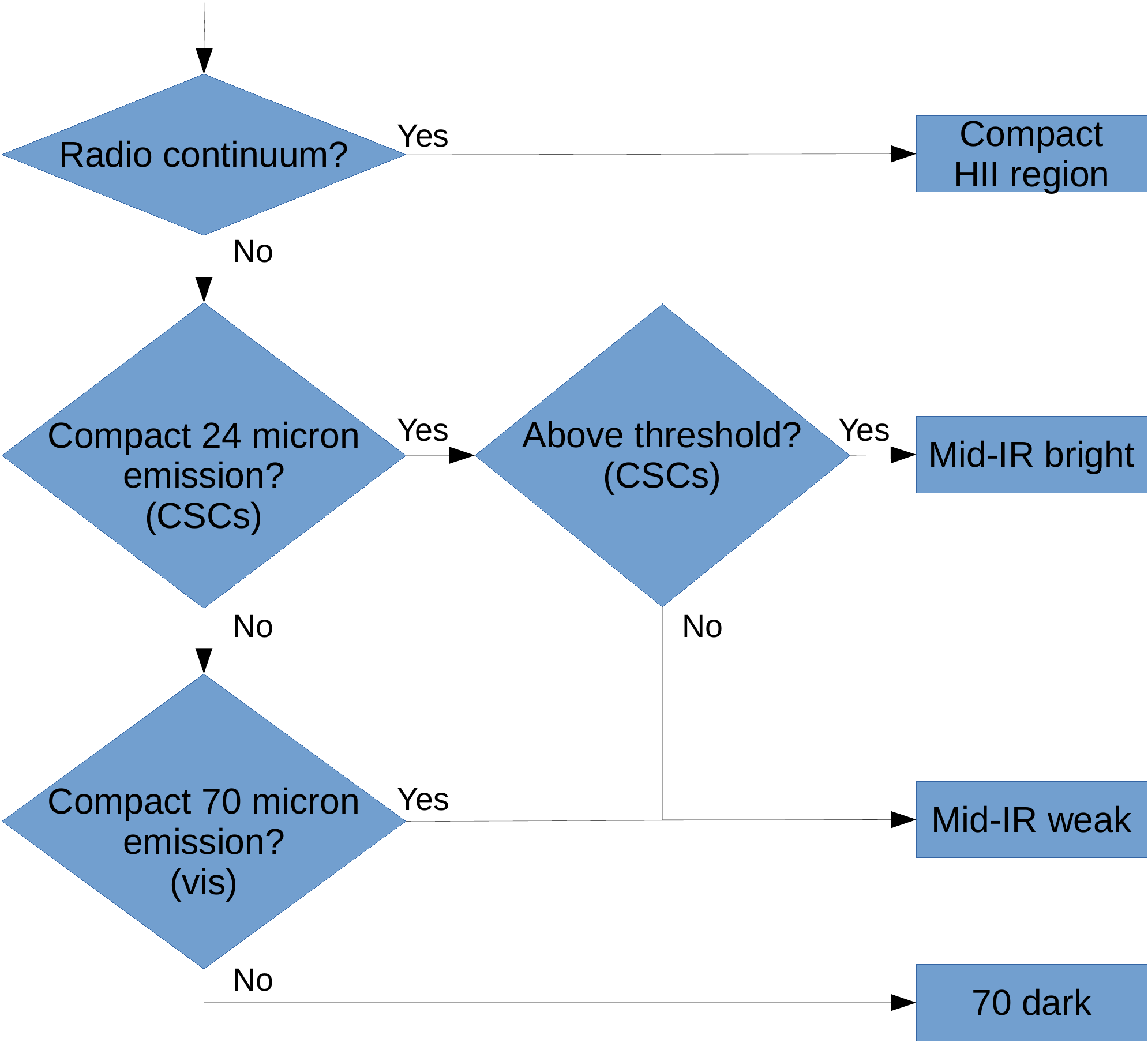}
\caption{Classification process for the Top100 sample.}
\label{fig.classification}
\end{center} 
\end{figure}

A schematic diagram of the classification process is shown in Figure \ref{fig.classification}. First, the sources are checked for radio continuum emission using CORNISH survey \citep{hoare2012,purcell2013}, the RMS survey (\citealt{urquhart2007,urquhart2009}) or the targeted observations towards methanol masers reported by \citet{walsh1999}. When radio continuum emission is found at either 4 or 8\,GHz within 10\arcsec\  of the ATLASGAL peak, the source is considered to be a compact \hii\ region.\footnote{Comparing the angular offsets between the peak of the submillimetre emission and a number of massive star formation tracers \citet{Urquhart2014} determined that $\sim$85\,per\,cent of compact embedded objected were located within 10\arcsec\ of each other and we have therefore adopted this radius as our association criterion.} Furthermore, we refined the distinction between mid-infrared weak and mid-infrared bright sources by inspecting the emission in the 21\,\mum\ MSX \citep{Price2001} and 24\,\mum\ MIPSGAL \citep{Carey2009} images. Looking for signs of star-formation, a source is considered mid-infrared bright if there is a compact mid-infrared source associated with the submm emission peak and the flux reported in the compact source catalogs \citep{Egan2003, Gutermuth2015} is above 2.6\,Jy, corresponding to a 4, 8 or 15\,M$_\odot$ star at 1, 4 and 20 kpc, respectively \citep{Heyer2016}. Accordingly a source is considered to be mid-infrared weak, when the compact mid-infrared emission in the 21/24\,\mum\ band is below 2.6\,Jy or no compact source is associated with the peak. Finally, the sources in the starless/pre-stellar phase are identified from a visual inspection of Hi-GAL PACS\,70\,\mum\ images. Showing no compact emission at 70\,\mum\ within 10\arcsec\ of the submm emission peak a source is considered 70\,\mum\ weak. This means that sources showing diffuse emission at 70\,\mum\, or compact 70\,\mum\ sources offset from the dust peak are still considered as 70\,\mum\ weak clumps.

In Figure\,\ref{fig.3color} we show example three color images for each class and below we briefly describe the observed feature of each phase and give the number of source classified in each:

\begin{figure}[htp!]
\begin{center}
\includegraphics[width=0.71\linewidth]{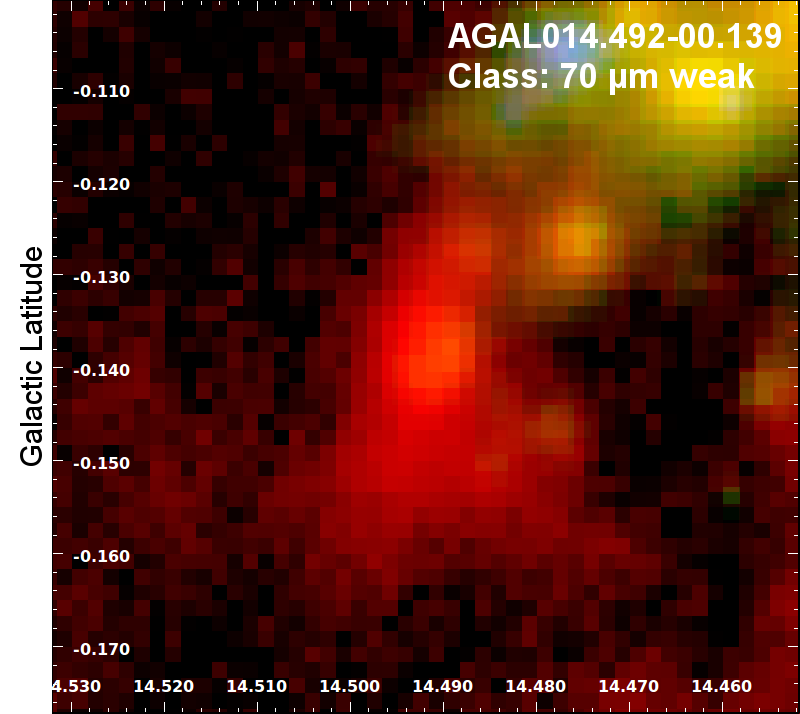}\\
\vspace{0.02cm}
\includegraphics[width=0.71\linewidth]{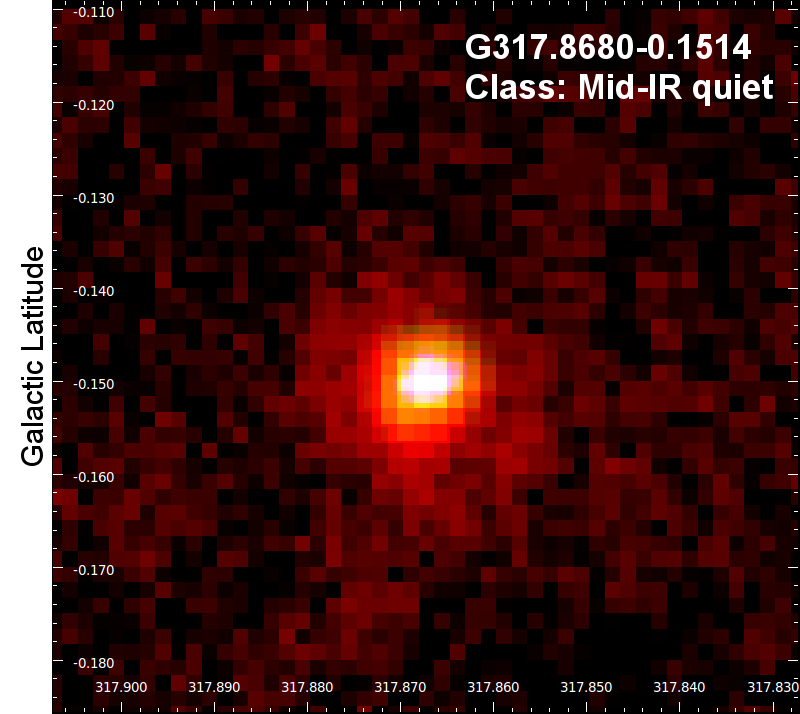}\\
\vspace{0.02cm}
\includegraphics[width=0.71\linewidth]{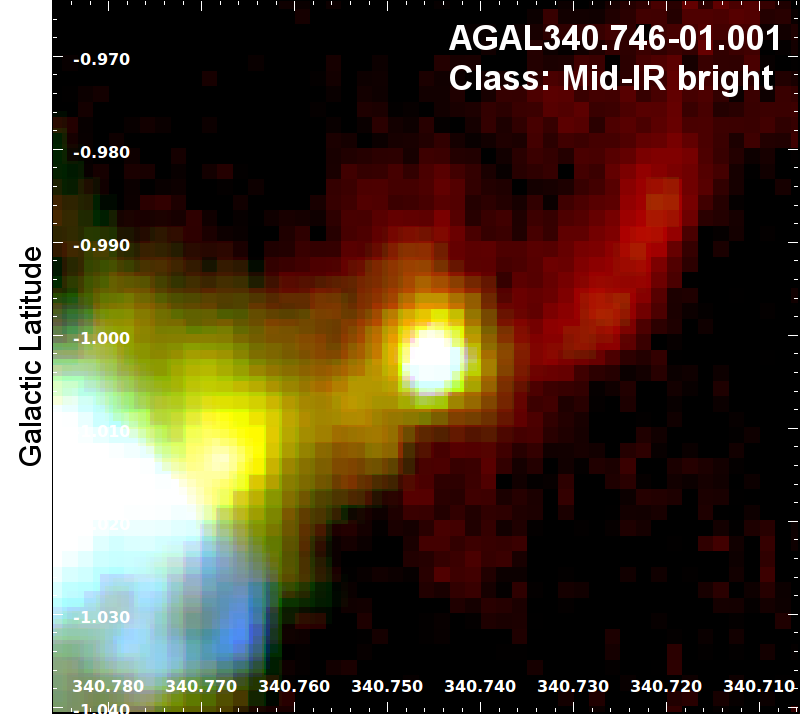}\\
\vspace{0.02cm}
\includegraphics[width=0.71\linewidth]{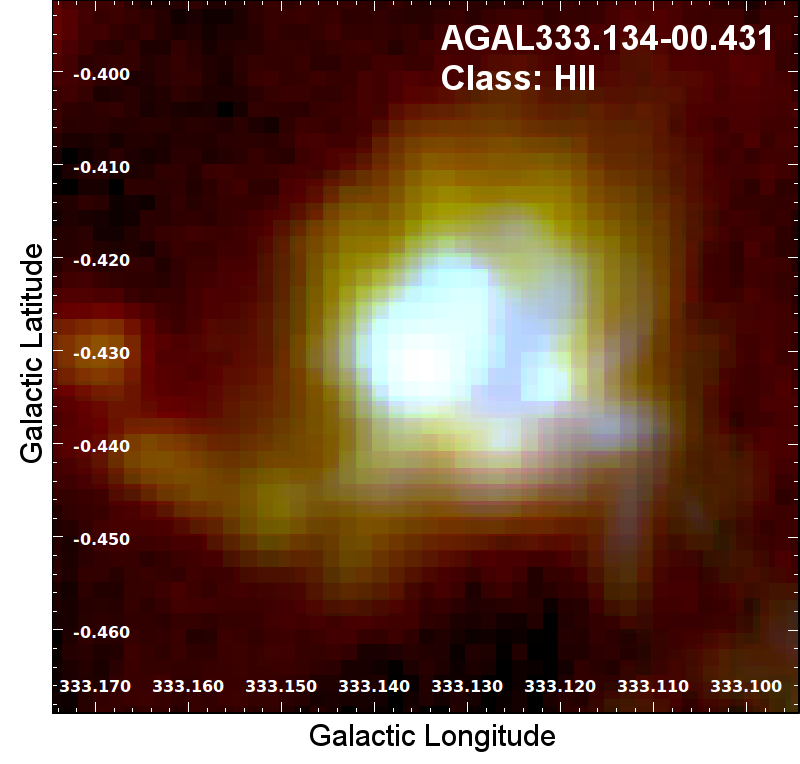}

   	\caption{Three color images of sample sources for each class sorted from youngest (top) to most evolved (bottom). Size: 5'$\times$5'; red: ATLASGAL\,870\,\mum; green: PACS\,160\,\mum; blue: PACS\,70\,\mum.}
	\label{fig.3color}
   \end{center} 
\end{figure}

\begin{itemize}

\item Starless/pre-stellar stage (16 sources): A quiescent phase, which represents the earliest stage of massive star formation (Fig.\,\ref{fig.3color}, top panel). These clumps are either mostly devoid of any embedded pointlike sources in the Hi-GAL 70 micron images or only show weak emission. They are likely to be the coldest and least luminous sources of the whole sample, and may already be collapsing but no protostellar object has yet formed \citep[e.g.][]{Motte2010,Elia2013,Traficante2015}. This class is called ``70\,\mum\ weak'' from here on.
 
 \item Protostellar stage (33 sources): compact point sources are clearly seen in the  70\,\mum\ Hi-GAL image and so protostellar objects are present (Fig.\,\ref{fig.3color}, second panel). The embedded 70\,\mum\ sources are either not associated with any mid-infrared counterparts within 10\arcsec\ of the peak emission or the associated compact emission is below our threshold of 2.6\,Jy, which indicates that the star formation is at an early stage and that the clumps are likely to be dominated by cold gas. We call this class ``Mid-IR weak'' throughout the paper.
 
  \item High-mass protostellar stage (36 sources): this phase is characterized by strong compact mid-infrared emission seen in 8 and 24 micron images and is one of the most active stages of massive star formation (Fig.\,\ref{fig.3color}, third panel). These clumps are likely undergoing collapse in the absence of a strong magnetic field (\citealt{Urquhart2014,Urquhart2015}), show signs of infall \citep{Wyrowski2016} and are likely to be driving strong outflows (Navarete et al. in prep.). Due to the infall, outflows, and already active young stellar objects, these sources are also likely to be significantly hotter than the sources in the quiescent phase, giving rise to the bright emission at mid-IR wavelengths (called ``Mid-IR bright'' from here on).
  
  \item Compact \hii\ region phase (25 sources): In the latest evolutionary phase we define here, the sources have just begun to disperse their natal envelope and are ionizing their local environment, creating compact H\textsc{ii} regions (Fig.\,\ref{fig.3color}, bottom panel). These sources are associated with bright mid-infrared emission and compact radio continuum emission arising from the ionization of their environment, making them easily distinguishable from the earlier evolutionary phases. We refer to this class as ``H\textsc{ii} regions'' from here on.
\end{itemize}

\subsection{Distances}\label{data.distance}

We have determined distances for 109 of the 110 sources of the sample. These distances have been drawn from the literature and supplemented with our own kinematic distances \citep{Wienen2015}. The distances given in Table\,\ref{tbl.shortdata} are based on those given by \citet{Giannetti2014} but incorporate the results of the latest maser parallax measurements reported by \citet{Reid2014}; this has resulted in the distances for 6 sources changing by $\sim$2\,kpc. For a small number of sources the distances adopted by \citet{Giannetti2014} disagreed with distances reported in the literature. Given that the distances extracted from the literature are kinematic in nature and have been determined using the same \hi\ data and comparable radial velocity measurements these variations likely result from slight differences in the method applied and the sensitivity and transition of the line surveys used in different studies.

There are 8 sources where the literature distance and distance adopted by \citet{Giannetti2014} disagree by more than a few kpc. For one source (i.e., AGAL330.954$-$00.182\footnote{We use source names from \citet{Contreras2013}}.) higher resolution \hi\ data have become available from targeted follow-up observations made toward UC\,\hii\ regions \citep{Urquhart2012} and we have adopted the distance obtained from the analysis of this data, which utilizes the presence or non-presence of H\textsc{i} self absorption as a distance indicator. We have examined the mid-infrared images for the other 7 sources and find 3 to be coincident with localized areas of extinction, which would suggest that a near distance is more likely (i.e., AGAL008.684$-$00.367, AGAL008.706$-$00.414, AGAL353.066$+$00.452). Examination of the \hi\ data towards AGAL316.641$-$00.087 reveals absorption at the source velocity and so we place this source at the near distance. The lack of an absorption feature in the \hi\ data or evidence of extinction would suggest that the three remaining sources are located at the far distance and indeed for three sources this is likely to be the case \citep[i.e. AGAL337.704$-$00.054 and AGAL337.176$-$00.032, AGAL337.258$-$00.101;][]{Giannetti2015}.

For one source (i.e. AGAL008.831$-$00.027) we did not obtain a distance, as its LSR velocity is  close to zero ($\mathrm{V}_\mathrm{LSR}=0.53\,\mathrm{km\,s^{-1}}$), hence rendering a kinematic distance unreliable.

\subsubsection{Source Distribution}

\begin{figure}
\begin{center}
\includegraphics[width=0.49\textwidth]{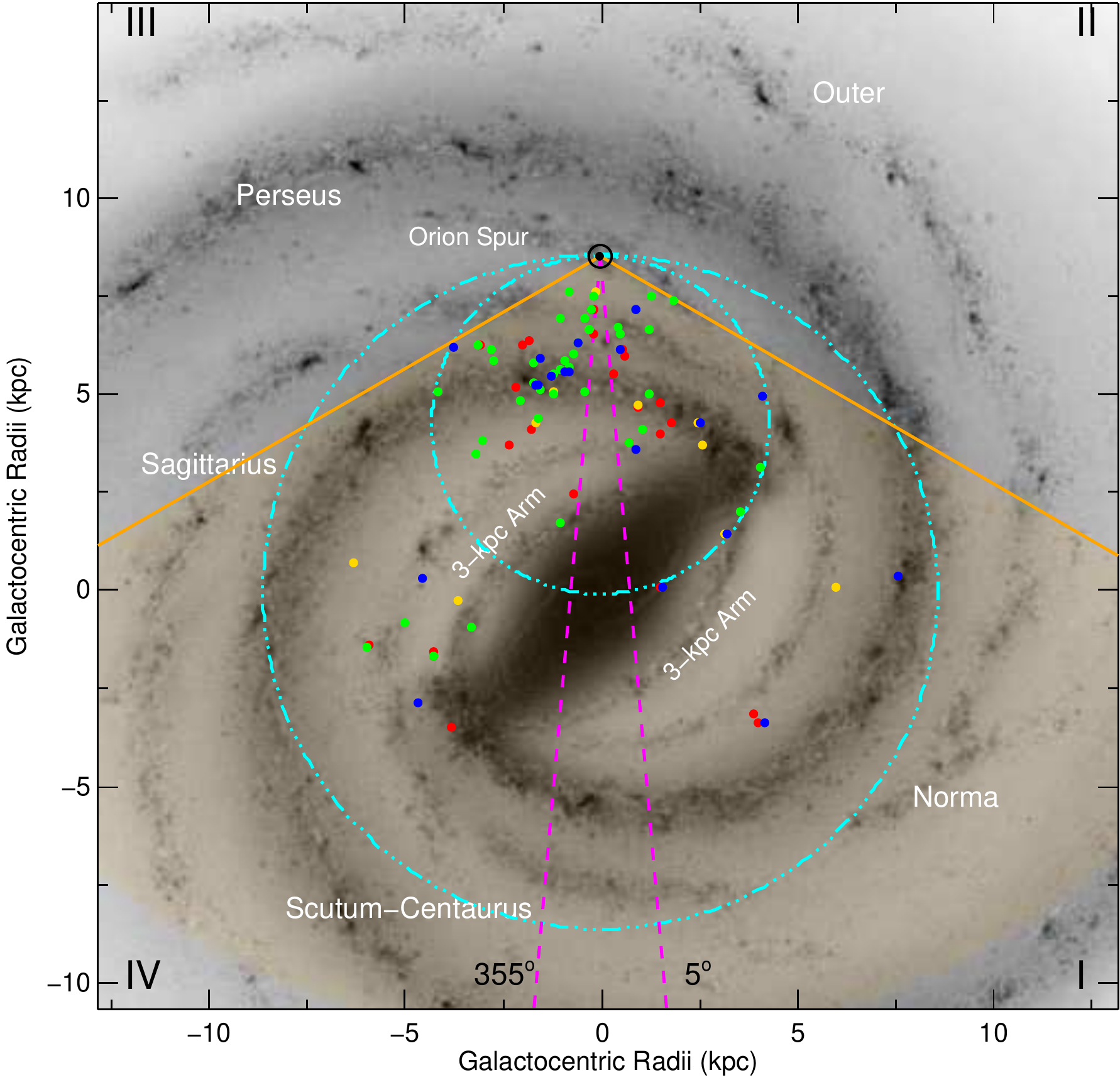}
\caption{\label{fig:galactic_distribution} Galactic distribution of the ATLASGAL Top100 sample. The positions of the \hii\ regions, mid-infrared bright, mid-infrared weak and 70~$\mu$m weak sources are indicated by the blue, green and red and yellow filled circles, respectively. The orange shaded area indicates the region of the Galactic plane covered by the ATLASGAL survey to a distance of 20\,kpc, within which the survey is complete for compact clumps with masses $>$1000\,\msun. The background image is a schematic of the Galactic disc as viewed from the Northern Galactic Pole (courtesy of NASA/JPL-Caltech/R. Hurt (SSC/Caltech)). The Sun is located at the apex of the wedge and is indicated by the $\odot$ symbol. The smaller of the two cyan dot-dashed circles represents the locus of tangent points, while the larger circle traces the solar circle. The spiral arms are labeled in white and Galactic quadrants are given by the roman numerals in the corners of the image. The magenta line shows the innermost region toward the Galactic centre where distances are not reliable.}
\end{center}
\end{figure}

In Fig.\,\ref{fig:galactic_distribution} we present the distribution of the Top100 on a schematic diagram of the Milky Way that includes many of the key elements of Galactic structure, such as the location of the spiral arms and the Galactic long and short bars \citep{Churchwell2009}.

\begin{figure}
\begin{center}
\includegraphics[width=0.49\textwidth, trim= 0 0 0 0]{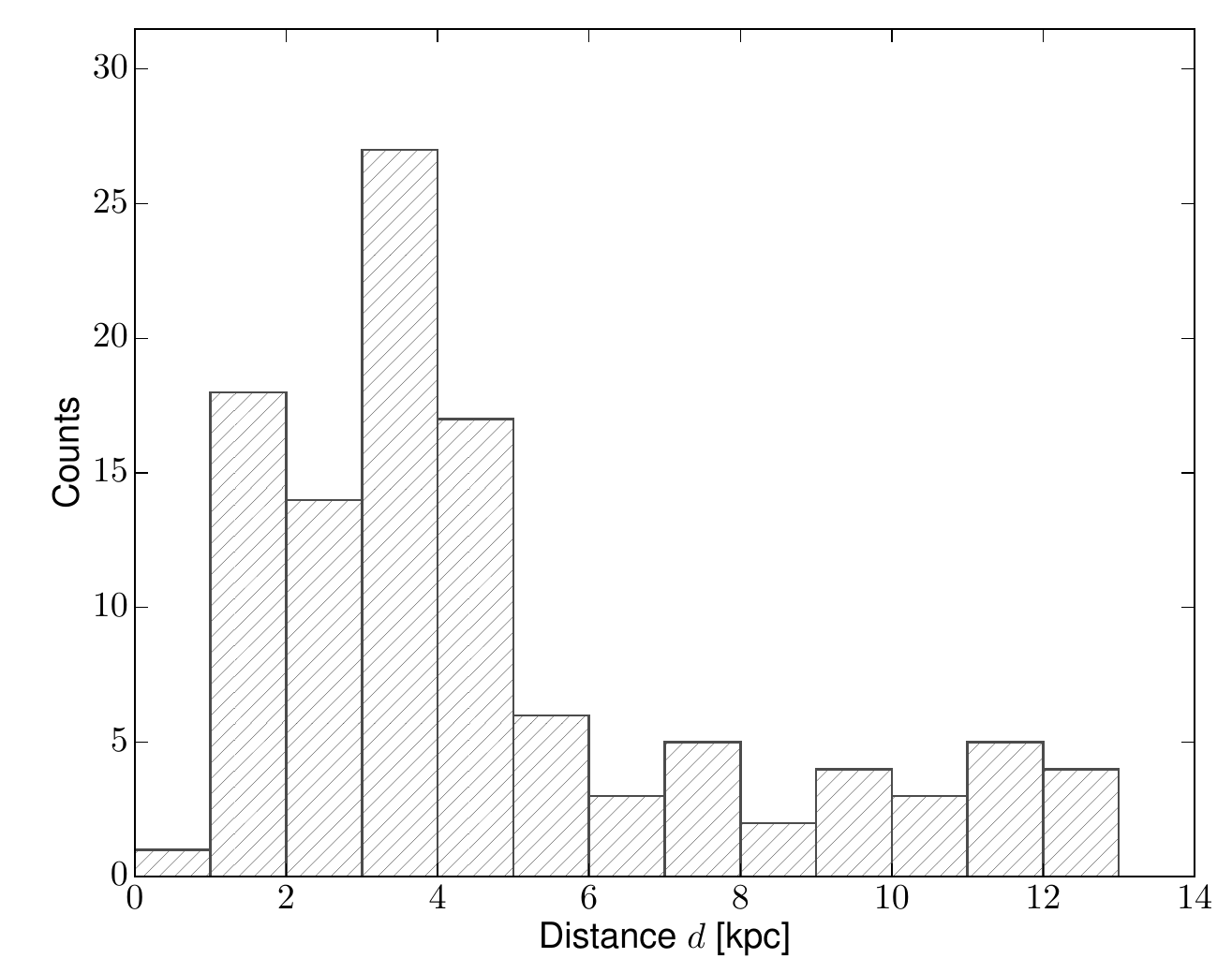}
\caption{\label{fig:distance_hist} Heliocentric distance distribution of all ATLASGAL Top100 sources for which a distance has been unambiguously determined. The bin size is 1\,kpc.}
\end{center}
\end{figure}

In Fig.\,\ref{fig:distance_hist} we show the distance distribution of the Top100 sources, which features two distinct peaks at $\sim$2\,kpc and $\sim$4\,kpc. These peaks correlate with the near parts of the \Sag\ and \Scu\ arms, respectively, as seen in Fig.\,\ref{fig:galactic_distribution}. We note that we only show the histogram for the full sample, as the sources in all four phases have a similar distance distribution. This is confirmed by Anderson-Darling tests\footnote{We chose the Anderson-Darling test over a Kolmogorov-Smirnov test, as it is more sensitive to subtle differences in smaller samples \citep[compare e.g.][]{Razali2011}.} \citep{Stephens1974} performed for the different classes yielding $p$-values higher than 0.02 for all classes, confirming the null hypothesis of two samples being drawn from the same distribution at the $3\sigma$ significance level. Second, there is a drop-off in the number of sources beyond a distance of 6\,kpc. As the whole sample is selected such that the brightest sources in the ATLASGAL catalog were selected for each category, a general distance bias toward closer distances and being located in the closest spiral arms is not surprising.

\section{Spectral energy distributions}\label{data.atlasgal}\label{data.higal}\label{sec.photometry}

The evolutionary scheme for high-mass star formation described in Sect.\,2 is primarily based on the visual examination of infrared images and the presence or absence of an infrared embedded point source and is therefore largely a phenomenological classification scheme. To robustly test this evolutionary sequence we need to determine the sample members' physical properties and search for trends in these parameters that support this. Key elements of this are the source temperatures and bolometric luminosities. These parameters can be determined from model fits to the source's spectral energy distributions (SEDs). 

To do this we have extracted multi-wavelength continuum data (mid-infrared to submillimeter wavelengths) and performed aperture photometry to reconstruct the dust continuum SEDs. The fluxes obtained from the photometry were fitted with a simple model to derive the dust temperature and integrated flux, which were subsequently used to estimate the bolometric luminosities and total masses. The data used and methods supplied will be described in detail in the following subsections.

\subsection{Dust continuum surveys}\label{sect.continuumdata}

The mid-infrared wavelength regime is covered using archival data from either the MSX \citep{Egan2003} or the WISE \citep{Wright2010} surveys. The far-infrared spectrum is covered by the two Herschel \citep{Pilbratt2010} instruments SPIRE \citep{Griffin2010} and PACS \citep{Poglitsch2010}, covering the wavelength range from 70\,$\mu$m up to 500\,$\mu$m with 5 different bands centered at 70, 160, 250, 350 and 500\,$\mu$m, respectively. These bands are especially well suited to determine the peak of the SED for cool dust. The level 2.5 maps from the Herschel Infrared Galactic Plane Survey \citep[Hi-GAL,][]{molinari2010} were retrieved from the Herschel Science Archive (HSA)\footnote{\url{http://herschel.esac.esa.int/Science_Archive.shtml}}, and were downloaded in version 11 of the Standard Product Generation (SPGv11) pipeline. To obtain the flux for the longest wavelength (submillimeter) entry of the SED, the 870~$\mu$m ATLASGAL maps are used.

For all of these data sets we extracted $5\times5$ arcminute sized images centered on the source positions. Where necessary the map units were converted to Jy\,pixel$^{-1}$ while keeping the original resolution. For MSX and WISE images this meant converting from W\,m$^{-2}$\,sr$^{-1}$ and digital numbers (DN) to Jy\,pixel$^{-1}$, respectively.\footnote{For more information see \url{http://irsa.ipac.caltech.edu/applications/MSX/MSX/imageDescriptions.htm} and \url{http://wise2.ipac.caltech.edu/docs/release/prelim/expsup/sec2_3f.html\#tbl1}}

\subsection{Aperture photometry}

\begin{figure*}[tp!]
\centering \includegraphics[angle=0,width=17.8cm]{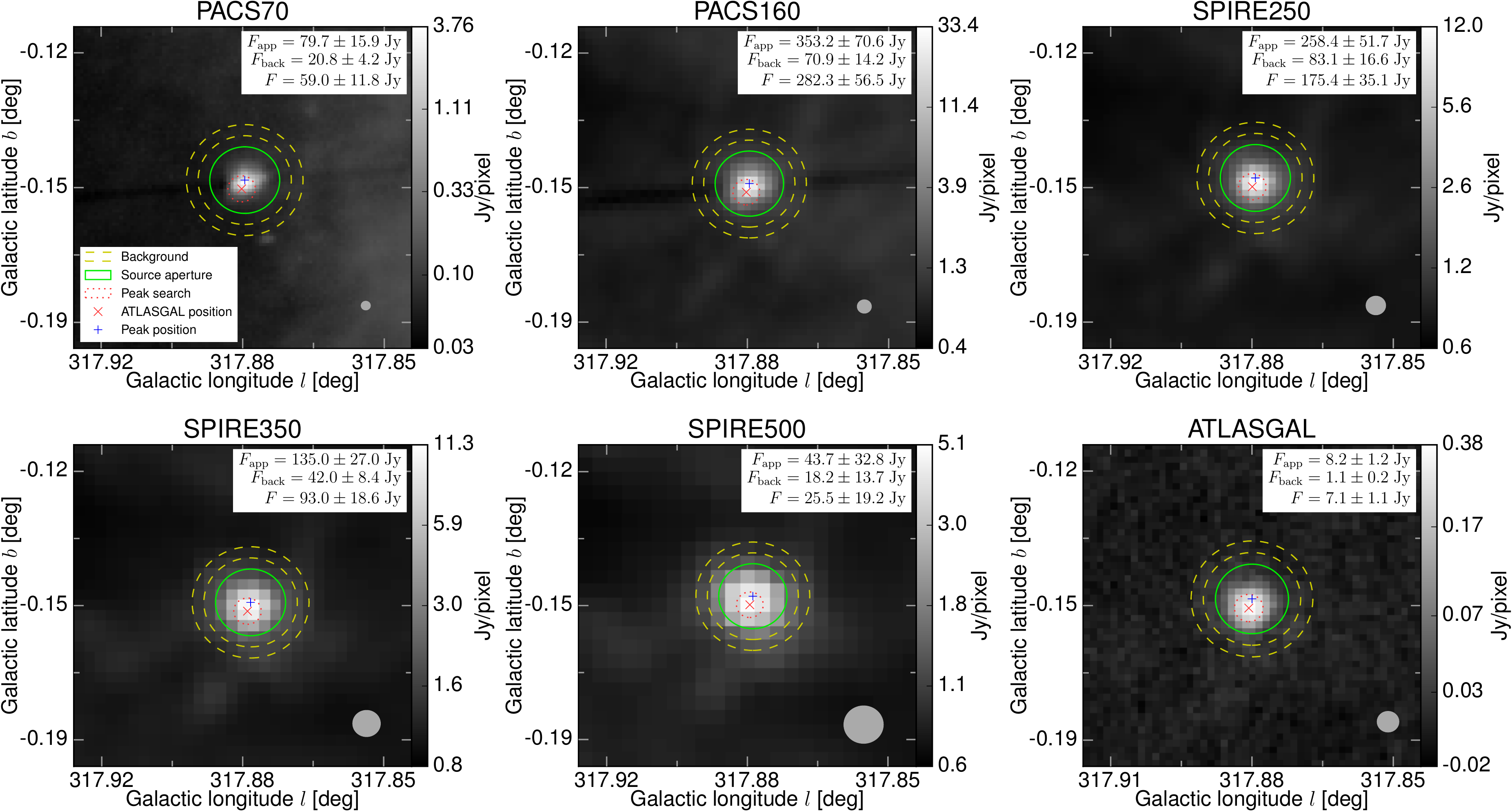}
   	\caption{Images of a single source (AGAL317.867$-$00.151) seen in the different bands, showing the aperture (green circle), background annulus (yellow dashed circle), ATLASGAL and peak flux position (red and blue crosses, respectively) and the peak pixel search area (red dotted circle). The beamsize is indicated as a grey circle in the lower right.}
	\label{fig.photometry}
\end{figure*}

Aperture photometry was used to extract fluxes in a consistent way from the mid-infrared and submillimetre maps. The flux density $F_\mathrm{aper}$ was integrated over a circular aperture centered on the peak flux pixel position. The peak position was identified in either the 250\,$\mu$m, 160\,$\mu$m or the 870\,$\mu$m band, depending on whether the band was suffering from saturation, following the order of the bands as previously stated. Assuming a Gaussian shaped source brightness profile with a FWHM as reported in \citet{Csengeri2014}, we use an aperture size for each source with a radius of $3 \sigma$, where $\sigma = \mathrm{FWHM} / (2\sqrt{2\ln{2}})$ to obtain most of the flux ($> 99$\%) of a source. With a minimum aperture size of 55.1\arcsec\ the apertures were also selected such that they are resolved by the lowest resolution data (i.e. 36.6\arcsec\ for SPIRE 500~$\mu$m). We performed tests that revealed that smaller aperture sizes underestimate the flux, while a larger aperture size might cut into some other emission nearby, as the source confusion for some of the clumps of our sample is significant, since these are associated with some of the most active star forming sites in the Galaxy. Subsequently, the background flux density $F_\mathrm{bg}$ obtained from the median pixel value of a circular annulus around the same center position as the aperture was subtracted from the aperture flux to obtain the background-corrected source flux $F$.

When fluxes could be successfully extracted from the MSX maps, we preferred these over the WISE fluxes, as the MSX maps have a resolution of 18\arcsec\ similar to the longer wavelength bands (e.g. 19.2\arcsec\ for ATLASGAL). In addition, MSX suffers less from saturation than the WISE data (compare e.g. Cutri et al. 2012, Chapter VI.3 and Robitaille et al. 2007, Table 1). In cases where the photometry extracted from the MSX images is a non-detection we turned to the higher sensitivity WISE data to determined the flux.

An example for the photometry extraction in the far-infrared to submm bands is presented in Fig.\,\ref{fig.photometry}, showing the position and size of the aperture and annulus used to estimate the background contribution for this source. In total, we found one source to be saturated at 70\,$\mu$m, one source at 160\,$\mu$m, 39 sources are saturated at 250\,$\mu$m and 14 sources at 350\,$\mu$m. Note that when one or more bands in the far-infrared to submm range suffers from saturation, always the SPIRE\,250\,\mum\ band is one of them. Moreover we note that the fraction of sources being saturated in at least one band increases through the classes: from no source suffering from saturation for the 70\,\mum\ weak sample, 6 sources (i.e. 16\%) of the mid-infrared weak class, 10 in the mid-infrared bright sample (i.e. 31\%) and 23 (i.e. 92\%) for the HII regions. For the photometry, the saturated pixels were set to the maximum pixel value of the image, and the flux is only taken as a lower limit for the SED fitting (see Section \ref{method.sed}). To ensure good fitting results we require at least 3 bands in the far-IR to submm regime to be free from saturation.

\begin{figure}
\centering
\includegraphics[width=0.95\linewidth]{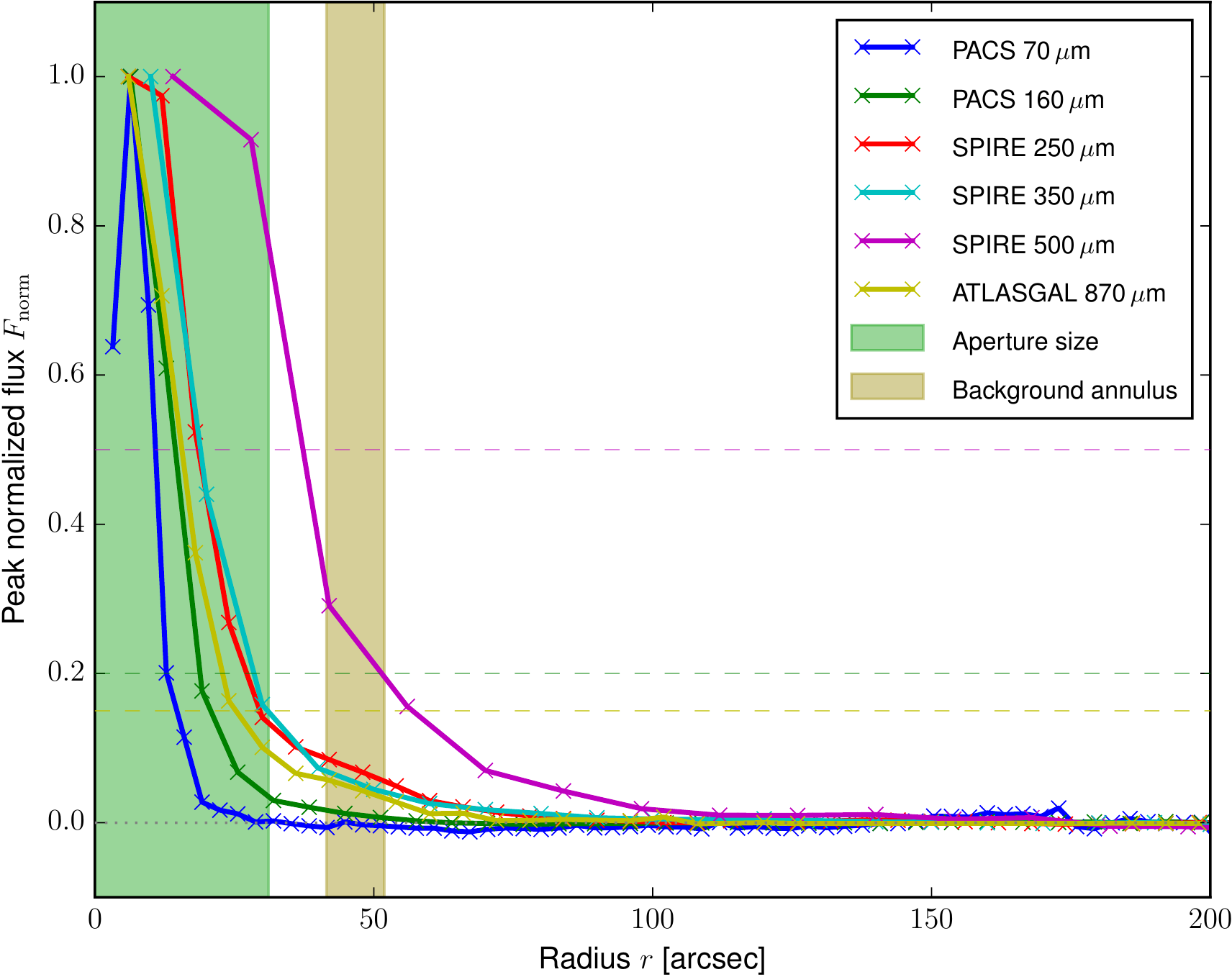}
   	\caption{Emission profiles of a single source (AGAL317.867$-$00.151) for the different bands, showing the aperture (green shaded area) and background annulus (ochre shaded area). The horizontal dashed lines shown in yellow, magenta and green indicate the flux uncertainties we assume for the ATLASGAL, SPIRE 500\,\mum\ and remaining Herschel bands, respectively. Note that the 500\,\mum\ band traces the more extended emission and we therefore assume a measurment uncertainty of 50\% in this band when fitting the SEDs.}
	\label{fig.structure}
\end{figure}

In Figure \ref{fig.structure} we show the emission profiles of the submm bands for a single source (AGAL317.867$-$00.151). To better emphasize the structure and make it comparable between the different bands, for each wavelength we subtracted the median background emission as determined far away (i.e. 5 times the aperture size) from the source where the profile gets flat and then normalized the flux to the peak emission in that band. As can be seen from the plot, the SPIRE 500\,\mum\ band traces the more extended emission, falling off to the background plateau level rather slowly compared to the other bands. Conversely, the other bands trace the peak of the emission within the aperture, but the background aperture still cuts into some local plateau associated with the cloud. The contribution of this is lower than 10\,percent for all bands except the SPIRE 500\,\mum. Accordingly we assume a rather large measurement uncertainty for the 500\,\mum\ band of 50\,percent to account for the added uncertainty from the background correction as well as to account for the the large pixel size of 15\arcsec\ in this band. For the other Herschel bands we assume a measurement uncertainty of 20\% bands and a measurement uncertainty of 15\% for the flux estimate of the ATLASGAL band. Finally, the absolute calibration uncertainties are added to the intrinsic measurement error in quadrature to obtain the uncertainties of the (non-saturated) flux densities.


\subsection{SED Models}\label{method.sed}\label{sect.sed}

\begin{figure}[tp!]
\begin{center}
\includegraphics[width=0.83\linewidth]{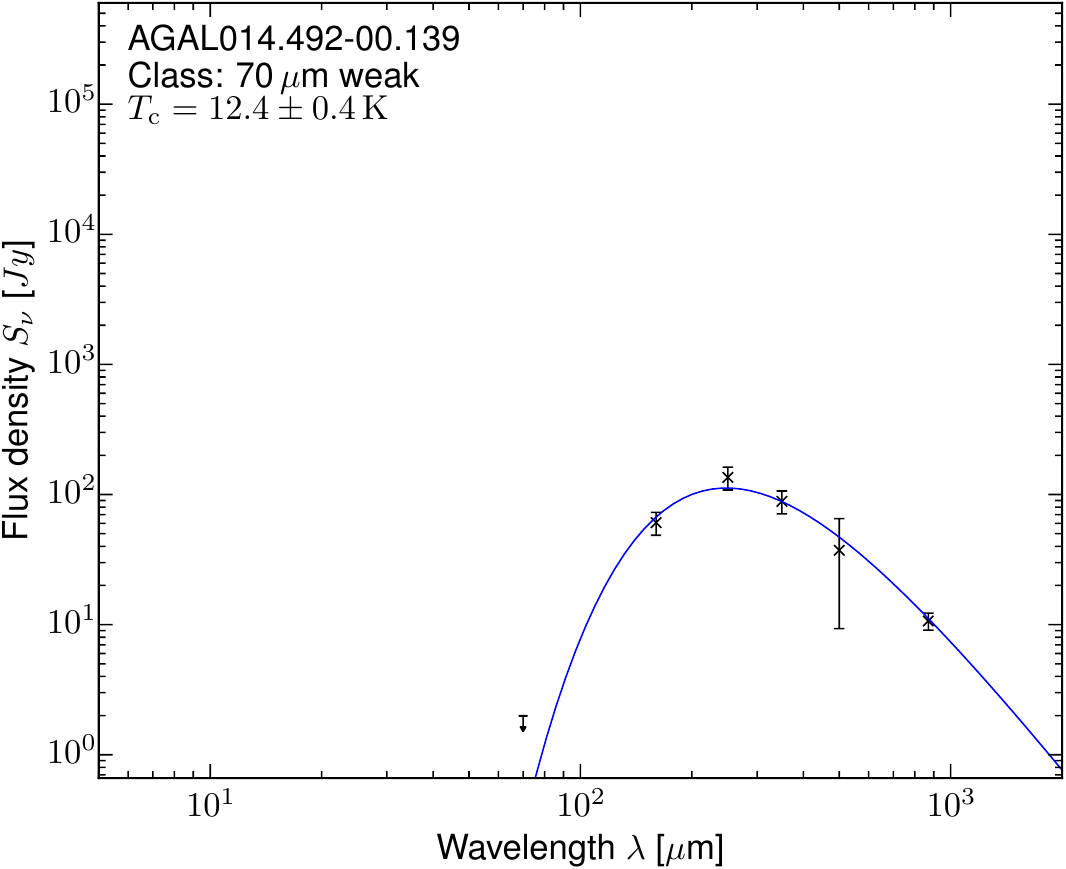}\\
\vspace{-0.665cm}
\includegraphics[width=0.83\linewidth]{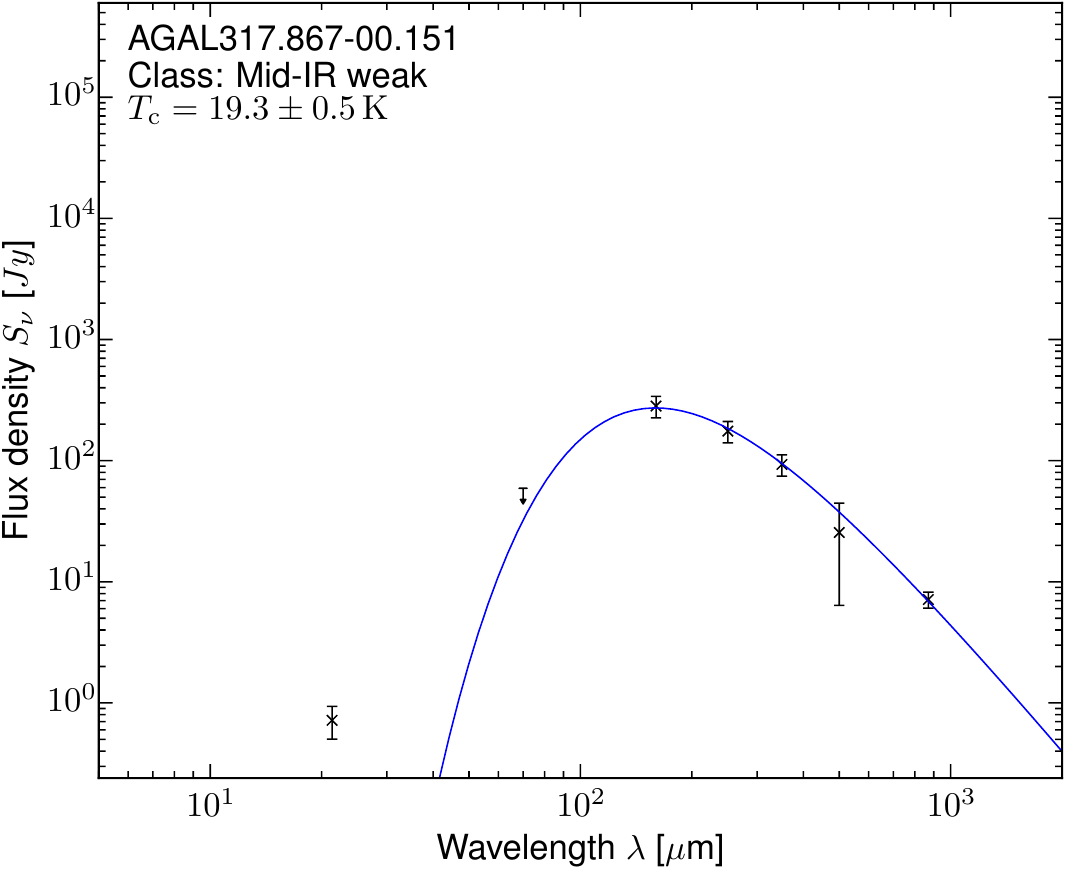}\\
\vspace{-0.665cm}
\includegraphics[width=0.83\linewidth]{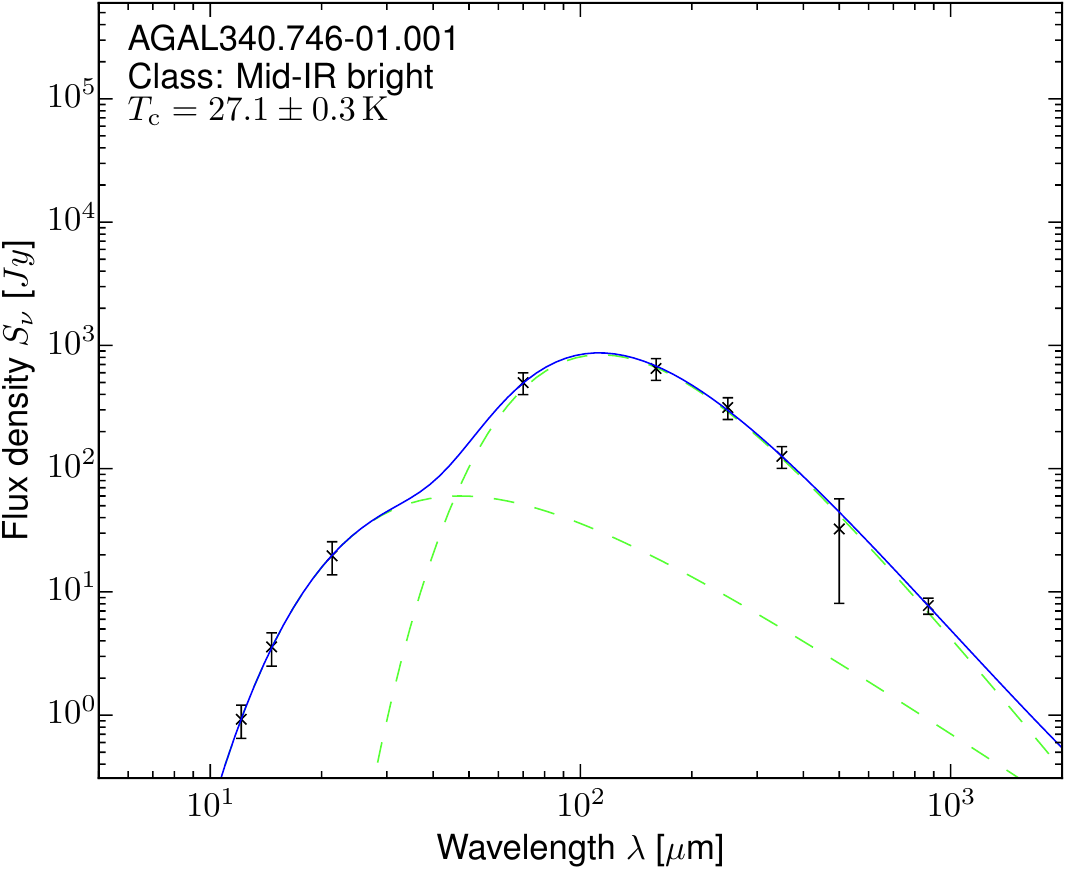}\\
\vspace{-0.665cm}
\includegraphics[width=0.83\linewidth]{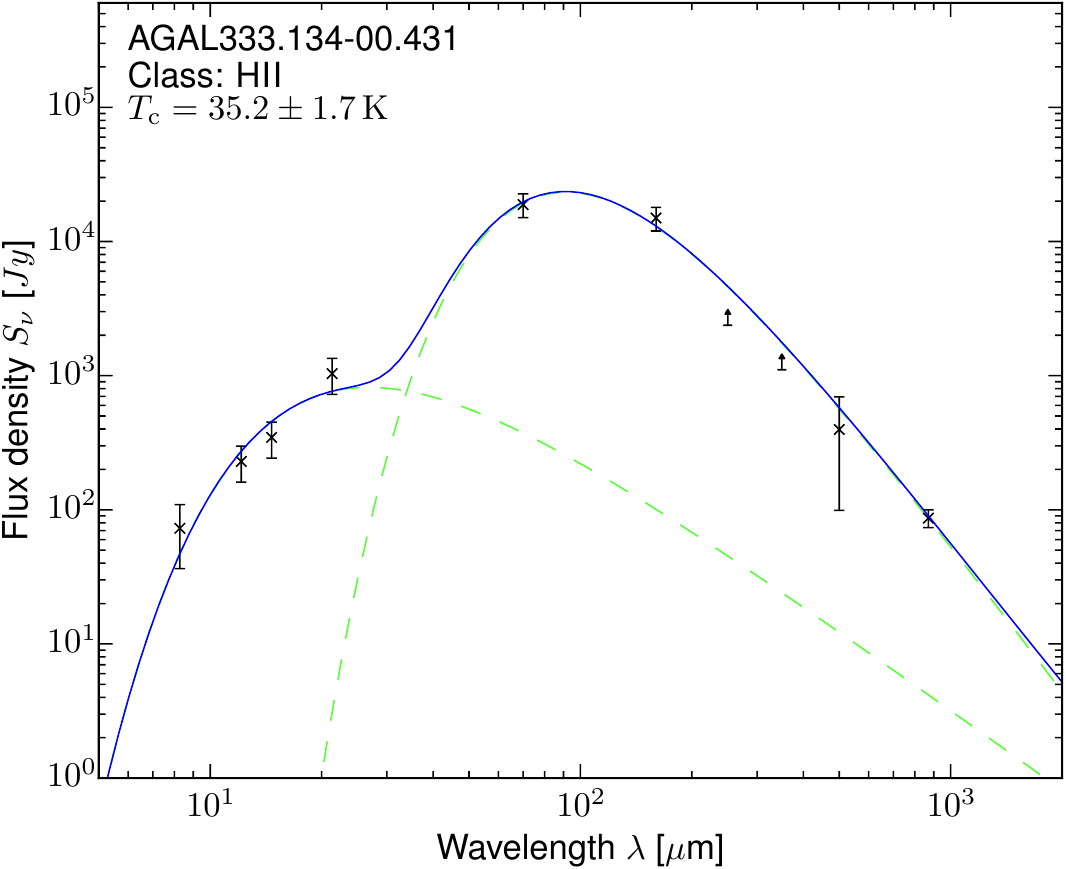}

   	\caption{Sample SEDs for all four evolutionary classes sorted from youngest to most evolved sources from top to bottom. A single component greybody is fitted for the upper two, whereas a two-component fit is used for the later two sources. The green dashed lines in the lower panels show the greybody and blackbody components of the fit.}
	\label{fig.sed}
   \end{center} 
\end{figure}

The multi-wavelength photometic data obtained were fitted using standard methods to obtain the dust temperature and bolometric luminosity of each source.

For sources for which no mid-infrared emission is detected (or a flux density measurement is available for only one mid-infrared band), we follow the method of \citet{Elia2010} and \citet{Motte2010}, fitting a single greybody model to the flux densities measured for the cold dust envelope. When fitting only the cold component, we use the 70~$\mu$m flux density as an upper limit for the dust emission. Assuming the measured flux density at 70~$\mu$m is always a combination from the cold dust envelope and a young embedded object, taking it as an upper limit for the dust emission avoids overestimating the dust temperature. For the fitting of the greybody we leave the dust spectral index $\beta$ fixed to a value of 1.75, computed as the mean value from the dust opacities over all dust models of \citet{Ossenkopf1994} for the submm regime. This also facilitates comparison of the envelope masses calculated here with previous results presented in the literature \citep[e.g.][]{Thompson2004, NguyenLuong2011}.

Where at least two flux density measurements are available at the (different) shorter wavelengths, a two-component model consisting of a blackbody and a greybody is fitted to the SED \citep[e.g.][]{Beuther2010}. As the 70~$\mu$m flux has a significant influence on the general quality of the fit \citep{Mottram2011a} and hence on the temperature estimate of the cold dust, the blackbody being added to the model quantitatively constrains the contribution of a deeply embedded, hot component to the 70~$\mu$m flux density. Note that for the mid-infrared weak sources a flux density measurement at 21\mum\ might be available, but as at least two measurements are necessary for the hot component to be fitted, only the greybody is taken into account, emphasizing the importance of the 70~$\mu$m flux being taken as an upper limit in such cases (compare Figure \ref{fig.sed}, second tile).

We were able to fit the SEDs for all 110 sources in our sample. A single greybody component fit was used to model the emission for 38 of the  mid-infrared dark sources while the remaining 73 sources were fitted using the two component fit of a greybody plus the blackbody as described in the previous section. In Fig.\,\ref{fig.sed} we show sample SEDs for all four classes and the model fits to the data. The top panels show a single component greybody fit to a 70\,\mum\ weak and to a mid-infrared weak source whilst the lower panels show the results of a two component fit to sources of the class of mid-infrared weak sources and the \hii\,regions. A comparison of the two-component model with the radiative transfer model used by \citet{Robitaille2007a} is presented in Sect.\,\ref{results.luminosities}.

\begin{table*}
\begin{center}
\caption{Source parameters for the first 15 sources.\label{tbl.shortdata}}
\begin{tabular}{ccccccccccccc}
\hline\hline
Name & $d$ & $d_\mathrm{ref}$ & $V_\mathrm{lsr}$ & $l_\mathrm{app}$ & $b_\mathrm{app}$ & $D_\mathrm{app}$ & Class & $T$ & $\Delta T$ & $L_\mathrm{bol}$ & $M_\mathrm{clump}$ & $\alpha_\mathrm{vir}$ \\
 & (kpc) &  & (km/s) & ($\deg$) & ($\deg$) & ($''$) &  & (K) & (K) & $(\mathrm{L}_\odot)$ & $(\mathrm{M}_\odot)$ &  \\
\hline

AGAL006.216$-$00.609 & $2.9$ & $(22)$ & 18.5 & 6.216 & $-$0.609 & 71.6 & IRw & 16.0 & 0.2 & $7.3\times10^{2}$ & $4.5\times10^{2}$ & $-$ \\
AGAL008.671$-$00.356 & $4.8$ & $(19)$ & 35.1 & 8.668 & $-$0.355 & 69.1 & H\textsc{ii} & 26.3 & 1.4 & $8.6\times10^{4}$ & $3.0\times10^{3}$ & $-$ \\
AGAL008.684$-$00.367 & $4.8$ & $(19)$ & 38.0 & 8.682 & $-$0.367 & 66.1 & IRw & 24.2 & 0.2 & $2.7\times10^{4}$ & $1.4\times10^{3}$ & $0.69$ \\
AGAL008.706$-$00.414 & $4.8$ & $(19)$ & 39.4 & 8.704 & $-$0.412 & 92.1 & IRw & 11.8 & 0.3 & $5.0\times10^{2}$ & $1.6\times10^{3}$ & $0.22$ \\
AGAL008.831$-$00.027 & $-$ & $-$ & 0.5 & 8.831 & $-$0.027 & 70.9 & IRw & 24.1 & 1.5 & $-$ & $-$ & $-$ \\
AGAL010.444$-$00.017 & $8.6$ & $(1)$ & 75.9 & 10.442 & $-$0.016 & 65.5 & IRw & 20.7 & 0.3 & $1.1\times10^{4}$ & $1.6\times10^{3}$ & $0.40$ \\
AGAL010.472+00.027 & $8.6$ & $(1)$ & 67.6 & 10.472 & +0.028 & 55.1 & H\textsc{ii} & 30.5 & 2.4 & $4.6\times10^{5}$ & $1.0\times10^{4}$ & $0.22$ \\
AGAL010.624$-$00.384 & $5.0$ & $(1)$ & $-$2.9 & 10.623 & $-$0.382 & 65.0 & H\textsc{ii} & 34.5 & 3.6 & $4.2\times10^{5}$ & $3.7\times10^{3}$ & $0.42$ \\
AGAL012.804$-$00.199 & $2.4$ & $(28)$ & 36.2 & 12.805 & $-$0.197 & 84.4 & H\textsc{ii} & 35.1 & 2.3 & $2.4\times10^{5}$ & $1.8\times10^{3}$ & $0.78$ \\
AGAL013.178+00.059 & $2.4$ & $(25)$ & 50.4 & 13.176 & +0.062 & 87.6 & 70w & 24.2 & 0.8 & $8.3\times10^{3}$ & $3.6\times10^{2}$ & $0.96$ \\
AGAL013.658$-$00.599 & $4.5$ & $(22)$ & 48.4 & 13.656 & $-$0.596 & 67.0 & IRb & 27.4 & 1.3 & $2.0\times10^{4}$ & $5.6\times10^{2}$ & $0.53$ \\
AGAL014.114$-$00.574 & $2.6$ & $(3)$ & 20.8 & 14.112 & $-$0.572 & 81.0 & IRw & 22.4 & 0.8 & $3.1\times10^{3}$ & $3.5\times10^{2}$ & $0.65$ \\
AGAL014.194$-$00.194 & $3.9$ & $(3)$ & 39.2 & 14.194 & $-$0.191 & 72.0 & IRw & 18.2 & 0.6 & $2.7\times10^{3}$ & $8.2\times10^{2}$ & $0.50$ \\
AGAL014.492$-$00.139 & $3.9$ & $(3)$ & 39.5 & 14.491 & $-$0.137 & 89.3 & 70w & 12.4 & 0.4 & $7.5\times10^{2}$ & $1.9\times10^{3}$ & $0.55$ \\
AGAL014.632$-$00.577 & $1.8$ & $(27)$ & 18.5 & 14.631 & $-$0.576 & 83.5 & IRw & 22.5 & 0.4 & $2.7\times10^{3}$ & $2.5\times10^{2}$ & $0.84$ \\
\hline
\end{tabular}
\tablefoot{The Columns are as follows: Name: the ATLASGAL catalog source name. $d$: distance. $d_\mathrm{ref}$ Distance reference (see list of references below table). $V_\mathrm{lsr}$: source velocity. $l_\mathrm{app}$: Galactic longitude of aperture center. $b_\mathrm{app}$: Galactic latitude of aperture center. $D_\mathrm{app}$: aperture diameter. Class: class of the source (H\textsc{ii}: H\textsc{ii} region, IRb: mid-infrared bright, 24d: mid-infrared weak (c: confused within the aperture), 70d: 70~$\mu$m weak). $T$: Dust temperature $\Delta T$: Error of the dust temperature. $L_\mathrm{bol}$: Bolometric luminosity. $M_\mathrm{clump}$: Clump mass. $\alpha_\mathrm{vir}$: virial parameter.\\All data will be available on the ATLASGAL website (\url{http://atlasgal.mpifr-bonn.mpg.de}) and CDS (\url{http://cdsweb.u-strasbg.fr}).}
\end{center}Distance references: (1): \citet{Sanna2014}; (2): \citet{Zhang2014}; (3): \citet{Giannetti2014}; (4): \citet{Zhang2013}; (5): \citet{Brunthaler2009}; (6): \citet{Giannetti2015}; (7): \citet{Giannetti2014}; (8): \citet{Sato2014}; (9): \citet{Xu2011}; (10): \citet{Sato2010}; (11): \citet{Urquhart2012}; (13): \citet{moises2011}; (14): \citet{busfield2006}; (15): \citet{snell1990}; (16): \citet{Davies2012}; (17): Tangent Point; (18): \citet{roman2009}; (19): \citet{Green2011}; (20): \citet{Kurayama2011}; (21): \citet{Zhang2009}; (22): \citet{Wienen2015}; (23): \citet{Urquhart2014a}; (24): \citet{Xu2009}; (25): \citet{Immer2013}; (26): \citet{caswell1975}; (27): \citet{Wu2014}; (28): \citet{immer2012} \\Note: Only the data for 15 sources are given here. The full table can be found in Table \ref{tbl.fulldata}.
\end{table*}

\begin{table*}
\begin{center}
\caption{Overview of the different classes and their parameters. Given are the mean, median, minimum, maximum, standard deviation and standard error for each parameter.}
\label{tbl.classdata}
\begin{tabular}{l|cccccccc}
\hline
Class/Parameter & Mean & Median & Min & Max & Standard deviation & Standard error  \\
\hline
\\
\textbf{70\,\mum\ weak} \\
\hline
$T_\mathrm{dust}/\mathrm{K}$ & $16.4$ & $16.9$ & $10.7$ & $24.24$ & $3.35$ & $0.8$ \\
$L_\mathrm{bol}/\mathrm{L}_\odot$ & $3.2\times10^{3}$ & $2.2\times10^{3}$ & $4.3\times10^{2}$ & $9.1\times10^{3}$ & $2.9\times10^{3}$ & $7.3\times10^{2}$ \\
$M_\mathrm{env}/\mathrm{M}_\odot$ & $2.1\times10^{3}$ & $1.3\times10^{3}$ & $1.2\times10^{2}$ & $1.0\times10^{4}$ & $2.6\times10^{3}$ & $6.4\times10^{2}$ \\
$L_\mathrm{bol}/M_\mathrm{env} \cdot \mathrm{L}_\odot / \mathrm{M}_\odot$ & $3.4$ & $2.6$ & $0.2$ & $22.6$ & $5.1$ & $1.3$ \\
$r/\mathrm{pc}$ & $1.1$ & $1.0$ & $0.4$ & $2.74$ & $0.63$ & $0.2$ \\
$N_\mathrm{H_2}/\mathrm{cm}^{-2}$ & $5.6\times10^{22}$ & $4.8\times10^{22}$ & $2.4\times10^{22}$ & $1.3\times10^{23}$ & $2.7\times10^{22}$ & $6.8\times10^{21}$ \\
\\
\textbf{Mid-IR weak} \\
\hline
$T_\mathrm{dust}/\mathrm{K}$ & $19.9$ & $21.4$ & $11.7$ & $26.18$ & $4.26$ & $0.7$ \\
$L_\mathrm{bol}/\mathrm{L}_\odot$ & $2.2\times10^{4}$ & $5.7\times10^{3}$ & $5.7\times10^{1}$ & $2.2\times10^{5}$ & $4.1\times10^{4}$ & $7.1\times10^{3}$ \\
$M_\mathrm{env}/\mathrm{M}_\odot$ & $2.2\times10^{3}$ & $1.3\times10^{3}$ & $1.8\times10^{1}$ & $1.1\times10^{4}$ & $2.5\times10^{3}$ & $4.4\times10^{2}$ \\
$L_\mathrm{bol}/M_\mathrm{env} \cdot \mathrm{L}_\odot / \mathrm{M}_\odot$ & $8.3$ & $7.2$ & $0.3$ & $25.7$ & $7.3$ & $1.3$ \\
$r/\mathrm{pc}$ & $1.0$ & $0.7$ & $0.2$ & $2.49$ & $0.64$ & $0.1$ \\
$N_\mathrm{H_2}/\mathrm{cm}^{-2}$ & $1.1\times10^{23}$ & $7.6\times10^{22}$ & $3.7\times10^{22}$ & $7.6\times10^{23}$ & $1.3\times10^{23}$ & $2.3\times10^{22}$ \\
\\
\textbf{Mid-IR bright} \\
\hline
$T_\mathrm{dust}/\mathrm{K}$ & $28.1$ & $28.2$ & $21.9$ & $34.53$ & $3.65$ & $0.6$ \\
$L_\mathrm{bol}/\mathrm{L}_\odot$ & $5.0\times10^{4}$ & $2.2\times10^{4}$ & $9.9\times10^{2}$ & $2.5\times10^{5}$ & $5.4\times10^{4}$ & $9.1\times10^{3}$ \\
$M_\mathrm{env}/\mathrm{M}_\odot$ & $1.5\times10^{3}$ & $6.0\times10^{2}$ & $1.8\times10^{1}$ & $9.1\times10^{3}$ & $2.3\times10^{3}$ & $3.8\times10^{2}$ \\
$L_\mathrm{bol}/M_\mathrm{env} \cdot \mathrm{L}_\odot / \mathrm{M}_\odot$ & $51.2$ & $37.6$ & $7.5$ & $150.0$ & $38.1$ & $6.4$ \\
$r/\mathrm{pc}$ & $0.6$ & $0.6$ & $0.2$ & $2.45$ & $0.41$ & $0.1$ \\
$N_\mathrm{H_2}/\mathrm{cm}^{-2}$ & $1.7\times10^{23}$ & $7.4\times10^{22}$ & $3.2\times10^{22}$ & $7.8\times10^{23}$ & $1.8\times10^{23}$ & $3.0\times10^{22}$ \\
\\
\textbf{H\textsc{ii} regions} \\
\hline
$T_\mathrm{dust}/\mathrm{K}$ & $31.7$ & $31.8$ & $22.8$ & $41.12$ & $3.97$ & $0.8$ \\
$L_\mathrm{bol}/\mathrm{L}_\odot$ & $4.6\times10^{5}$ & $2.0\times10^{5}$ & $3.0\times10^{3}$ & $3.8\times10^{6}$ & $7.7\times10^{5}$ & $1.5\times10^{5}$ \\
$M_\mathrm{env}/\mathrm{M}_\odot$ & $6.0\times10^{3}$ & $2.1\times10^{3}$ & $2.8\times10^{2}$ & $4.3\times10^{4}$ & $9.0\times10^{3}$ & $1.8\times10^{3}$ \\
$L_\mathrm{bol}/M_\mathrm{env} \cdot \mathrm{L}_\odot / \mathrm{M}_\odot$ & $86.7$ & $75.5$ & $10.8$ & $358.8$ & $67.4$ & $13.5$ \\
$r/\mathrm{pc}$ & $0.8$ & $0.7$ & $0.2$ & $1.89$ & $0.47$ & $0.1$ \\
$N_\mathrm{H_2}/\mathrm{cm}^{-2}$ & $3.6\times10^{23}$ & $2.6\times10^{23}$ & $1.3\times10^{23}$ & $1.1\times10^{24}$ & $2.1\times10^{23}$ & $4.2\times10^{22}$ \\
\\
\textbf{all} \\
\hline
$T_\mathrm{dust}/\mathrm{K}$ & $24.7$ & $24.7$ & $10.7$ & $41.12$ & $6.85$ & $0.7$ \\
$L_\mathrm{bol}/\mathrm{L}_\odot$ & $1.3\times10^{5}$ & $1.6\times10^{4}$ & $5.7\times10^{1}$ & $3.8\times10^{6}$ & $4.1\times10^{5}$ & $3.9\times10^{4}$ \\
$M_\mathrm{env}/\mathrm{M}_\odot$ & $2.8\times10^{3}$ & $1.2\times10^{3}$ & $1.8\times10^{1}$ & $4.3\times10^{4}$ & $5.1\times10^{3}$ & $4.9\times10^{2}$ \\
$L_\mathrm{bol}/M_\mathrm{env} \cdot \mathrm{L}_\odot / \mathrm{M}_\odot$ & $39.7$ & $22.5$ & $0.2$ & $358.8$ & $50.7$ & $4.8$ \\
$r/\mathrm{pc}$ & $0.9$ & $0.7$ & $0.2$ & $2.74$ & $0.56$ & $0.1$ \\
$N_\mathrm{H_2}/\mathrm{cm}^{-2}$ & $1.8\times10^{23}$ & $8.6\times10^{22}$ & $2.4\times10^{22}$ & $1.1\times10^{24}$ & $1.9\times10^{23}$ & $1.8\times10^{22}$ \\

\end{tabular}
\end{center}
\end{table*}

\section{Results}\label{results}

\begin{figure*}[t]
	\centering
	\includegraphics[width=0.4\linewidth]{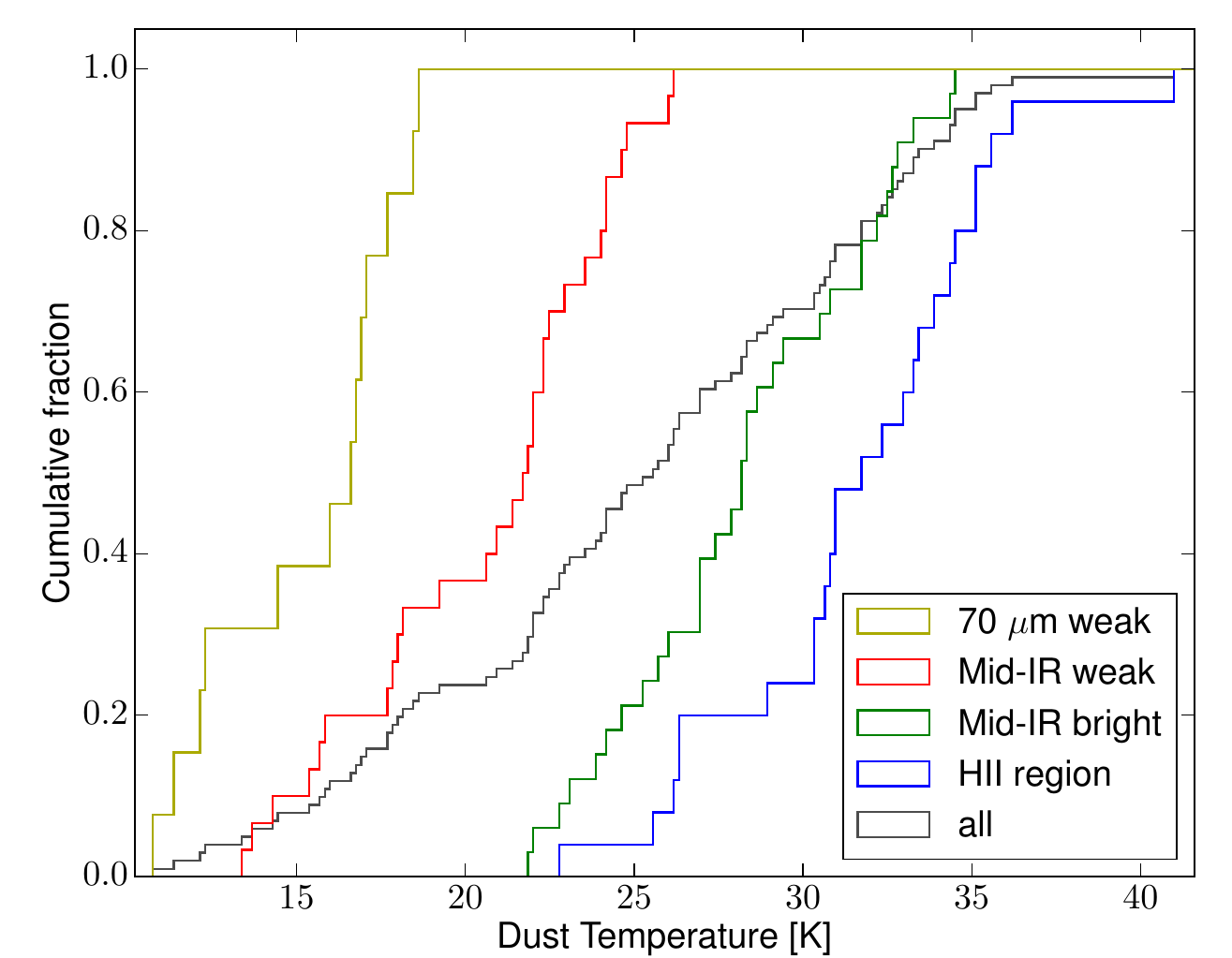}
    \includegraphics[width=0.4\linewidth]{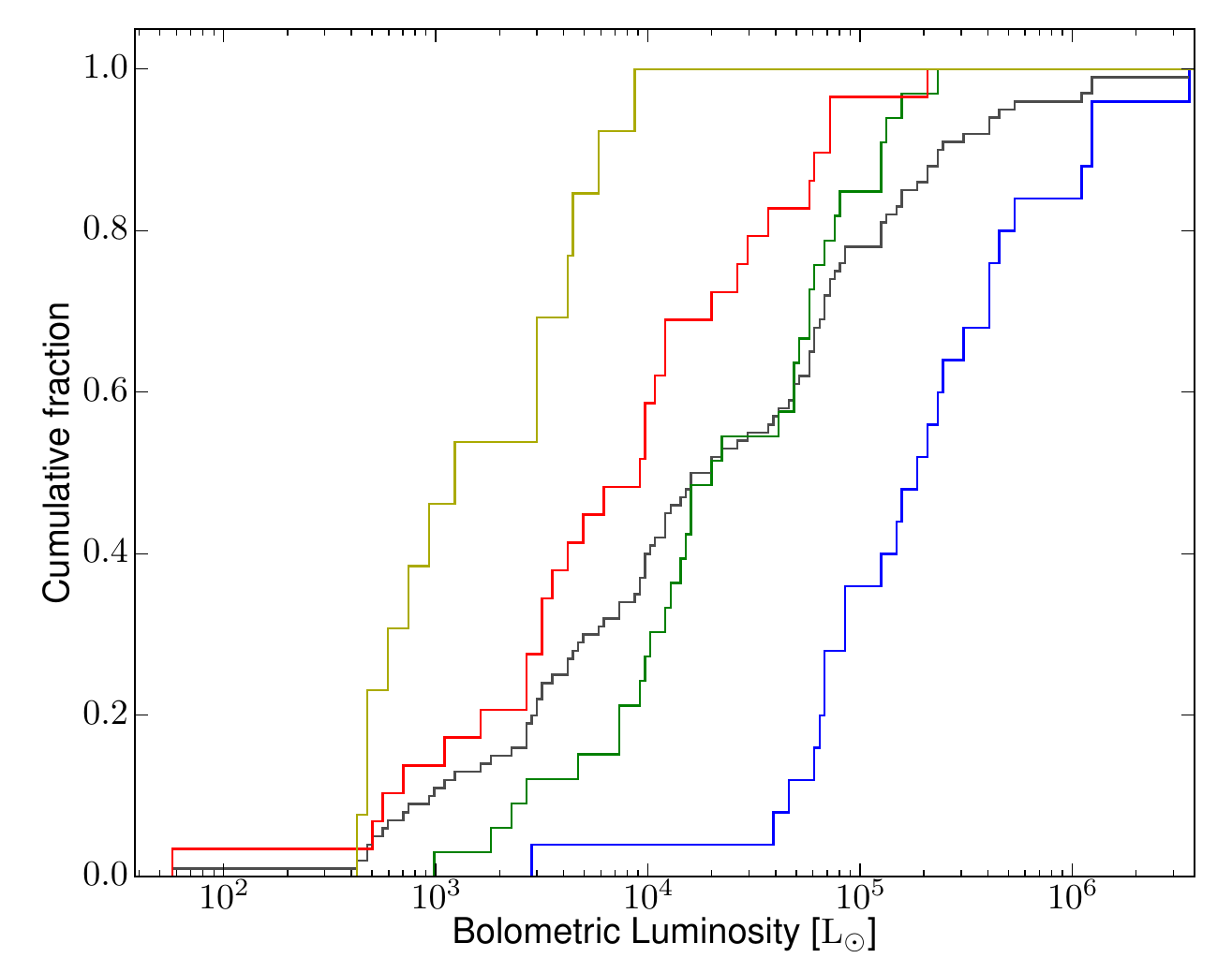}\\
    \includegraphics[width=0.4\linewidth]{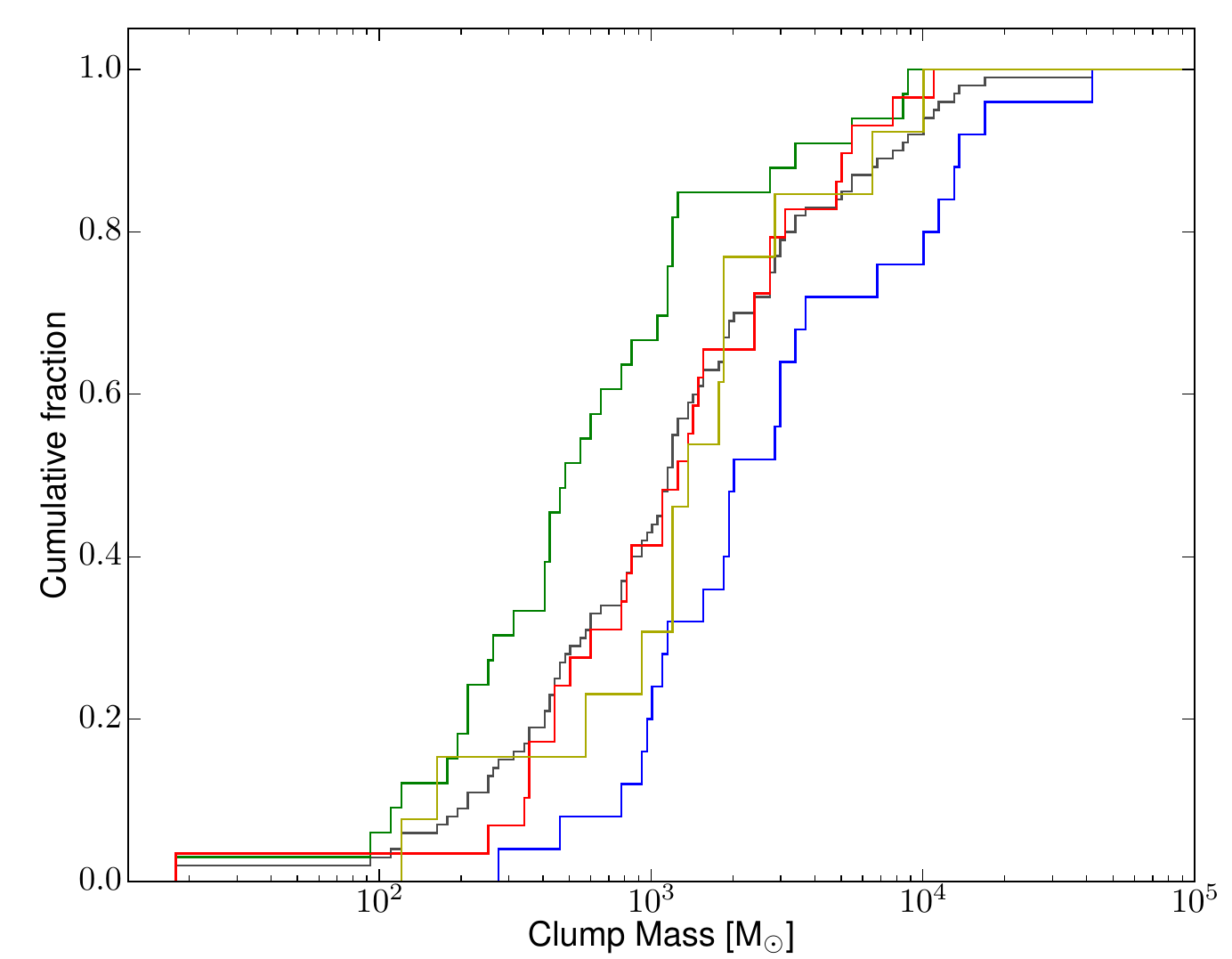}
    \includegraphics[width=0.4\linewidth]{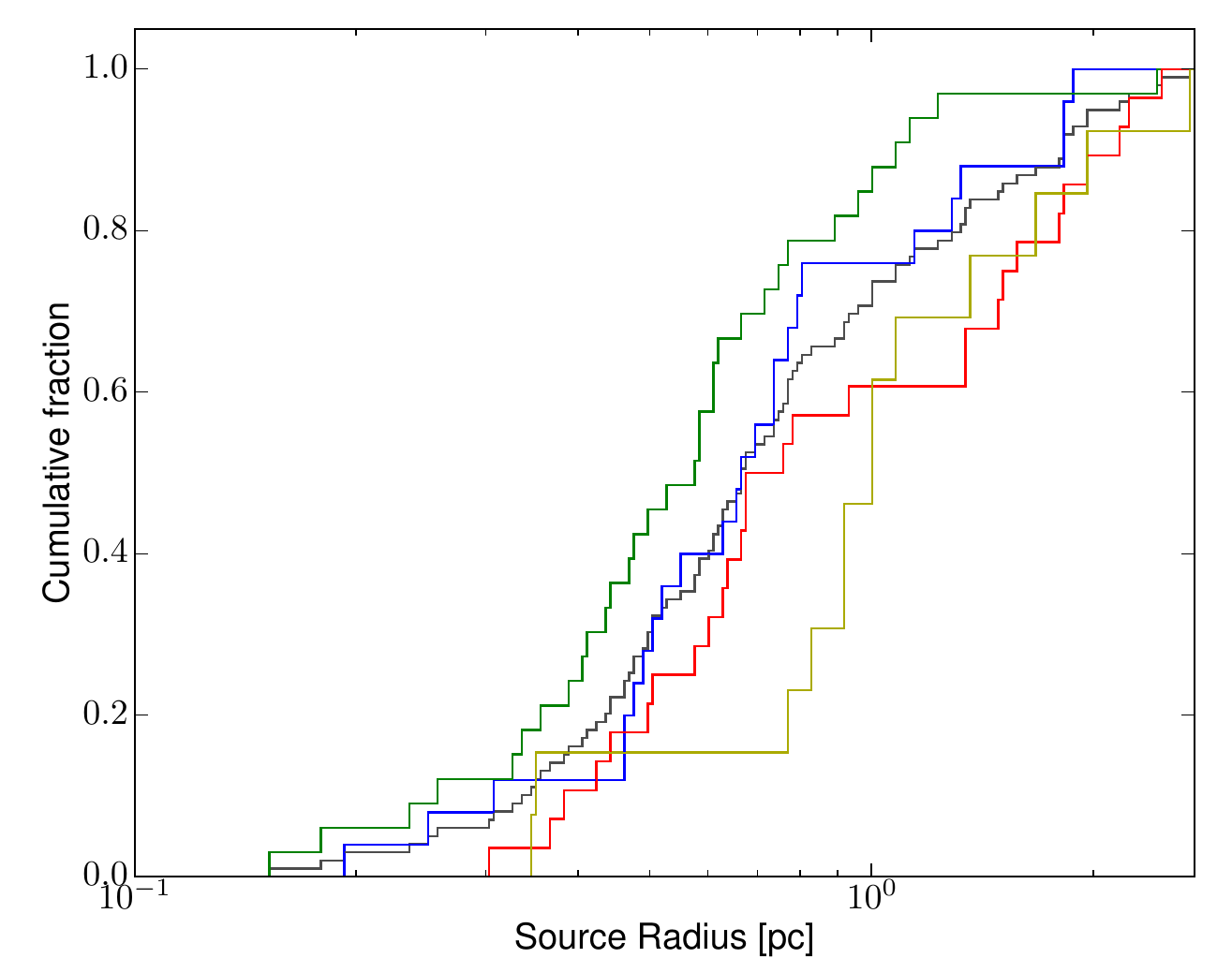}\\
      \includegraphics[width=0.4\linewidth]{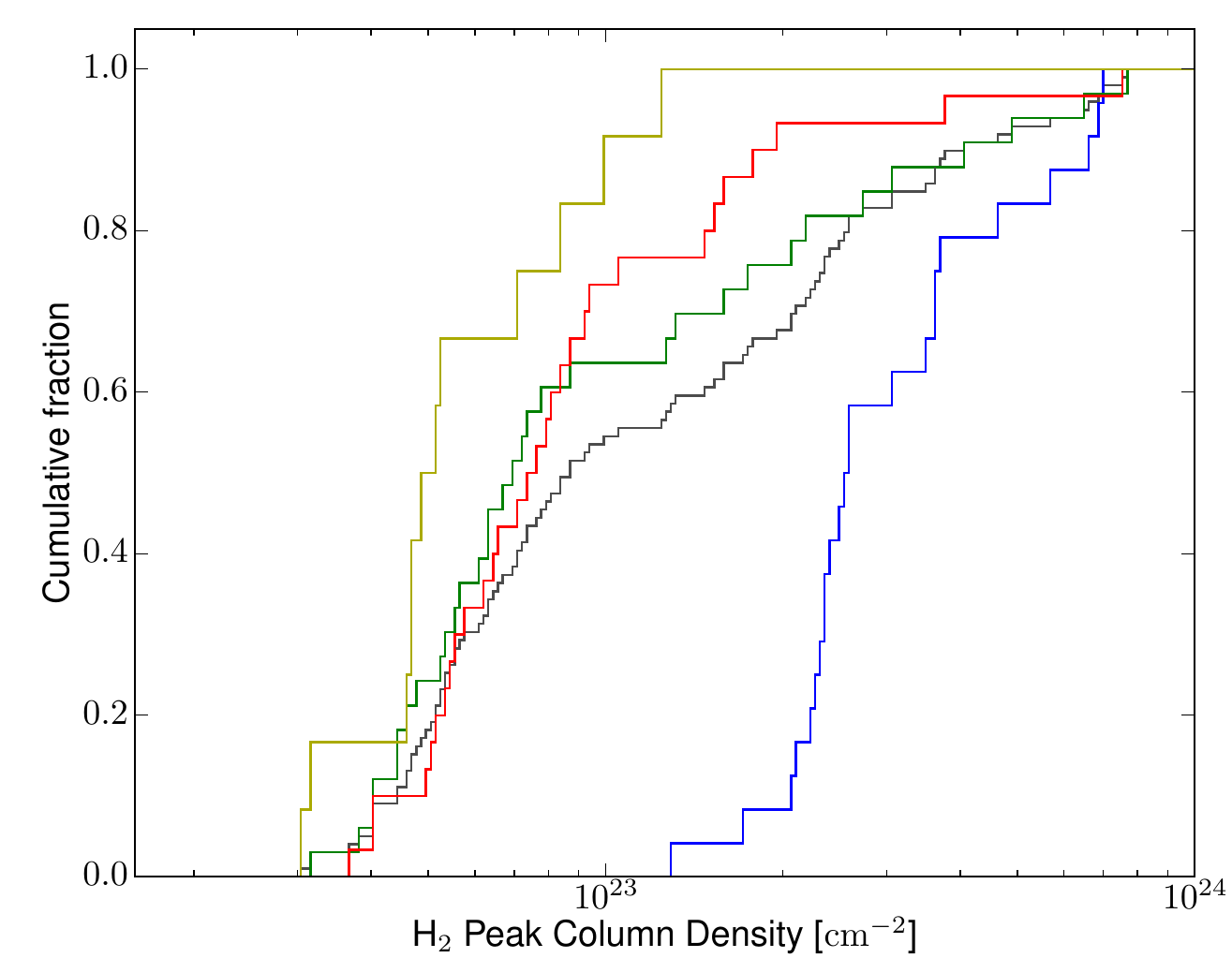}
      \includegraphics[width=0.4\linewidth]{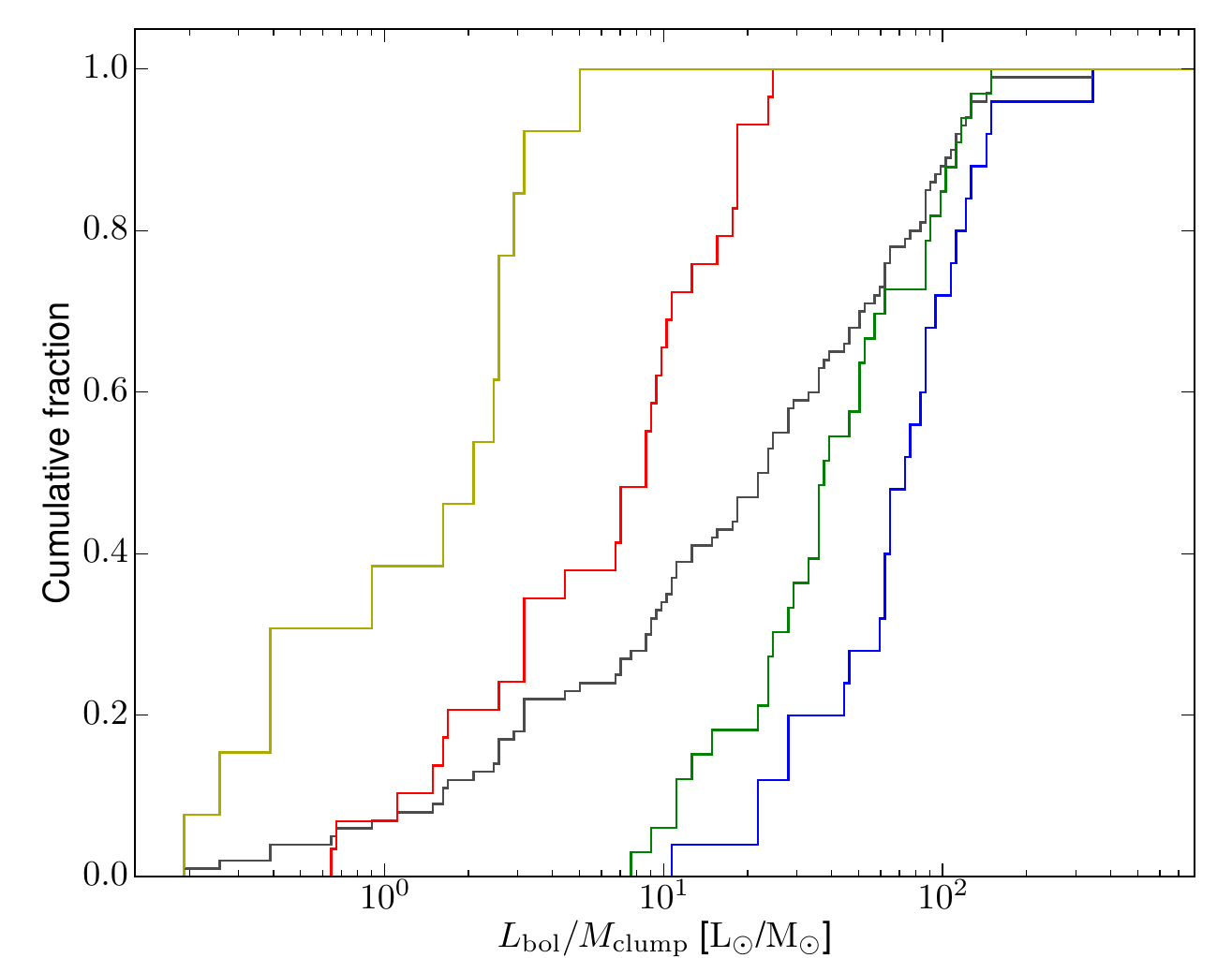}
	\caption{Cumulative distribution plots showing the range of values for the derived parameters for the whole sample (grey curve) and the four evolutionary sub-samples (see legend for colours).}
	\label{fig.parameter_Hist}
\end{figure*}

\noindent We have obtained estimates for the bolometric dust temperatures and fluxes for the whole sample. Combining these parameters with the distances discussed in Sect.\,2 we have calculated the source masses, luminosities and column densities for all but one source. In Tables\,\ref{tbl.shortdata} and \ref{tbl.classdata} we give the parameters for each source and a summary of the derived parameters for each class, respectively. Fig.\,\ref{fig.parameter_Hist} presents cumulative distribution functions for the dust temperatures, bolometric luminosities, masses, linear source sizes and column densities.

\label{sect.results}
We excluded 9 sources from the analysis for which the SEDs or classification were unreliable, as they are located in rather complex regions. For 4 sources (AGAL024.651$-$00.169, AGAL305.209$+$00.206, AGAL338.926$+$00.554, AGAL354.944$-$00.537) the aperture is cutting into some nearby source and hence overestimating the background corrected flux. For 3 sources (AGAL008.706$-$00.414, AGAL022.376$+$00.447, AGAL351.161$+$00.697) the background aperture is picking up some broad emission, leading the source flux to be underestimated. For AGAL028.564$-$00.236 the region within the aperture is too complex to be interpreted as a single source and AGAL013.178$+$00.059 is located on a background with a strong gradient leading to an unreliable flux estimate. This reduces the sample size for the analysis of evolutionary trends to 102 sources.

Comparing the average values of the parameters for the different source classifications and the cumulative distributions presented in Table\,\ref{tbl.classdata} and Fig.\,\ref{fig.parameter_Hist} reveals evidence for evolutionary trends with increases in temperatures, bolometric luminosities and column densities as a function of the different stages. We also find that the mean masses and linear sizes of the subsamples are similar and so we conclude that the initial conditions were also similar, assuming the clumps not to accrete from a much larger mass reservoir. 

\subsection{Dust temperature and bolometric luminosity}

The dust temperature is a fitted parameter of the modified blackbody model. Values range from 11\,K for the 70\,$\mu$m weak sources to 41\,K for the \hii\ regions' dust envelopes. The lowest temperatures are close to the temperature expected for quiescent clumps, which is determined by cosmic ray heating \citep[$\sim$10\,K;][]{Urban2009}, consistent with the proposed starless nature of the 70\,\mum\ weak sources. Take note that these sources are likely to have even lower temperatures in the inner part as indicated by \citet{Bernard2010}. The mean temperature of the dust envelope increases through the different categories from 16.4\,K for the 70\,$\mu$m weak sources up to a mean temperature of 31.7\,K for the H\textsc{ii} regions, with an average dust temperature of 24.7 K for the whole sample. \label{sect.temperature} We point out that the dust temperatures for the H\textsc{ii} regions should be taken with care (and possibly as lower limits) as the emission might be optically thick for a portion of the SED \citep[see e.g.][]{Elia2016}.

As can be seen from the cumulative histogram in Fig. \ref{fig.parameter_Hist} (upper left panel), the dust temperature increases with the evolutionary stage and the different classes (and hence evolutionary stages) are well separated with regard to its value. This is confirmed by Anderson-Darling tests yielding $p$-values lower than $2.7\times10^{-3}$ for all classes rejecting the null hypothesis that these are drawn from the same distribution at the $3\sigma$ significance level.

\label{results.luminosities}

To estimate the bolometric luminosities $L_\mathrm{bol}$ for each source we combine the distance and integrated flux determined from the fitted SED model using the equation

\begin{eqnarray}
L_\mathrm{bol} = 4 \pi d^2 \int S_{\lambda}\,{\rm{d}}\lambda,
\end{eqnarray} 

\noindent where $d$ is the distance to the source (Sect.\,\ref{data.distance}) and $\int S_{\lambda}\,{\rm{d}}\lambda$ is the integrated flux. The bolometric luminosities for the whole sample range from $57$\,$\mathrm{L}_\odot$ for the least luminous source to $3.8\times10^6$\,$\mathrm{L}_\odot$ for the most luminous \hii\,region, with a mean value of $1.3\times10^5$\,$\mathrm{L}_\odot$ and a median value $1.5\times10^4$\,$\mathrm{L}_\odot$. The cumulative distribution function of the luminosities shows they increase with the evolutionary stage yielding $p$-values $<2.4\times10^{-5}$ for the Anderson-Darling test. However, these tests reveal no significant difference between the mid-infrared weak and mid-infrared bright samples ($p$-value = $1.2\times10^{-2}$) as well as between the mid-infrared weak and the 70~$\mu$m weak sources ($p$-value = $5.6\times10^{-3}$).

\begin{figure}[tp!]
   \centering
   \includegraphics[width=\hsize]{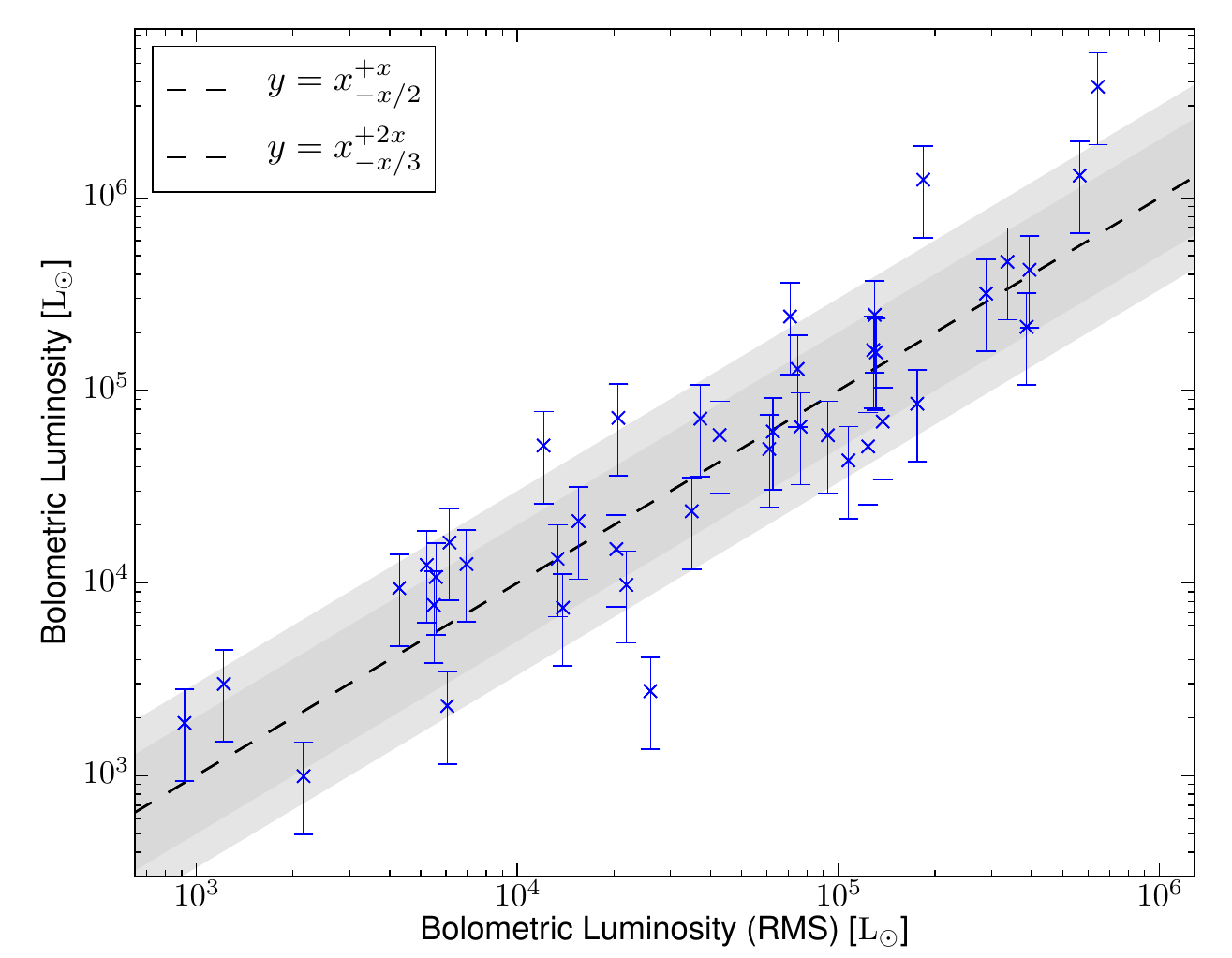}
      \caption{Comparison of our luminosities to those of \citet{Urquhart2014a} for sources common in both studies. The grey shaded areas indicate the area of of an agreement within a factor of two (dark) and three (light). Most of the luminosities agree within a factor of three, showing that our simplified two-component model is sufficient to get an estimate for the luminosity.}
         \label{fig.rmsLum}
\end{figure}

To verify that the extraction of the photometry and using a relatively simple two-component model to fit the SED produces reliable results, we compare our derived luminosities with those reported for a subsample of the same sources reported by the RMS team \citep{Urquhart2014a}. This other research has used fluxes drawn from the Hi-GAL point source catalogue \citep{Molinari2016} and the more complex radiative transfer models determined by \citet{Whitney2005} and the fitting tool developed by \citet{Robitaille2007a}. In Figure\,\ref{fig.rmsLum} we show this comparison of the bolometric luminosities for 41 matching sources of the RMS sample. It is clear from this plot that there is good agreement between these two very different methods.

\subsection{Clump mass, size and column density}\label{sect.mass}

To estimate the masses of the sources we followed the procedure of \citet{Hildebrand1983} for an optically thin emission and a single temperature, as the derived optical depth at 870\,$\mu$m is $\tau_{870}\ll1$ for all sources:

\begin{eqnarray}
M_\mathrm{clump} = \frac{d^2 S_{870}R}{B_{870}(T_\mathrm{d})  \kappa_{870}}
\label{eq.mass}
\end{eqnarray}

\noindent where $d$ is the distance to the cloud, $S_{870}$ is the integrated 870\,$\mu$m flux density obtained from the aperture photometry, $R$ the gas-to-dust ratio assumed to be 100, $B_{870}(T_\mathrm{d})$ the intensity of the blackbody at 870\,$\mu$m at the dust envelope temperature $T_\mathrm{d}$, and $\kappa_{870}=1.85$\,cm$^2$g$^{-1}$ the dust opacity at 870\,$\mu$m calculated as the average of all dust models from \citet{Ossenkopf1994} for the dust emissivity index of 1.75 that we use for the fitting of the SEDs.

The derived clump masses range from 18\,$\mathrm{M}_\odot$ up to $4.3\times10^4$\,$\mathrm{M}_\odot$ with a mean value of $2.8\times10^3$\,$\mathrm{M}_\odot$ and a median value of $1.2\times10^3$\,$\mathrm{M}_\odot$ for the whole sample. Two sources are found to have comparatively low masses (i.e. AGAL316.641$-$00.087 and AGAL353.066$+$00.452), making them unlikely to form any massive stars. The mass cumulative distribution plot shown in Figure\,\ref{fig.parameter_Hist} (mid left panel) reveals little variation between the different classes and is therefore relatively independent of the evolutionary phase. This is confirmed by Anderson-Darling tests with only the mid-infrared bright and H\textsc{ii} regions yielding a $p$-value of 1.0$\times10^{-4}$ yielding them unlikely to be drawn from the same distribution. We note that this difference is likely an effect of the dust temperatures being lower limits for the compact H\textsc{ii} regions (see Section \ref{sect.temperature}) and hence the envelope masses being upper limits.

The linear size of a source (i.e. its radius $r$) in pc can be calculated from the distance $D$ in pc and the deconvolved source size $\theta_\mathrm{deconv}$ as

  \begin{eqnarray}
r = d \cdot \tan(\theta_\mathrm{deconv}/2)
\label{eq.size}
\end{eqnarray}
\noindent with the deconvolved size of the source being calculated as 

\begin{eqnarray}
\theta_\mathrm{deconv} = \sqrt{\mathrm{D}_\mathrm{ap}^2 - \theta_\mathrm{beam}^2}
\end{eqnarray}

\noindent where $\mathrm{D}_\mathrm{ap}$ is the source size as used for the aperture and $\theta_\mathrm{beam}$ is the FWHM size of the beam (i.e. 19.2\arcsec). The cumulative distribution of the linear source size is presented in the middle right panel of Fig.\,\ref{fig.parameter_Hist} with a mean value of 0.9\,pc and a median value of 0.7\,pc over all classes. We find mostly no significant differences between the different evolutionary classes with Anderson-Darling tests all yielding $p$-values $> 0.003$, except between the mid-infrared bright and the 70\,\mum\ weak sources ($p$-value $=2.6\times10^{-3})$.


We also calculate the beam averaged column density according to \citet{schuller2009}, assuming the dust emission at 870\,$\mu$m is optically thin:

\begin{eqnarray}
N_\mathrm{H_2} = \frac{F_{870}  R}{B_{870}(T_\mathrm{d}) \; \Omega_\mathrm{app} \; \kappa_{870} \; \mu_\mathrm{H_2} \; m_\mathrm{H}},
\label{eq.columndensity}
\end{eqnarray}

\noindent where $F_{870}$ being the peak flux density, the beam solid angle $\Omega_\mathrm{app}$, and $\mu_\mathrm{H_2}=2.8$ as the mean molecular weight of the interstellar medium with respect to a hydrogen molecule according to \citet{Kauffmann2008} and $m_\mathrm{H}$ the mass of a hydrogen atom. The other parameters are as previously defined.

In Fig. \ref{fig.parameter_Hist} (lower left panel) we show the histogram of the beam averaged peak column densities. The mean column density increases from $2.4\times10^{22}$\,cm$^{-2}$ for the 70\,\mum\ weak sources to column densities in excess of $10^{24}$\,cm$^{-2}$ for the H\textsc{ii} regions. This is to be expected, as the clumps increase their density in the process of collapse. We note that the class of compact H\textsc{ii} regions is well distinguished from the other classes, being confirmed by Anderson-Darling tests ($p$-values $<2.4\times10^{-5}$). We point out that the column densities are effected by the dust temparatures being lower limits for the compact H\textsc{ii} regions as discussed in section \ref{sect.temperature}, hence rendering the column densities upper limits. Furthermore, this nicely shows that the young H\textsc{ii} regions have not dispersed the envelope significantly yet, as we would expect for the sources being selected such as to be the brightest in ATLASGAL, hence making them the densest phase before dispersing the dust envelopes. For the other classes the null-hypothesis of the samples being drawn from the same distribution can not be rejected with $p$-values $>1.1\times10^{-2}$. However, taking a closer look at the cumulative distribution of the column density in Fig. \ref{fig.parameter_Hist} (lower left plot) and taking into account the small number of only 16 source of the 70\,\mum\ weak sample, we speculate that this subclass might be well distinguishable for a larger sample.

\section{Discussion}\label{discussion}

\subsection{Comparison with massive star formation relations}

 

\begin{figure}[t]
	\centering
	\includegraphics[width=\hsize]{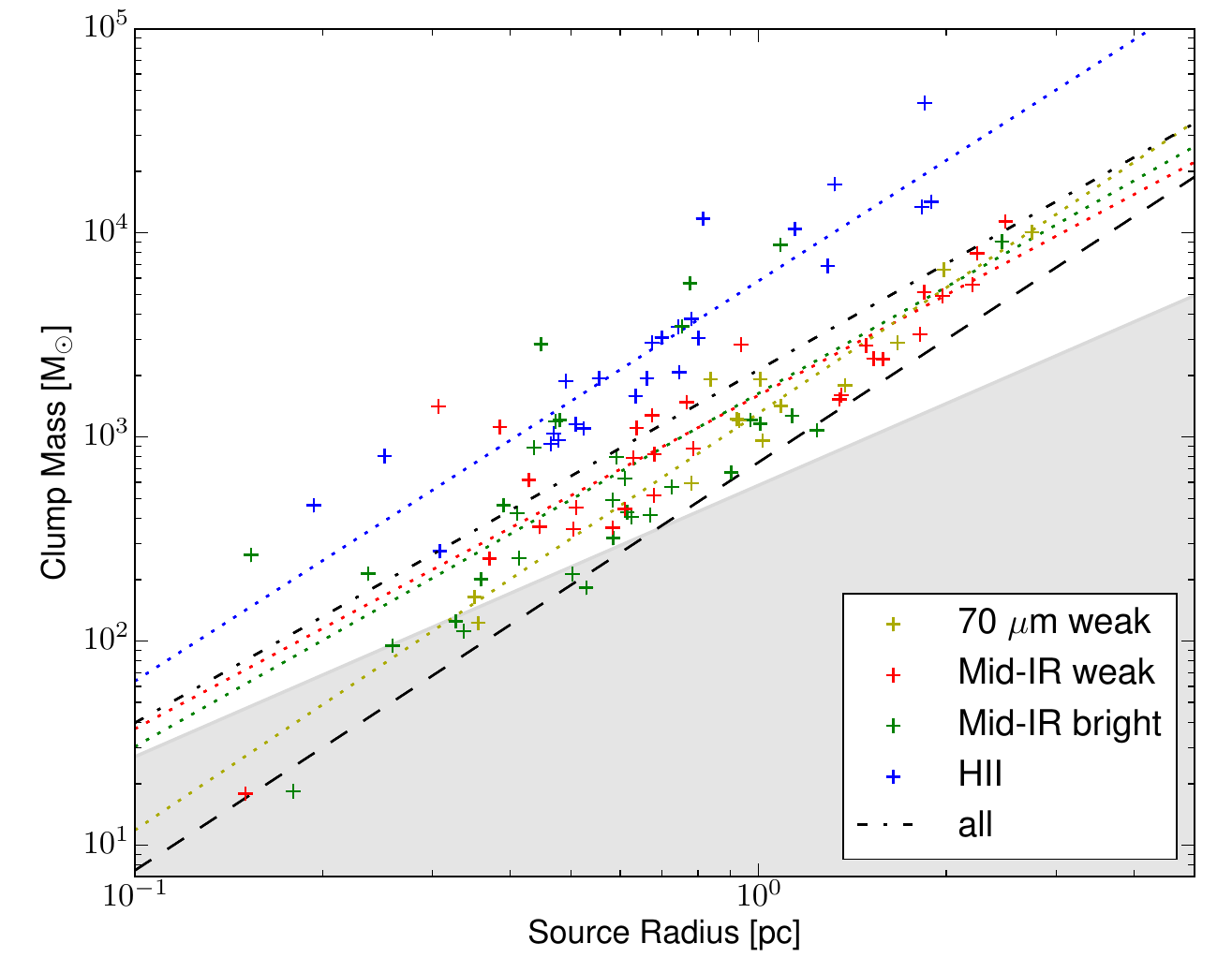}
	\caption{Mass-size relationship of the ATLASGAL Top100. The lower right grey shaded area highlights the region where it is unlikely that high-mass stars are formed according to \citet{kauffmann2010a} and the black dashed line indicates a surface density of 0.05 g cm$^{-2}$ determined as the lower limit for efficient massive star formation by \citet{Urquhart2014}. Note that the slopes of the linear log-log fits of all classes agree with each other within the margin of error of the fitted power laws.}
\label{fig.MvsR}
\end{figure}

In Figure \ref{fig.MvsR} we present the mass-size relation for the sample. According to \citet{kauffmann2010a} a lower limit for high-mass star formation is given by $M(r) \geq 580\,\mathrm{M}_\odot \cdot (r/\mathrm{pc})^{1.33}$ when reducing the mass coefficient by a factor of 1.5 from 870 to 580 to match with the dust absorption coefficient used in the present paper. Taking this threshold as the lower limit for massive star formation we find that 93 (i.e. 85\%) of our sources have the potential to form at least one massive star. Similarly assuming the threshold found by \citet{Urquhart2014}, indicating that sources with surface densities above 0.05\,g\,cm$^{-2}$ can efficiently form high mass stars, up to 99 sources (i.e. 98\%\ of the sources with good SEDs and classification) have already formed at least one high mass star or are likely to do so in the future.

\begin{figure}[t]
   \centering
   \includegraphics[width=\hsize]{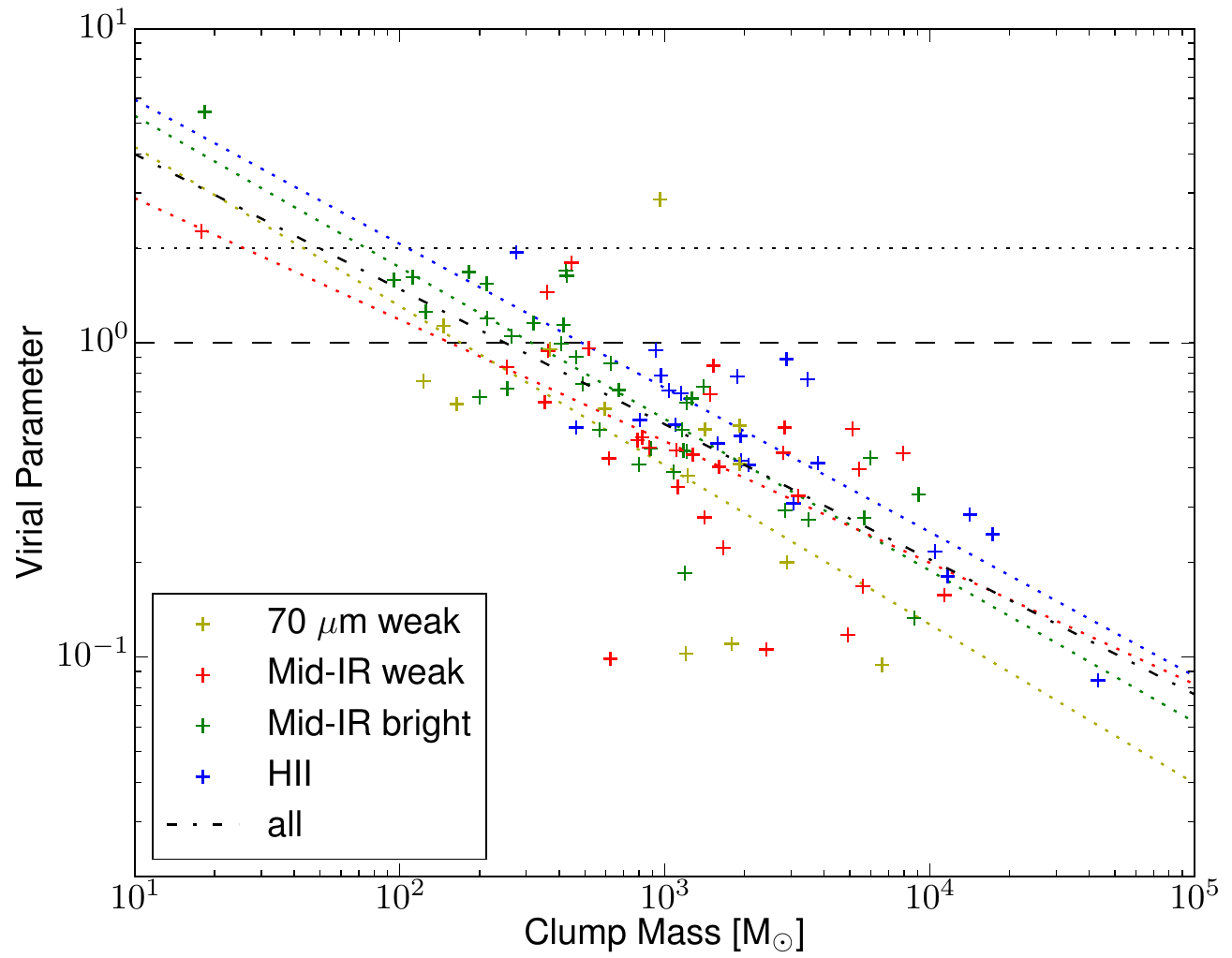}
   \caption{Clump mass versus virial parameter $\alpha_\mathrm{vir}$. The dashed and dotted lines represent $M_\mathrm{vir} = M_\mathrm{\odot}$ and $M_\mathrm{vir} = 2 M_\mathrm{\odot}$, respectively. Sources below these lines are likely to be unstable and and hence collapsing with and without a magnetic field being present for the dashed and dotted lines, respectively.}
	\label{fig.a_vir}
\end{figure}

\label{sect.avir}To estimate the stability of the clumps, we calculate the virial parameter for the 102 sources of \citet{Giannetti2014}, but correct for the revised distances $d$. Revising the distances affects the linear source sizes $r_\mathrm{FWHM}$ for the regions the velocity dispersion was measured for (i.e. approximately the FWHM of the source). We calculate the virial parameter according to \citet{Bertoldi1992}:

\begin{eqnarray}
\alpha_\mathrm{vir} = \frac{5}{G}  \cdot \left(\frac{\Delta V}{2\sqrt{2\ln{2}}}\right)^2 \cdot r_\mathrm{FWHM} \cdot M_\mathrm{clump}^{-1}
\end{eqnarray}

where we adopted the C$^{17}$O\,(3--2) linewidth $\Delta V$ in km/s from \citet{Giannetti2014}, $G$ the gravitaional constant and $M_\mathrm{clump}$ the clump mass estimated from the flux given by \citet{Csengeri2014} using Equation \ref{eq.mass}. Here the linewidth and the envelope mass are estimated using the FWHM source sizes. The result is shown in Figure \ref{fig.a_vir}.

In general, we see a similar anti-correlation of the clump mass with the virial parameter as reported by other work \citep[e.g.][]{Kauffmann2013,Urquhart2014,Giannetti2014}, indicating that stability of clumps decreases as their mass increases. Fitting a power law to the data for all classes, we find the trend of all classes to be very similar to each other with a general slope of $s=-0.43\pm0.04$ for the whole sample (compare Fig.\,\ref{fig.a_vir}).

\citet{Kauffmann2013} adopted a critical value of $\alpha_\mathrm{crit}=2$ for a Bonnor-Ebert sphere \citep{Bonnor1955} not being supported by a magnetic field. Under this assumption a clump is likely to be unstable if $\alpha_\mathrm{vir} < 2$. From this we conclude that at least 99 sources (i.e. 97\%) are likely to be unstable. 

For the whole sample only three sources show a virial parameter of $\alpha_\mathrm{vir} > 2$ (AGAL316.641$-$00.087, AGAL338.066$+$00.044, AGAL353.066+00.452) and are therefore unlikely to be collapsing, if external pressure does not confine them. Two of these sources (AGAL316.641$-$00.087, AGAL353.066+00.452) with a virial parameter of $\alpha_\mathrm{vir} \approx 5.4$ and $2.3$, respectively, were found to have masses lower than 19\,M$_\odot$ making them unlikely to form a massive star.

In case of a magnetic field being present, \citet{Bertoldi1992} showed that with the magnetic field and kinetic support being equal, the critical value of the virial parameter is lower ($\alpha_\mathrm{crit}=1$). Taking this lower critical parameter as the threshold for collapse, we find that 84 of the sources (i.e. 82\%) are gravitationally unstable and in the absence of strong magnetic fields are likely to be collapsing.

To further investigate the minimum magnetic field strength needed to support the clumps against gravitational collapse, we follow the procedure of \citet{Kauffmann2013}, calculating the critical field strength as

\begin{eqnarray}
B_\mathrm{crit} = 81\, \mathrm{\mu G}\cdot \frac{M_\Phi}{M_\mathrm{BE}} \left(\frac{\Delta V}{2\sqrt{2\ln{2}}}\right)^2 \frac{1}{r_\mathrm{FWHM}}
\end{eqnarray}

where we substitute

\begin{eqnarray}
\frac{M_\Phi}{M_\mathrm{BE}} = \frac{2}{\alpha_\mathrm{vir}} -1
\end{eqnarray}

as calculated from the virial parameter $\alpha_\mathrm{vir}$. We find the median critical field strength to range from 0.8\,mG for the 70\,\mum\ weak class up to 6.2\,mG for the compact H\textsc{ii} regions. \citet{Crutcher2012} proposes a density-dependant upper limit for the magnetic field strength in a molecular cloud of

\begin{eqnarray}
B_\mathrm{max} = 10\,\mu\mathrm{G}\; \left(\frac{n_{\mathrm{H}_2}}{150\,\mathrm{cm}^{-3}}\right)^{0.65},
\label{eq.crutcher}
\end{eqnarray}

implying that if $B_\mathrm{crit} > B_\mathrm{max}$ no existing magnetic field can stabilize the cloud. To apply our data, we further transform Equation \ref{eq.crutcher} as shown by \citet[][C.4]{Kauffmann2013} into

\begin{eqnarray}
B_\mathrm{max} = 336\, \mathrm{\mu G}\; \left(\frac{M_\mathrm{clump}}{10\,\mathrm{M}_\odot}\right)^{0.65} \left(\frac{r_\mathrm{FWHM}}{0.1\,\mathrm{pc}}\right)^{-1.95}.
\end{eqnarray}

\begin{figure}[t]
   \centering
   \includegraphics[width=\hsize]{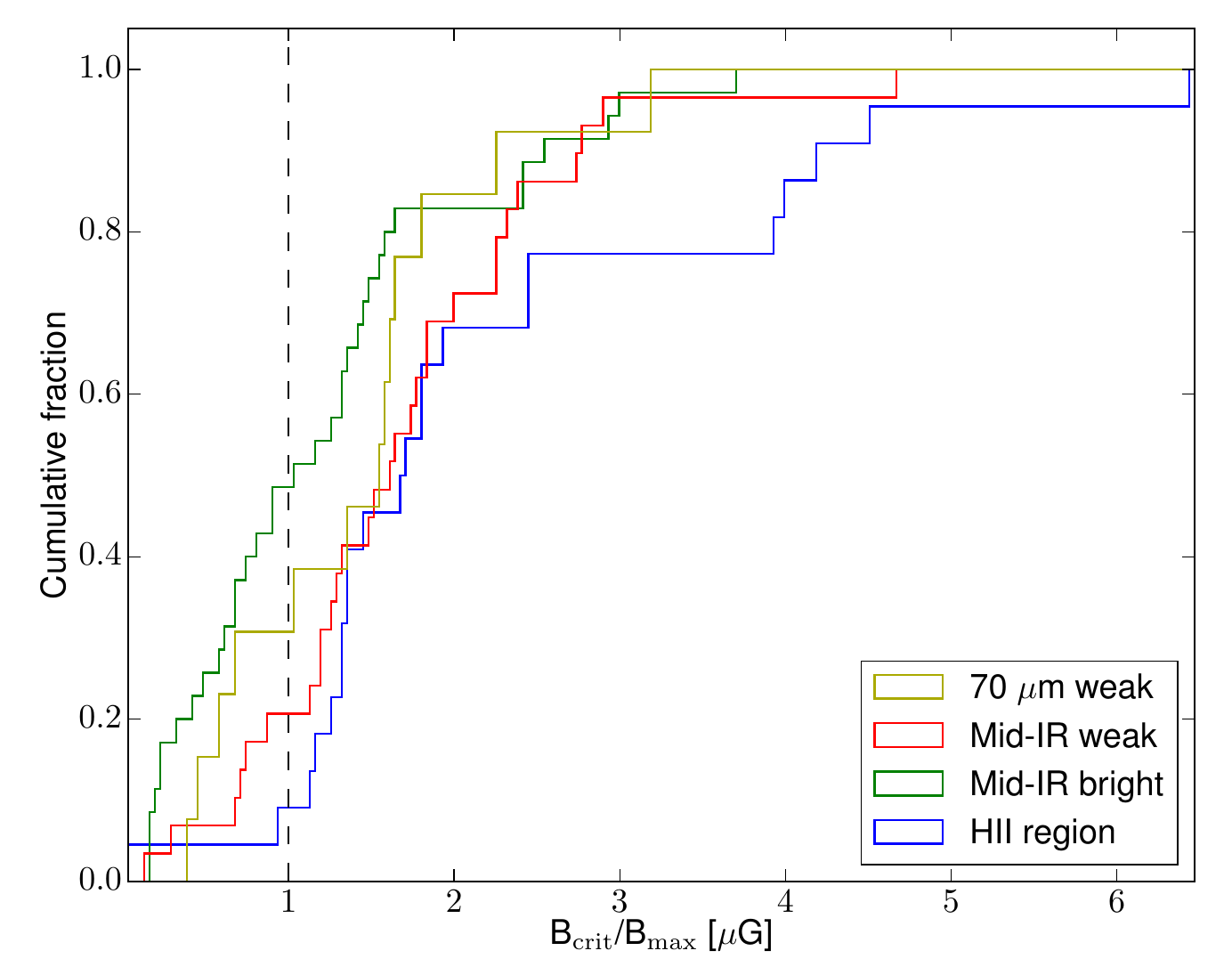}
   \caption{Ratio of the minimum critical magnetic field strength needed to stabilize our sources $B_\mathrm{crit}$ to the upper limit magnetic field strength $B_\mathrm{max}$ for a cloud as suggested by \citet{Crutcher2012}. Sources right of the dashed line ($B_\mathrm{crit} = B_\mathrm{max}$) can not build a magnetic field strong enough to prevent the clumps from collapsing.}
	\label{fig.bcrit}
\end{figure}

Comparing $B_\mathrm{crit}$ to $B_\mathrm{max}$ (see Figure \ref{fig.bcrit}) we find the minimum magnetic field strength needed to stabilize our sources to be higher than the upper limit suggested by \citet{Crutcher2012} for 69\% of our sample. From this we conclude that the majority of our sources can either not be stabilized or would need extreme magnetic fields to support them against gravitational collapse.

Our analysis has revealed that there is no significant difference in the global properties of the sample with similar mass and size distributions found for all of the sub-samples, hence likely having had similar initial physical conditions. We have also found that the majority of the sample satisfies the mass-size threshold criterion required for massive star formation and that they are unstable to gravity and likely to collapse in the absence of a strong magnetic field, clearly confirming that the sources of the ATLASGAL Top100 are a sample of massive star forming regions.

\subsection{Evolutionary Sequence of the Sample}

The analysis presented in the previous subsection found little difference in the masses and stability of the various sub-samples and revealed them all to be good candidates for massive star formation. The results obtained from the SED fitting show trends for increasing temperatures and bolometric luminosity that are consistent with the proposed evolutionary sequence. In this subsection we will extend this analysis in order to robustly test the evolutionary scheme of the ATLASGAL Top100.

\begin{figure}[t]
   \centering
   \includegraphics[width=\hsize]{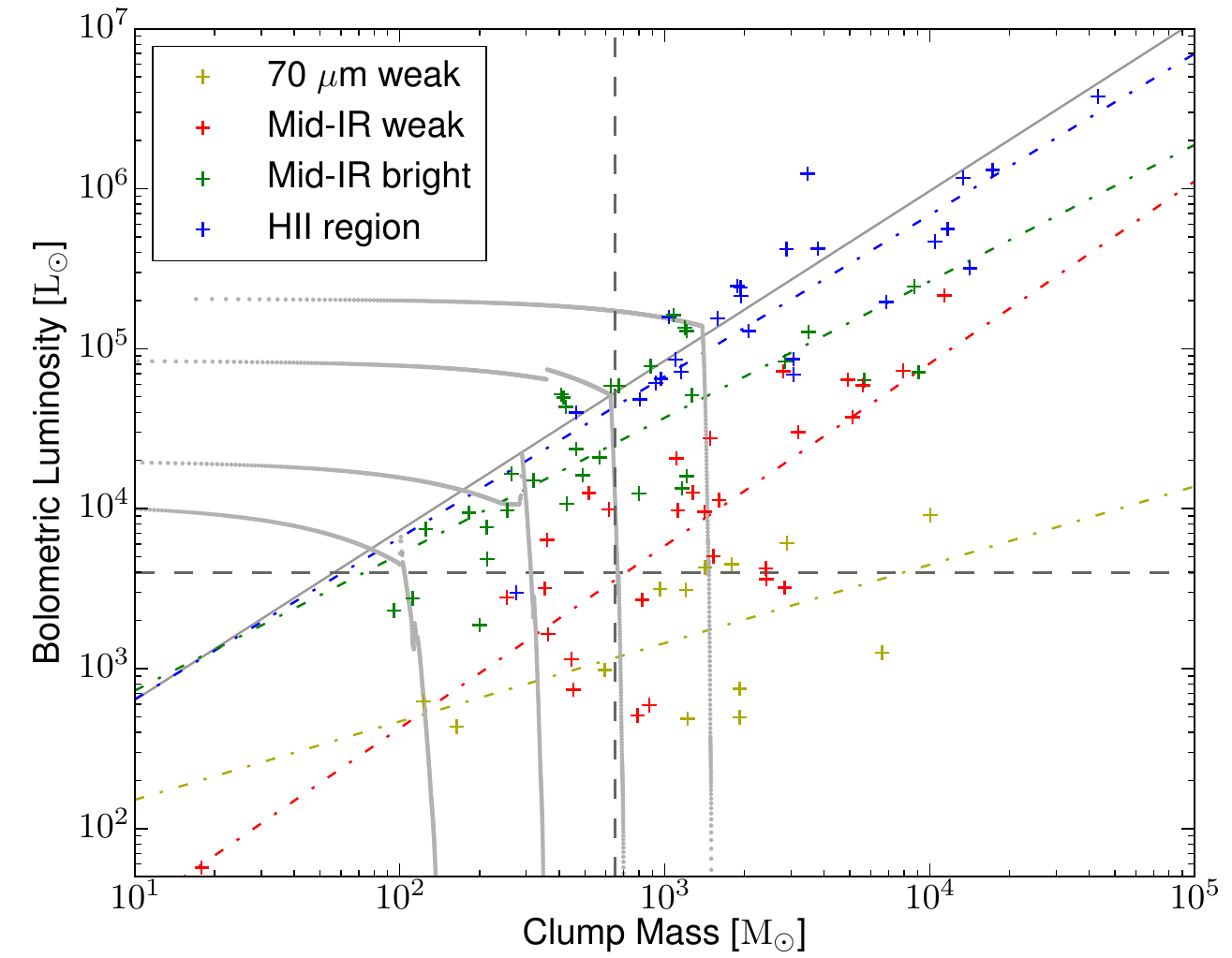}
   \caption{Mass-luminosity distribution of the whole sample. The grey tracks are evolutionary models used by \citet{Molinari2008}. The dash-dotted colored lines show a linear fit to the corresponding class of sources. The horizontal dashed line shows the luminosity of a B1.5 star as calculated by \citet[Table 1]{Mottram2011}, whereas the vertical dashed line shows the threshold of 650\,M$_\odot$ beyond which clumps are likely to host massive dense clumps or high-mass protostars \citep{Csengeri2014}.}
	\label{fig.MvsL}
\end{figure}

Following the procedure of \citet{Molinari2008}, we compare here the clump mass versus the bolometric luminosity (Figure \ref{fig.MvsL}). In such a plot it is expected that clumps first gain luminosity, moving upward in the plot until the embedded stars reach the zero age main sequence (ZAMS, solid diagonal line) and then lose mass, as the dust envelope gets dispersed, moving leftward in the plot (compare with \citealt{Molinari2008}). It can be seen, that the earlier evolutionary stages (i.e. the 70\,$\mu$m and mid-infrared weak sources) follow this trend clearly toward the zero age main sequence. The later stages (i.e. the mid-infrared bright and H\textsc{ii} regions) are not as well separated, making it difficult to distinguish their evolutionary stage from dust emission alone, consistent with the findings of \citet{Urquhart2014}.

Fitting a power law to the data for all four classes, we find that for the H\textsc{ii} regions, mid-infrared bright and weak sources the trend is very similar, yielding slopes of $s=1.01\pm0.13$, $s=0.85\pm0.13$ and $s=1.14\pm0.16$, respectively. These values are lower than those reported by \citet{Molinari2008} and \citet{Urquhart2014}, but still in agreement within $3\sigma$. Only the sample of 70~$\mu$m weak sources yields a slightly lower slope of $s=0.49\pm0.24$, which, however, is still in agreement with the slopes of the other samples. The low number of sources in this sub-sample could result in the fit being significantly affected by incomplete sampling and/or outliers. Nevertheless, the positions of the sub-samples seen in the mass-luminosity plot are consistent with the expected evolutionary tracks and nicely illustrate that our sample covers a wide range of evolutionary stages.

Given that early B-type stars have a minimum luminosity of $\sim$ $10^{3.6}$\,$\mathrm{L}_\odot$ \citep[Table 1]{Mottram2011}, we can see from the mass-luminosity distribution (Fig.\,\ref{fig.MvsL}) that a large proportion (80 sources, i.e. 73\%) of our sample already has luminosities in excess of this threshold. As the luminosity of a cluster is dominated by the most massive star, it is likely that these clumps have already formed at least one high-mass star. \citet{Csengeri2014} take a mass limit of 650\,M$_\odot$ for clumps to likely form high-mass protostars, and 73 sources (i.e. 66\%) are found above this limit. Taking into account both aforementioned thresholds (L$_\mathrm{bol}>10^{3.6}$\,$\mathrm{L}_\odot$ and M$_\mathrm{clump}>$650\,M$_\odot$, ), it is therefore likely that at least 93 sources (i.e. 85\% of our sample) have either already formed or will likely go on to form a cluster consisting of at least one massive star in the future.

To further investigate the evolutionary stage of the sources, we present the cumulative histogram of the bolometric luminosity to clump mass ratio in Fig.\,\ref{fig.parameter_Hist} (lower right panel). This ratio is a crude proxy for the evolutionary stage and we would therefore expect this to increase as the embedded objects evolve and become more luminous. This plot clearly shows a trend for increasing $L_\mathrm{bol}/M_\mathrm{clump}$ ratio with evolution, increasing continuously from a minimum ratio of 0.2 for the 70\,\mum\ weak sources up to a ratio of 358.8 for the H\textsc{ii} regions. Performing Anderson-Darling tests, we find that almost all evolutionary stages are clearly distinct from each other, yielding $p$-values $<3.6\times10^{-4}$, rejecting the null-hypothesis of the samples being drawn from the same distribution at the $3\sigma$ significance level. We only find the mid-infrared bright and H\textsc{ii} region phases to be similar enough, as to be drawn from the same distribution with a $p$-value of $2.6\times10^{-2}$.

\subsection{Dust continuum emission as evolutionary stage indicator}

\begin{table*}
\begin{center}
\caption{Overview of different physical parameters and whether or not the Anderson-Darling test is able to distinguish the different classes of sources.} \label{tbl.parameteroverview}
\begin{tabular}{l|c|c|c|c|c|c}
\hline\hline
\multirow{2}{*}{Parameter} & H\textsc{ii} & H\textsc{ii} & H\textsc{ii} & Mid-IR bright & Mid-IR bright & Mid-IR quiet \\
 & Mid-IR bright & Mid-IR quiet & 70\,\mum\ weak & Mid-IR quiet & 70\,\mum\ weak & 70\,\mum\ weak \\
\hline

$T$ & + & + & + & + & + & + \\
\hline
$L_\mathrm{bol}$ & + & + & + & - & + & - \\
\hline
$M_\mathrm{env}$ & + & - & - & - & - & - \\
\hline
$L_\mathrm{bol}/M_\mathrm{env}$ & - & + & + & + & + & + \\
\hline
$r$ & - & - & - & - & + & -\\
\hline
$N_\mathrm{H_2}$ & + & + & + & - & - & - \\
\hline
\end{tabular}
\end{center}
\end{table*}

As we have seen in the previous sections, the physical parameters and derived quantities obtained from dust continuum emission namely temperature, bolometric luminosity, clump mass, $L_\mathrm{bol}/M_\mathrm{env}$, linear source size or column density are differently well suited to distinguish between the evolutionary phases of massive star formation.

In Table\,\ref{tbl.parameteroverview} we give an overview of whether or not the Anderson-Darling tests are able to distinguish two evolutionary stages of our sample using the given parameter. We note that except for the temperature no single parameter is suited to discriminate between all evolutionary stages alone. However, all classes can be well distinguished by a combination of at least two different parameters and the evolutionary sequence is well reflected in the physical parameters obtained from the dust emission. But although there are clear trends seen in the distributions of some of the derived parameters there is also significant overlap between them (compare Table\,\ref{tbl.classdata}). In turn this makes it impossible to assign a given source to a single evolutionary stage just from the dust parameters and further criteria are required.

The optical depth might limit the usability of the dust emission in extreme cases, where the emission being optically thick at far-infrared wavelengths would underestimate the fluxes of the SED. Similarly the estimated dust temperature might be underestimated for sources for which the emission longward of $\lambda>21$\,\mum\ becomes optically thick \citep{Elia2016}. Both effects likely play an important role for the more evolved sources, making the derived physical parameters less reliable for the densest and most evolved sources.

Comparison of extinction properties of low-mass and high-mass clouds and cores have shown that only modest extinctions are required for a core to manifest as infrared dark \citep{Pillai2016}. In a forthcoming work, we show that at the distance of a few kpc, dust emission alone cannot reveal populations of low mass protostars (Class 0 and higher) embedded within clouds and line observations are crucial in distinguishing such cores from those that are genuinely quiescent (Pillai et al. in prep.). This is further supported by recent analysis of \citet{Feng2016} and \citet{Tan2016}, who have reported a bipolar outflow of a high-mass protostar associated with a 70\,\mum\ dark source. As outflows are associated with ongoing star formation, this source as well as the the analysis presented by \citet{Pillai2016} are good examples of the possible limitations of the dust classification when no source is detected at 70\,\mum.

Accordingly, the ATLASGAL team is investigating a wide range of molecular line tracers as well as different masers for their suitability as evolutionary stage indicators to supplement the rather rough discrimination obtained from the dust continuum parameters. An overview of these projects will be given in the next section.



\subsection{Complementary molecular observations}
Molecular lines can be used to independently derive and investigate the physical and chemical properties of the gas in star-forming regions. In the radio (centimeter to millimeter wavelengths) regime, extinction is in general negligible and it is therefore easier to see directly the effect of protostellar activity (e.g., outflows, gas warm-up and hot cores) in molecular lines.

Here we discuss the effects of mechanical and thermal feedback as probed by different tracers available for the Top100 sample, synthesizing the results of various studies of methanol and water masers as well as thermal emission from high density molecular tracers in terms of temperature evolution and probes of star formation activity.

\paragraph{Thermal feedback:} 
CO depletion as function of evolution was investigated by \citet{Giannetti2014}. The abundance of CO is correlated with the evolutionary stages in the low-mass regime \citep{Caselli1998,Bacmann2002}. In the Top100 we find the same trend: CO depletion decreases in more evolved clumps as a function of $L/M$, indirectly showing that the sources become warmer with time. The revised classification proposed here does not affect these earlier results. We also find that the C$^{17}$O(3--2) linewidth used in Section \ref{sect.avir} increases with evolutionary stage from a median value of 2.2\,km\,s$^{-1}$ for the 70\,\mum\ weak class to 5.3\,km\,s$^{-1}$ for the compact H\textsc{ii} regions under the new classification, confirming the earlier results and further bolstering the evolutionary sequence presented here.

Acetonitrile (CH$_3$CN) and methyl acetylene (CH$_3$CCH) are reliable thermometers of the gas.
Acetonitrile and methyl acetylene, as well as methanol will be discussed in detail by Giannetti et al. (in prep.), where preliminary results give further support for our revised classification scheme.


\setlength{\tabcolsep}{3pt}
\begin{table}
\caption{Association with Class\,II methanol and H$_2$O masers.}\label{masers} 
\begin{center}
\begin{tabular}{lccccc}
  \hline\hline 
  Class & Number$^a$ & MMB & MMB & HOPS & HOPS\\
     & of sources & assoc.  &ratio&  assoc. & ratio\\
  \hline

70\,\mum\ weak	& 16/10	 & 1 & 	0.06 & 1 & 0.10\\
Mid-IR weak & 34/27	 & 17 & 	0.50 & 15&  0.56 \\
Mid-IR bright & 36/17	 & 28 & 	0.78 & 14&  0.82\\
\hii\ region	& 25/17	 & 20 & 	0.80 & 11& 0.65 \\
  \hline
\end{tabular}
\end{center}
$^a$ The first number gives the number of sources within the region covered by the MMB while the second number gives the number of sources covered by HOPS.

\end{table}  
\setlength{\tabcolsep}{6pt}

\begin{figure}[tp!]
\begin{center}
\includegraphics[width=0.71\linewidth]{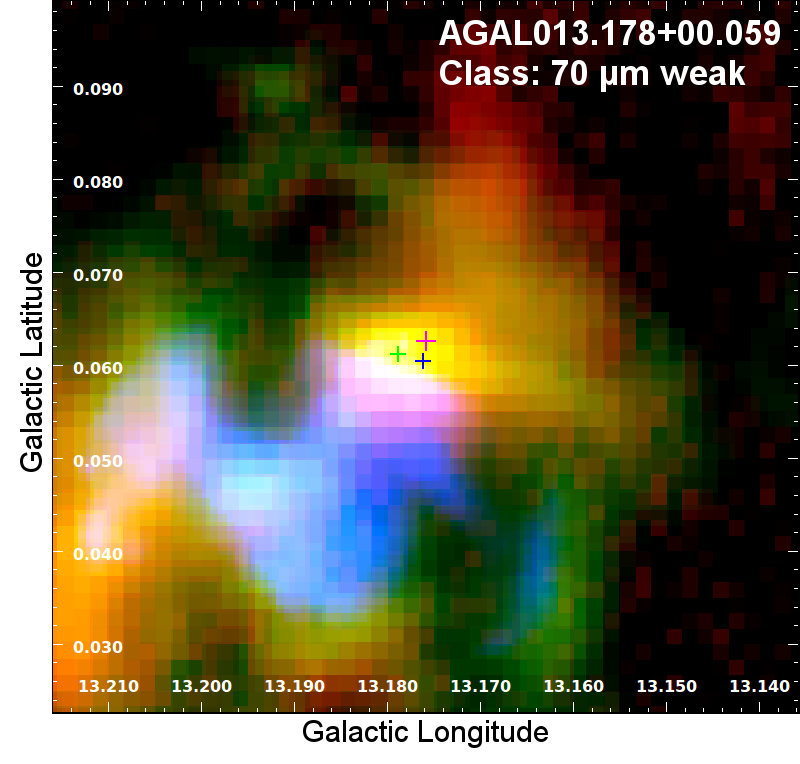}\\

   	\caption{Three color image of the 70\,\mum\ weak source AGAL013.178$+$00.059. Note the strong heating from the south east. The magenta, green and blue crosses mark the positions of the source peak, the associated methanol and water maser, respectively. Size: $\sim$3'$\times$3'; red: ATLASGAL\,870\,\mum; green: PACS\,160\,\mum; blue: PACS\,70\,\mum.}
	\label{fig.3color_g013}
   \end{center} 
\end{figure}

Furthermore, we searched for Class II methanol maser associations in the MMB survey \citep{Urquhart2013}. Class II methanol masers are radiatively pumped and therefore trace the radiation field from the central object. We found associations for 66 clumps, with 7 clumps being associated with 2 or more methanol masers. A detailed study of the properties of methanol maser associated clumps for the whole ATLASGAL survey is presented in \citet{Urquhart2013}, which includes 55 of the Top100 clumps. The remaining 12 clumps matched with MMB sources are new and have resulted from comparison with the final part of the MMB catalogue recently published in \citet{Breen2015}. The overall association rate between the Top100 and the MMB catalogue is $\sim$63\%, however, the association rate is significantly higher for the more evolved clump classifications as can be seen in Table\,\ref{masers}. The association ratio is similar for the mid-infrared bright and compact H\textsc{ii} region classes but drops to $\sim$50\% for the mid-infrared weak sources and close to zero for the 70\,\mum\ weak sources. We find  $\sim$72\% (i.e. 48/66) of all sources associated with a methanol maser to be either mid-infrared bright clumps or compact H\textsc{ii} regions.
The only 70\,\mum\ weak source with maser activity is AGAL013.178+00.059 (Fig. \ref{fig.3color_g013}). This source is removed from the SED analysis as discussed in Section \ref{sect.results}. AGAL013.178+00.059 is located near the edge of an evolved H\textsc{ii} region and is likely being externally heated and undergoing compression on one side. The external heating has resulted in some extended 70\,\mum\ emission that has made a definitive classification more difficult. Although care needs to be taken in the interpretation of the association between this clump and the masers, it does suggest that star formation may already be underway in some of these 70\,\mum\ weak clumps. This is further supported by the fact that a water maser is detected in the source (see discussion below).

\citet{Gallaway2013} examined the mid-infrared emission towards a larger sample of 776 methanol masers and found a similar fraction associated with mid-infrared emission. This suggests that methanol masers are associated with more evolved stages where the embedded objects are already producing significant luminosity and have started to warm their local environments. This also ties in nicely with a recent study of outflows towards methanol masers \citep{deVillers2014, deVillers2015} which found that the dynamical age of outflows associated with methanol masers are older than for the general population of molecular outflows reported in the literature.

\paragraph{Mechanical feedback:}
Standard tracers of outflow activity are for example high-velocity line-wings in CO and SiO lines, as well as water masers.



From the ATLASGAL Top100 sample, 36 sources lie in the I$^{st}$ Galactic quadrant, and are part of the sources investigated by \citet{Csengeri2016}. All of them are detected in the SiO(2--1) and (5--4) transitions except for two mid infrared weak sources. Line-wings are detected altogether towards 21 sources, including three clumps which are dark at 70\,\mum, demonstrating that molecular lines can be a more sensitive probe of star formation activity than dust alone. In a forthcoming paper Navarete et al. will extend the analysis on the content of molecular outflows in the Top100 sample, using observations in the Mid-$J$ CO lines to further test the classification scheme.

We have also searched for position coincidences with water masers identified by the H$_2$O Southern Galactic Plane Survey (HOPS; \citealt{walsh2011,walsh2014}). This survey covers Galactic longitudes $300\degr\ < \ell < 30\degr$ and Galactic latitudes $|b| < 0.5\degr$ and so only includes 71 of the Top100 sources.  In Table\,\ref{masers} we give the number of sources, the number of these associated with water masers and the fractional association rates for each evolutionary type. There is a clear trend for increasing frequency of association with water masers with the first three evolutionary stages. This is in agreement with the evolutionary scheme based on the dust analysis. Comparing the water maser association rates with the methanol maser association rates for the different stages we find them to be similar for the first earliest three stages, but the water maser association rate is noticeably lower for the more evolved \hii\ region stage. As water masers can be collisionaly excited and hence are an indicator for shocks and outflows, the H$_2$O maser found for AGAL013.178+00.059 (Figure \ref{fig.3color_g013}) can either be associated with material ejection or with compression from the south east, making this an interesting source to investigate possibly in more detail.

To summarize, the results of the molecular line observations as well as the associations between Class\,II methanol and H$_2$O masers support the evolutionary classification scheme outlined in Sect.\,2. In particular, the lack of Class\,II masers towards the 70\,$\mu$m-dark sources points to the star formation associated with these clumps being in a very early evolutionary phase, which is supported by the SED analysis. Despite the limited sensitivity of the SED analysis to the distinct evolutionary phases, our comprehensive programme of follow-up molecular line observations will be essential to refine the scheme and derive the physical properties of each stage. The results of these follow-up observations will be fully discussed in a number of forthcoming papers (e.g. Navarete et al. (in prep.), Giannetti et al. (in prep.), Urquhart et al. (in prep.)).

\section{Summary}\label{conclusion}

In this paper we characterize the properties of 110 massive star forming regions that have been selected to cover important evolutionary stages in the formation of massive stars. This sample includes examples of the coldest pre-stellar stages through to the formation of the ultra-compact \hii\ regions. Using multi-wavelength data, this sample has been classified into four distinct stages: 70\,$\mu$m weak, mid-infrared weak, mid-infrared bright and \hii\ regions.  Distances for the sources presented in this paper have been revised, incorporating the latest maser parallax measurements. 

Exploiting the dust continuum SEDs from mid-infrared to submm wavelengths, we derived dust temperatures and integrated fluxes; these are subsequently used to estimate reliable bolometric luminosities, clump masses and peak column densities for the whole sample. Comparing the physical properties of the clumps we found no significant differences between the distances, clump masses or physical sizes for the different evolutionary phases. 

\begin{enumerate}

\item The SED analysis provided useful constraints for the dust temperature and integrated emission. We find the dust temperatures to increase from 11\,K for the coldest sources to 41\,K for the H\textsc{ii} regions with an average value of 24.7\,K for the whole sample. A similar trend is seen in the mean bolometric luminosity increasing from $3.2\times10^3$\,L$_\odot$ for the 70\,\mum\ weak sample to $4.6\times10^5$\,L$_\odot$ for the compact H\textsc{ii} regions. The classification of the sample is further verified by the continuous increase in the bolometric luminosity to clump mass ratio from an average ratio of 3.4 for the youngest to 86.7 for the most evolved class, all of which is consistent with the proposed evolutionary scheme described in Sect.\,2. Although the SED analysis reveals significant differences between the different classes, we also find that there is a large overlap between the different evolutionary phases and additional information is required to confirm and refine the evolutionary scheme. \\

\item We have found that the majority of the sample satisfy the size-mass criterion for massive star formation and so have the potential to form massive stars in the future if not already currently doing so. Evaluating the clump stability we find that the vast majority are unstable against gravity and likely to be undergoing global collapse in the absence of significant magnetic support. It therefore seems likely that the majority of the sample will form a cluster that includes at least one massive star. Furthermore, the masses of some clumps exceed $4\times10^4$\,M$_\odot$ and have bolometric luminosities in excess of $3\times10^6$\,L$_\odot$ and are therefore likely to be forming the most massive, earliest O-type stars. \\

\item We also give an overview of complementary molecular line observations that are being conducted by the ATLASGAL team to classify the source in a more robust way. The association rates of methanol and water masers as well as molecular outflows reveals significant differences between the different phases. In addition thermal emission lines and SiO line profiles are used to discriminate the evolutionary stages. Examining a combination of these molecular line and continuum observation we have found strong support for the proposed evolutionary scheme and verified that the ATLASGAL Top100 sample represents a statistically significant catalog of massive star-forming clumps covering a range of evolutionary phases from the coldest and youngest 70~$\mu$m weak to the most evolved clumps hosting \hii\ regions still embedded in their natal environment.\\

\end{enumerate}

Using well established standard methods we have investigated the full evolutionary sequence from the earliest pre-stellar to the latest embedded \hii\ region phases on a well-selected sample of (almost) exclusively high-mass star forming clumps drawn from an unbiased dust survey covering the whole inner Galaxy for the first time. Furthermore, as the properties of this sample have been determined in a consistent way and the majority ($\sim$70\,\%) is located within 5\,kpc these sources are ideal for high resolution follow-up observations to investigate processes such as fragmentation, infall and outflows. This sample therefore provides a solid foundation for more detailed studies to investigate changes of the physical properties and kinematics of the gas through the process of massive star-formation in the future.


\begin{acknowledgements}

This work was partially carried out within the Collaborative Research Council 956, sub-project A6, funded by the Deutsche Forschungsgemeinschaft (DFG) and partially funded by the ERC Advanced Investigator Grant GLOSTAR (247078).

J. Urquhart and A. Giannetti acknowledge support from the \emph{Deut\-sche For\-schungs\-ge\-mein\-schaft, DFG\/}, via the Collaborative Research Centre (SFB) 956 'Conditions and Impact of Star Formation'.

T. Csengeri acknowledges support from the \emph{Deut\-sche For\-schungs\-ge\-mein\-schaft, DFG\/}, via the SPP (priority programme) 1573 'Physics of the ISM'.

This research made use of information from the ATLASGAL database at \url{http://atlasgal.mpifr-bonn.mpg.de/cgi-bin/ATLASGAL_DATABASE.cgi} supported by the MPIfR, Bonn, as well as information from the RMS database at
\url{http://rms.leeds.ac.uk/cgi-bin/public/RMS_DATABASE.cgi} which
was constructed with support from the Science and Technology Facilities Council of the UK.

This research made use of Astropy\footnote{\url{http://www.astropy.org}}, a community-developed core Python package for Astronomy \citep{AstropyCollaboration2013} and the Astropy affilate software package photutils\footnote{\url{https://github.com/astropy/photutils}}.

This research made use of Montage, funded by the National Aeronautics and Space Administration's Earth Science Technology Office, Computation Technologies Project, under Cooperative Agreement Number NCC5-626 between NASA and the California Institute of Technology. Montage is maintained by the NASA/IPAC Infrared Science Archive. This publication also makes use of data products from the Wide-field Infrared Survey Explorer, which is a joint project of the University of California, Los Angeles, and the Jet Propulsion Laboratory/California Institute of Technology, funded by the National Aeronautics and Space Administration.
\end{acknowledgements}

\bibliography{library}

\begin{thebibliography}{102}
\expandafter\ifx\csname natexlab\endcsname\relax\def\natexlab#1{#1}\fi

\bibitem[{{Aguirre} {et~al.}(2011){Aguirre}, {Ginsburg}, {Dunham}, {Drosback},
  {Bally}, {Battersby}, {Bradley}, {Cyganowski}, {Dowell}, {Evans}, {Glenn},
  {Harvey}, {Rosolowsky}, {Stringfellow}, {Walawender}, \&
  {Williams}}]{aguirre2011}
{Aguirre}, J.~E., {Ginsburg}, A.~G., {Dunham}, M.~K., {et~al.} 2011, \apjs,
  192, 4

\bibitem[{{Astropy Collaboration} {et~al.}(2013){Astropy Collaboration},
  {Robitaille}, {Tollerud}, {Greenfield}, {Droettboom}, {Bray}, {Aldcroft},
  {Davis}, {Ginsburg}, {Price-Whelan}, {Kerzendorf}, {Conley}, {Crighton},
  {Barbary}, {Muna}, {Ferguson}, {Grollier}, {Parikh}, {Nair}, {Unther},
  {Deil}, {Woillez}, {Conseil}, {Kramer}, {Turner}, {Singer}, {Fox}, {Weaver},
  {Zabalza}, {Edwards}, {Azalee Bostroem}, {Burke}, {Casey}, {Crawford},
  {Dencheva}, {Ely}, {Jenness}, {Labrie}, {Lim}, {Pierfederici}, {Pontzen},
  {Ptak}, {Refsdal}, {Servillat}, \& {Streicher}}]{AstropyCollaboration2013}
{Astropy Collaboration}, {Robitaille}, T.~P., {Tollerud}, E.~J., {et~al.} 2013,
  \aap, 558, A33

\bibitem[{{Bacmann} {et~al.}(2002){Bacmann}, {Lefloch}, {Ceccarelli},
  {Castets}, {Steinacker}, \& {Loinard}}]{Bacmann2002}
{Bacmann}, A., {Lefloch}, B., {Ceccarelli}, C., {et~al.} 2002, \aap, 389, L6

\bibitem[{{Benjamin et al.} \& {et al.}(2003)}]{benjamin2003}
{Benjamin et al.}, R.~A. \& {et al.} 2003, \pasp, 115, 953

\bibitem[{{Bernard} {et~al.}(2010){Bernard}, {Paradis}, {Marshall}, {Montier},
  {Lagache}, {Paladini}, {Veneziani}, {Brunt}, {Mottram}, {Martin},
  {Ristorcelli}, {Noriega-Crespo}, {Compi{\`e}gne}, {Flagey}, {Anderson},
  {Popescu}, {Tuffs}, {Reach}, {White}, {Benedettini}, {Calzoletti},
  {Digiorgio}, {Faustini}, {Juvela}, {Joblin}, {Joncas}, {Mivilles-Deschenes},
  {Olmi}, {Traficante}, {Piacentini}, {Zavagno}, \& {Molinari}}]{Bernard2010}
{Bernard}, J.-P., {Paradis}, D., {Marshall}, D.~J., {et~al.} 2010, \aap, 518,
  L88

\bibitem[{{Bertoldi} \& {McKee}(1992)}]{Bertoldi1992}
{Bertoldi}, F. \& {McKee}, C.~F. 1992, \apj, 395, 140

\bibitem[{{Beuther} {et~al.}(2010){Beuther}, {Henning}, {Linz}, {Krause},
  {Nielbock}, \& {Steinacker}}]{Beuther2010}
{Beuther}, H., {Henning}, T., {Linz}, H., {et~al.} 2010, \aap, 518, L78

\bibitem[{{Bonnor}(1955)}]{Bonnor1955}
{Bonnor}, W.~B. 1955, \mnras, 115, 310

\bibitem[{{Breen} {et~al.}(2015){Breen}, {Fuller}, {Caswell}, {Green},
  {Avison}, {Ellingsen}, {Gray}, {Pestalozzi}, {Quinn}, {Richards}, {Thompson},
  \& {Voronkov}}]{Breen2015}
{Breen}, S.~L., {Fuller}, G.~A., {Caswell}, J.~L., {et~al.} 2015, \mnras, 450,
  4109

\bibitem[{{Brunthaler} {et~al.}(2009){Brunthaler}, {Reid}, {Menten}, {Zheng},
  {Moscadelli}, \& {Xu}}]{Brunthaler2009}
{Brunthaler}, A., {Reid}, M.~J., {Menten}, K.~M., {et~al.} 2009, \apj, 693, 424

\bibitem[{{Busfield} {et~al.}(2006){Busfield}, {Purcell}, {Hoare}, {Lumsden},
  {Moore}, \& {Oudmaijer}}]{busfield2006}
{Busfield}, A.~L., {Purcell}, C.~R., {Hoare}, M.~G., {et~al.} 2006, \mnras,
  366, 1096

\bibitem[{{Carey} {et~al.}(2009){Carey}, {Noriega-Crespo}, {Mizuno}, {Shenoy},
  {Paladini}, {Kraemer}, {Price}, {Flagey}, {Ryan}, {Ingalls}, {Kuchar},
  {Pinheiro Gon{\c c}alves}, {Indebetouw}, {Billot}, {Marleau}, {Padgett},
  {Rebull}, {Bressert}, {Ali}, {Molinari}, {Martin}, {Berriman}, {Boulanger},
  {Latter}, {Miville-Deschenes}, {Shipman}, \& {Testi}}]{Carey2009}
{Carey}, S.~J., {Noriega-Crespo}, A., {Mizuno}, D.~R., {et~al.} 2009, \pasp,
  121, 76

\bibitem[{{Caselli} {et~al.}(1998){Caselli}, {Walmsley}, {Terzieva}, \&
  {Herbst}}]{Caselli1998}
{Caselli}, P., {Walmsley}, C.~M., {Terzieva}, R., \& {Herbst}, E. 1998, \apj,
  499, 234

\bibitem[{{Caswell} {et~al.}(1975){Caswell}, {Murray}, {Roger}, {Cole}, \&
  {Cooke}}]{caswell1975}
{Caswell}, J.~L., {Murray}, J.~D., {Roger}, R.~S., {Cole}, D.~J., \& {Cooke},
  D.~J. 1975, \aap, 45, 239

\bibitem[{{Churchwell} {et~al.}(2009){Churchwell}, {Babler}, {Meade},
  {Whitney}, {Benjamin}, {Indebetouw}, {Cyganowski}, {Robitaille}, {Povich},
  {Watson}, \& {Bracker}}]{Churchwell2009}
{Churchwell}, E., {Babler}, B.~L., {Meade}, M.~R., {et~al.} 2009, \pasp, 121,
  213

\bibitem[{{Contreras} {et~al.}(2013){Contreras}, {Schuller}, {Urquhart},
  {Csengeri}, {Wyrowski}, {Beuther}, {Bontemps}, {Bronfman}, {Henning},
  {Menten}, {Schilke}, {Walmsley}, {Wienen}, {Tackenberg}, \&
  {Linz}}]{Contreras2013}
{Contreras}, Y., {Schuller}, F., {Urquhart}, J.~S., {et~al.} 2013, \aap, 549,
  A45

\bibitem[{{Crutcher}(2012)}]{Crutcher2012}
{Crutcher}, R.~M. 2012, \araa, 50, 29

\bibitem[{{Csengeri} {et~al.}(2016){Csengeri}, {Leurini}, {Wyrowski},
  {Urquhart}, {Menten}, {Walmsley}, {Bontemps}, {Wienen}, {Beuther}, {Motte},
  {Nguyen-Luong}, {Schilke}, {Schuller}, {Zavagno}, \& {Sanna}}]{Csengeri2016}
{Csengeri}, T., {Leurini}, S., {Wyrowski}, F., {et~al.} 2016, \aap, 586, A149

\bibitem[{{Csengeri} {et~al.}(2014){Csengeri}, {Urquhart}, {Schuller}, {Motte},
  {Bontemps}, {Wyrowski}, {Menten}, {Bronfman}, {Beuther}, {Henning}, {Testi},
  {Zavagno}, \& {Walmsley}}]{Csengeri2014}
{Csengeri}, T., {Urquhart}, J.~S., {Schuller}, F., {et~al.} 2014, \aap, 565,
  A75

\bibitem[{{Davies} {et~al.}(2012){Davies}, {Clark}, {Trombley}, {Figer},
  {Najarro}, {Crowther}, {Kudritzki}, {Thompson}, {Urquhart}, \&
  {Hindson}}]{Davies2012}
{Davies}, B., {Clark}, J.~S., {Trombley}, C., {et~al.} 2012, \mnras, 419, 1871

\bibitem[{{de Villiers} {et~al.}(2014){de Villiers}, {Chrysostomou},
  {Thompson}, {Ellingsen}, {Urquhart}, {Breen}, {Burton}, {Csengeri}, \&
  {Ward-Thompson}}]{deVillers2014}
{de Villiers}, H.~M., {Chrysostomou}, A., {Thompson}, M.~A., {et~al.} 2014,
  \mnras, 444, 566

\bibitem[{{de Villiers} {et~al.}(2015){de Villiers}, {Chrysostomou},
  {Thompson}, {Urquhart}, {Breen}, {Burton}, {Ellingsen}, {Fuller},
  {Pestalozzi}, {Voronkov}, \& {Ward-Thompson}}]{deVillers2015}
{de Villiers}, H.~M., {Chrysostomou}, A., {Thompson}, M.~A., {et~al.} 2015,
  \mnras, 449, 119

\bibitem[{{de Wit} {et~al.}(2004){de Wit}, {Testi}, {Palla}, {Vanzi}, \&
  {Zinnecker}}]{de-wit2004}
{de Wit}, W.~J., {Testi}, L., {Palla}, F., {Vanzi}, L., \& {Zinnecker}, H.
  2004, \aap, 425, 937

\bibitem[{{Egan} {et~al.}(2003){Egan}, {Price}, {Kraemer}, {Mizuno}, {Carey},
  {Wright}, {Engelke}, {Cohen}, \& {Gugliotti}}]{Egan2003}
{Egan}, M.~P., {Price}, S.~D., {Kraemer}, K.~E., {et~al.} 2003, VizieR Online
  Data Catalog, 5114, 0

\bibitem[{{Elia} {et~al.}(2013){Elia}, {Molinari}, {Fukui}, {Schisano}, {Olmi},
  {Veneziani}, {Hayakawa}, {Pestalozzi}, {Schneider}, {Benedettini}, {di
  Giorgio}, {Ikhenaode}, {Mizuno}, {Onishi}, {Pezzuto}, {Piazzo}, {Polychroni},
  {Rygl}, {Yamamoto}, \& {Maruccia}}]{Elia2013}
{Elia}, D., {Molinari}, S., {Fukui}, Y., {et~al.} 2013, \apj, 772, 45

\bibitem[{{Elia} \& {Pezzuto}(2016)}]{Elia2016}
{Elia}, D. \& {Pezzuto}, S. 2016, \mnras, 461, 1328

\bibitem[{{Elia} {et~al.}(2010){Elia}, {Schisano}, {Molinari}, {Robitaille},
  {Angl{\'e}s-Alc{\'a}zar}, {Bally}, {Battersby}, {Benedettini}, {Billot},
  {Calzoletti}, {di Giorgio}, {Faustini}, {Li}, {Martin}, {Morgan}, {Motte},
  {Mottram}, {Natoli}, {Olmi}, {Paladini}, {Piacentini}, {Pestalozzi},
  {Pezzuto}, {Polychroni}, {Smith}, {Strafella}, {Stringfellow}, {Testi},
  {Thompson}, {Traficante}, \& {Veneziani}}]{Elia2010}
{Elia}, D., {Schisano}, E., {Molinari}, S., {et~al.} 2010, \aap, 518, L97

\bibitem[{{Feng} {et~al.}(2016){Feng}, {Beuther}, {Zhang}, {Liu}, {Zhang},
  {Wang}, \& {Qiu}}]{Feng2016}
{Feng}, S., {Beuther}, H., {Zhang}, Q., {et~al.} 2016, \apj, 828, 100

\bibitem[{{Gallaway} {et~al.}(2013){Gallaway}, {Thompson}, {Lucas}, {Fuller},
  {Caswell}, {Green}, {Voronkov}, {Breen}, {Quinn}, {Ellingsen}, {Avison},
  {Ward-Thompson}, \& {Cox}}]{Gallaway2013}
{Gallaway}, M., {Thompson}, M.~A., {Lucas}, P.~W., {et~al.} 2013, \mnras, 430,
  808

\bibitem[{{Giannetti} {et~al.}(2014){Giannetti}, {Wyrowski}, {Brand},
  {Csengeri}, {Fontani}, {Walmsley}, {Nguyen Luong}, {Beuther}, {Schuller},
  {G{\"u}sten}, \& {Menten}}]{Giannetti2014}
{Giannetti}, A., {Wyrowski}, F., {Brand}, J., {et~al.} 2014, ArXiv e-prints
  [\eprint[arXiv]{1407.2215}]

\bibitem[{{Giannetti} {et~al.}(2015){Giannetti}, {Wyrowski}, {Leurini},
  {Urquhart}, {Csengeri}, {Menten}, {Bronfman}, \& {van der
  Tak}}]{Giannetti2015}
{Giannetti}, A., {Wyrowski}, F., {Leurini}, S., {et~al.} 2015, \aap, 580, L7

\bibitem[{{Green} \& {McClure-Griffiths}(2011)}]{Green2011}
{Green}, J.~A. \& {McClure-Griffiths}, N.~M. 2011, \mnras, 417, 2500

\bibitem[{{Griffin} {et~al.}(2010){Griffin}, {Abergel}, {Abreu}, {Ade},
  {Andr{\'e}}, {Augueres}, {Babbedge}, {Bae}, {Baillie}, {Baluteau}, {Barlow},
  {Bendo}, {Benielli}, {Bock}, {Bonhomme}, {Brisbin}, {Brockley-Blatt},
  {Caldwell}, {Cara}, {Castro-Rodriguez}, {Cerulli}, {Chanial}, {Chen},
  {Clark}, {Clements}, {Clerc}, {Coker}, {Communal}, {Conversi}, {Cox},
  {Crumb}, {Cunningham}, {Daly}, {Davis}, {de Antoni}, {Delderfield}, {Devin},
  {di Giorgio}, {Didschuns}, {Dohlen}, {Donati}, {Dowell}, {Dowell}, {Duband},
  {Dumaye}, {Emery}, {Ferlet}, {Ferrand}, {Fontignie}, {Fox}, {Franceschini},
  {Frerking}, {Fulton}, {Garcia}, {Gastaud}, {Gear}, {Glenn}, {Goizel},
  {Griffin}, {Grundy}, {Guest}, {Guillemet}, {Hargrave}, {Harwit}, {Hastings},
  {Hatziminaoglou}, {Herman}, {Hinde}, {Hristov}, {Huang}, {Imhof}, {Isaak},
  {Israelsson}, {Ivison}, {Jennings}, {Kiernan}, {King}, {Lange}, {Latter},
  {Laurent}, {Laurent}, {Leeks}, {Lellouch}, {Levenson}, {Li}, {Li},
  {Lilienthal}, {Lim}, {Liu}, {Lu}, {Madden}, {Mainetti}, {Marliani}, {McKay},
  {Mercier}, {Molinari}, {Morris}, {Moseley}, {Mulder}, {Mur}, {Naylor},
  {Nguyen}, {O'Halloran}, {Oliver}, {Olofsson}, {Olofsson}, {Orfei}, {Page},
  {Pain}, {Panuzzo}, {Papageorgiou}, {Parks}, {Parr-Burman}, {Pearce},
  {Pearson}, {P{\'e}rez-Fournon}, {Pinsard}, {Pisano}, {Podosek}, {Pohlen},
  {Polehampton}, {Pouliquen}, {Rigopoulou}, {Rizzo}, {Roseboom}, {Roussel},
  {Rowan-Robinson}, {Rownd}, {Saraceno}, {Sauvage}, {Savage}, {Savini},
  {Sawyer}, {Scharmberg}, {Schmitt}, {Schneider}, {Schulz}, {Schwartz},
  {Shafer}, {Shupe}, {Sibthorpe}, {Sidher}, {Smith}, {Smith}, {Smith},
  {Spencer}, {Stobie}, {Sudiwala}, {Sukhatme}, {Surace}, {Stevens}, {Swinyard},
  {Trichas}, {Tourette}, {Triou}, {Tseng}, {Tucker}, {Turner}, {Vaccari},
  {Valtchanov}, {Vigroux}, {Virique}, {Voellmer}, {Walker}, {Ward}, {Waskett},
  {Weilert}, {Wesson}, {White}, {Whitehouse}, {Wilson}, {Winter}, {Woodcraft},
  {Wright}, {Xu}, {Zavagno}, {Zemcov}, {Zhang}, \& {Zonca}}]{Griffin2010}
{Griffin}, M.~J., {Abergel}, A., {Abreu}, A., {et~al.} 2010, \aap, 518, L3

\bibitem[{{G{\"u}sten} {et~al.}(2006){G{\"u}sten}, {Nyman}, {Schilke},
  {Menten}, {Cesarsky}, \& {Booth}}]{Gusten2006}
{G{\"u}sten}, R., {Nyman}, L.~{\AA}., {Schilke}, P., {et~al.} 2006, \aap, 454,
  L13

\bibitem[{{Gutermuth} \& {Heyer}(2015)}]{Gutermuth2015}
{Gutermuth}, R.~A. \& {Heyer}, M. 2015, \aj, 149, 64

\bibitem[{{Heyer} {et~al.}(2016){Heyer}, {Gutermuth}, {Urquhart}, {Csengeri},
  {Wienen}, {Leurini}, {Menten}, \& {Wyrowski}}]{Heyer2016}
{Heyer}, M., {Gutermuth}, R., {Urquhart}, J.~S., {et~al.} 2016, \aap, 588, A29

\bibitem[{{Hildebrand}(1983)}]{Hildebrand1983}
{Hildebrand}, R.~H. 1983, \qjras, 24, 267

\bibitem[{{Hoare} {et~al.}(2012){Hoare}, {Purcell}, {Churchwell}, {Diamond},
  {Cotton}, {Chandler}, {Smethurst}, {Kurtz}, {Mundy}, {Dougherty}, {Fender},
  {Fuller}, {Jackson}, {Garrington}, {Gledhill}, {Goldsmith}, {Lumsden},
  {Mart{\'{\i}}}, {Moore}, {Muxlow}, {Oudmaijer}, {Pandian}, {Paredes},
  {Shepherd}, {Spencer}, {Thompson}, {Umana}, {Urquhart}, \&
  {Zijlstra}}]{hoare2012}
{Hoare}, M.~G., {Purcell}, C.~R., {Churchwell}, E.~B., {et~al.} 2012, \pasp,
  124, 939

\bibitem[{{Immer} {et~al.}(2013){Immer}, {Reid}, {Menten}, {Brunthaler}, \&
  {Dame}}]{Immer2013}
{Immer}, K., {Reid}, M.~J., {Menten}, K.~M., {Brunthaler}, A., \& {Dame}, T.~M.
  2013, \aap, 553, A117

\bibitem[{{Immer} {et~al.}(2012){Immer}, {Schuller}, {Omont}, \&
  {Menten}}]{immer2012}
{Immer}, K., {Schuller}, F., {Omont}, A., \& {Menten}, K.~M. 2012, \aap, 537,
  A121

\bibitem[{{Kauffmann} {et~al.}(2008){Kauffmann}, {Bertoldi}, {Bourke}, {Evans},
  \& {Lee}}]{Kauffmann2008}
{Kauffmann}, J., {Bertoldi}, F., {Bourke}, T.~L., {Evans}, II, N.~J., \& {Lee},
  C.~W. 2008, \aap, 487, 993

\bibitem[{{Kauffmann} {et~al.}(2013){Kauffmann}, {Pillai}, \&
  {Goldsmith}}]{Kauffmann2013}
{Kauffmann}, J., {Pillai}, T., \& {Goldsmith}, P.~F. 2013, \apj, 779, 185

\bibitem[{{Kauffmann} {et~al.}(2010){Kauffmann}, {Pillai}, {Shetty}, {Myers},
  \& {Goodman}}]{kauffmann2010a}
{Kauffmann}, J., {Pillai}, T., {Shetty}, R., {Myers}, P.~C., \& {Goodman},
  A.~A. 2010, \apj, 712, 1137

\bibitem[{{Kennicutt}(2005)}]{kennicutt2005}
{Kennicutt}, R.~C. 2005, in IAU Symposium, Vol. 227, Massive Star Birth: A
  Crossroads of Astrophysics, ed. R.~{Cesaroni}, M.~{Felli}, E.~{Churchwell},
  \& M.~{Walmsley}, 3--11

\bibitem[{{Kurayama} {et~al.}(2011){Kurayama}, {Nakagawa}, {Sawada-Satoh},
  {Sato}, {Honma}, {Sunada}, {Hirota}, \& {Imai}}]{Kurayama2011}
{Kurayama}, T., {Nakagawa}, A., {Sawada-Satoh}, S., {et~al.} 2011, \pasj, 63,
  513

\bibitem[{{Lucas} {et~al.}(2008){Lucas}, {Hoare}, {Longmore}, {Schr{\"o}der},
  {Davis}, {Adamson}, {Bandyopadhyay}, {de Grijs}, {Smith}, {Gosling},
  {Mitchison}, {G{\'a}sp{\'a}r}, {Coe}, {Tamura}, {Parker}, {Irwin}, {Hambly},
  {Bryant}, {Collins}, {Cross}, {Evans}, {Gonzalez-Solares}, {Hodgkin},
  {Lewis}, {Read}, {Riello}, {Sutorius}, {Lawrence}, {Drew}, {Dye}, \&
  {Thompson}}]{lucas2008}
{Lucas}, P.~W., {Hoare}, M.~G., {Longmore}, A., {et~al.} 2008, \mnras, 391, 136

\bibitem[{{Mois{\'e}s} {et~al.}(2011){Mois{\'e}s}, {Damineli}, {Figuer{\^e}do},
  {Blum}, {Conti}, \& {Barbosa}}]{moises2011}
{Mois{\'e}s}, A.~P., {Damineli}, A., {Figuer{\^e}do}, E., {et~al.} 2011,
  \mnras, 411, 705

\bibitem[{{Molinari} {et~al.}(2008){Molinari}, {Pezzuto}, {Cesaroni}, {Brand},
  {Faustini}, \& {Testi}}]{Molinari2008}
{Molinari}, S., {Pezzuto}, S., {Cesaroni}, R., {et~al.} 2008, \aap, 481, 345

\bibitem[{{Molinari} {et~al.}(2016){Molinari}, {Schisano}, {Elia},
  {Pestalozzi}, {Traficante}, {Pezzuto}, {Swinyard}, {Noriega-Crespo}, {Bally},
  {Moore}, {Plume}, {Zavagno}, {di Giorgio}, {Liu}, {Pilbratt}, {Mottram},
  {Russeil}, {Piazzo}, {Veneziani}, {Benedettini}, {Calzoletti}, {Faustini},
  {Natoli}, {Piacentini}, {Merello}, {Palmese}, {Del Grande}, {Polychroni},
  {Rygl}, {Polenta}, {Barlow}, {Bernard}, {Martin}, {Testi}, {Ali},
  {Andr{\'e}}, {Beltr{\'a}n}, {Billot}, {Carey}, {Cesaroni}, {Compi{\`e}gne},
  {Eden}, {Fukui}, {Garcia-Lario}, {Hoare}, {Huang}, {Joncas}, {Lim}, {Lord},
  {Martinavarro-Armengol}, {Motte}, {Paladini}, {Paradis}, {Peretto},
  {Robitaille}, {Schilke}, {Schneider}, {Schulz}, {Sibthorpe}, {Strafella},
  {Thompson}, {Umana}, {Ward-Thompson}, \& {Wyrowski}}]{Molinari2016}
{Molinari}, S., {Schisano}, E., {Elia}, D., {et~al.} 2016, \aap, 591, A149

\bibitem[{{Molinari} {et~al.}(2010){Molinari}, {Swinyard}, {Bally}, {Barlow},
  {Bernard}, {Martin}, {Moore}, {Noriega-Crespo}, {Plume}, {Testi}, {Zavagno},
  {Abergel}, {Ali}, {Andr{\'e}}, {Baluteau}, {Benedettini}, {Bern{\'e}},
  {Billot}, {Blommaert}, {Bontemps}, {Boulanger}, {Brand}, {Brunt}, {Burton},
  {Campeggio}, {Carey}, {Caselli}, {Cesaroni}, {Cernicharo}, {Chakrabarti},
  {Chrysostomou}, {Codella}, {Cohen}, {Compiegne}, {Davis}, {de Bernardis}, {de
  Gasperis}, {Di Francesco}, {di Giorgio}, {Elia}, {Faustini}, {Fischera},
  {Fukui}, {Fuller}, {Ganga}, {Garcia-Lario}, {Giard}, {Giardino}, {Glenn},
  {Goldsmith}, {Griffin}, {Hoare}, {Huang}, {Jiang}, {Joblin}, {Joncas},
  {Juvela}, {Kirk}, {Lagache}, {Li}, {Lim}, {Lord}, {Lucas}, {Maiolo},
  {Marengo}, {Marshall}, {Masi}, {Massi}, {Matsuura}, {Meny}, {Minier},
  {Miville-Desch{\^e}nes}, {Montier}, {Motte}, {M{\"u}ller}, {Natoli}, {Neves},
  {Olmi}, {Paladini}, {Paradis}, {Pestalozzi}, {Pezzuto}, {Piacentini},
  {Pomar{\`e}s}, {Popescu}, {Reach}, {Richer}, {Ristorcelli}, {Roy}, {Royer},
  {Russeil}, {Saraceno}, {Sauvage}, {Schilke}, {Schneider-Bontemps},
  {Schuller}, {Schultz}, {Shepherd}, {Sibthorpe}, {Smith}, {Smith},
  {Spinoglio}, {Stamatellos}, {Strafella}, {Stringfellow}, {Sturm}, {Taylor},
  {Thompson}, {Tuffs}, {Umana}, {Valenziano}, {Vavrek}, {Viti}, {Waelkens},
  {Ward-Thompson}, {White}, {Wyrowski}, {Yorke}, \& {Zhang}}]{molinari2010}
{Molinari}, S., {Swinyard}, B., {Bally}, J., {et~al.} 2010, \pasp, 122, 314

\bibitem[{{Motte} {et~al.}(2010){Motte}, {Zavagno}, {Bontemps}, {Schneider},
  {Hennemann}, {di Francesco}, {Andr{\'e}}, {Saraceno}, {Griffin}, {Marston},
  {Ward-Thompson}, {White}, {Minier}, {Men'shchikov}, {Hill}, {Abergel},
  {Anderson}, {Aussel}, {Balog}, {Baluteau}, {Bernard}, {Cox}, {Csengeri},
  {Deharveng}, {Didelon}, {di Giorgio}, {Hargrave}, {Huang}, {Kirk}, {Leeks},
  {Li}, {Martin}, {Molinari}, {Nguyen-Luong}, {Olofsson}, {Persi}, {Peretto},
  {Pezzuto}, {Roussel}, {Russeil}, {Sadavoy}, {Sauvage}, {Sibthorpe},
  {Spinoglio}, {Testi}, {Teyssier}, {Vavrek}, {Wilson}, \&
  {Woodcraft}}]{Motte2010}
{Motte}, F., {Zavagno}, A., {Bontemps}, S., {et~al.} 2010, \aap, 518, L77

\bibitem[{{Mottram} {et~al.}(2011{\natexlab{a}}){Mottram}, {Hoare}, {Davies},
  {Lumsden}, {Oudmaijer}, {Urquhart}, {Moore}, {Cooper}, \&
  {Stead}}]{Mottram2011}
{Mottram}, J.~C., {Hoare}, M.~G., {Davies}, B., {et~al.} 2011{\natexlab{a}},
  \apjl, 730, L33

\bibitem[{{Mottram} {et~al.}(2011{\natexlab{b}}){Mottram}, {Hoare}, {Urquhart},
  {Lumsden}, {Oudmaijer}, {Robitaille}, {Moore}, {Davies}, \&
  {Stead}}]{Mottram2011a}
{Mottram}, J.~C., {Hoare}, M.~G., {Urquhart}, J.~S., {et~al.}
  2011{\natexlab{b}}, \aap, 525, A149

\bibitem[{{Nguyen Luong} {et~al.}(2011){Nguyen Luong}, {Motte}, {Hennemann},
  {Hill}, {Rygl}, {Schneider}, {Bontemps}, {Men'shchikov}, {Andr{\'e}},
  {Peretto}, {Anderson}, {Arzoumanian}, {Deharveng}, {Didelon}, {di Francesco},
  {Griffin}, {Kirk}, {K{\"o}nyves}, {Martin}, {Maury}, {Minier}, {Molinari},
  {Pestalozzi}, {Pezzuto}, {Reid}, {Roussel}, {Sauvage}, {Schuller}, {Testi},
  {Ward-Thompson}, {White}, \& {Zavagno}}]{NguyenLuong2011}
{Nguyen Luong}, Q., {Motte}, F., {Hennemann}, M., {et~al.} 2011, \aap, 535, A76

\bibitem[{{Ossenkopf} \& {Henning}(1994)}]{Ossenkopf1994}
{Ossenkopf}, V. \& {Henning}, T. 1994, \aap, 291, 943

\bibitem[{{Pilbratt} {et~al.}(2010){Pilbratt}, {Riedinger}, {Passvogel},
  {Crone}, {Doyle}, {Gageur}, {Heras}, {Jewell}, {Metcalfe}, {Ott}, \&
  {Schmidt}}]{Pilbratt2010}
{Pilbratt}, G.~L., {Riedinger}, J.~R., {Passvogel}, T., {et~al.} 2010, \aap,
  518, L1

\bibitem[{{Pillai}(2016)}]{Pillai2016}
{Pillai}, T. 2016, in EAS Publications Series, Vol.~75, EAS Publications
  Series, 245--250

\bibitem[{{Poglitsch} {et~al.}(2010){Poglitsch}, {Waelkens}, {Geis},
  {Feuchtgruber}, {Vandenbussche}, {Rodriguez}, {Krause}, {Renotte}, {van
  Hoof}, {Saraceno}, {Cepa}, {Kerschbaum}, {Agn{\`e}se}, {Ali}, {Altieri},
  {Andreani}, {Augueres}, {Balog}, {Barl}, {Bauer}, {Belbachir}, {Benedettini},
  {Billot}, {Boulade}, {Bischof}, {Blommaert}, {Callut}, {Cara}, {Cerulli},
  {Cesarsky}, {Contursi}, {Creten}, {De Meester}, {Doublier}, {Doumayrou},
  {Duband}, {Exter}, {Genzel}, {Gillis}, {Gr{\"o}zinger}, {Henning},
  {Herreros}, {Huygen}, {Inguscio}, {Jakob}, {Jamar}, {Jean}, {de Jong},
  {Katterloher}, {Kiss}, {Klaas}, {Lemke}, {Lutz}, {Madden}, {Marquet},
  {Martignac}, {Mazy}, {Merken}, {Montfort}, {Morbidelli}, {M{\"u}ller},
  {Nielbock}, {Okumura}, {Orfei}, {Ottensamer}, {Pezzuto}, {Popesso},
  {Putzeys}, {Regibo}, {Reveret}, {Royer}, {Sauvage}, {Schreiber}, {Stegmaier},
  {Schmitt}, {Schubert}, {Sturm}, {Thiel}, {Tofani}, {Vavrek}, {Wetzstein},
  {Wieprecht}, \& {Wiezorrek}}]{Poglitsch2010}
{Poglitsch}, A., {Waelkens}, C., {Geis}, N., {et~al.} 2010, \aap, 518, L2

\bibitem[{Price {et~al.}(2001)Price, Egan, Carey, Mizuno, \&
  Kuchar}]{Price2001}
Price, S.~D., Egan, M.~P., Carey, S.~J., Mizuno, D.~R., \& Kuchar, T.~A. 2001,
  The Astronomical Journal, 121, 2819

\bibitem[{{Purcell} {et~al.}(2013){Purcell}, {Hoare}, {Cotton}, {Lumsden},
  {Urquhart}, {Chandler}, {Churchwell}, {Diamond}, {Dougherty}, {Fender},
  {Fuller}, {Garrington}, {Gledhill}, {Goldsmith}, {Hindson}, {Jackson},
  {Kurtz}, {Mart{\'{\i}}}, {Moore}, {Mundy}, {Muxlow}, {Oudmaijer}, {Pandian},
  {Paredes}, {Shepherd}, {Smethurst}, {Spencer}, {Thompson}, {Umana}, \&
  {Zijlstra}}]{purcell2013}
{Purcell}, C.~R., {Hoare}, M.~G., {Cotton}, W.~D., {et~al.} 2013, \apjs, 205, 1

\bibitem[{Razali(2011)}]{Razali2011}
Razali, Nornadiah;~Wah, Y.~B. 2011, Journal of Statistical Modeling and
  Analytics, 2, 21

\bibitem[{{Reid} {et~al.}(2014){Reid}, {Menten}, {Brunthaler}, {Zheng}, {Dame},
  {Xu}, {Wu}, {Zhang}, {Sanna}, {Sato}, {Hachisuka}, {Choi}, {Immer},
  {Moscadelli}, {Rygl}, \& {Bartkiewicz}}]{Reid2014}
{Reid}, M.~J., {Menten}, K.~M., {Brunthaler}, A., {et~al.} 2014, \apj, 783, 130

\bibitem[{{Robitaille} {et~al.}(2007){Robitaille}, {Whitney}, {Indebetouw}, \&
  {Wood}}]{Robitaille2007a}
{Robitaille}, T.~P., {Whitney}, B.~A., {Indebetouw}, R., \& {Wood}, K. 2007,
  \apjs, 169, 328

\bibitem[{{Roman-Duval} {et~al.}(2009){Roman-Duval}, {Jackson}, {Heyer},
  {Johnson}, {Rathborne}, {Shah}, \& {Simon}}]{roman2009}
{Roman-Duval}, J., {Jackson}, J.~M., {Heyer}, M., {et~al.} 2009, \apj, 699,
  1153

\bibitem[{{Sanna} {et~al.}(2014){Sanna}, {Reid}, {Menten}, {Dame}, {Zhang},
  {Sato}, {Brunthaler}, {Moscadelli}, \& {Immer}}]{Sanna2014}
{Sanna}, A., {Reid}, M.~J., {Menten}, K.~M., {et~al.} 2014, \apj, 781, 108

\bibitem[{{Sato} {et~al.}(2010){Sato}, {Reid}, {Brunthaler}, \&
  {Menten}}]{Sato2010}
{Sato}, M., {Reid}, M.~J., {Brunthaler}, A., \& {Menten}, K.~M. 2010, \apj,
  720, 1055

\bibitem[{{Sato} {et~al.}(2014){Sato}, {Wu}, {Immer}, {Zhang}, {Sanna}, {Reid},
  {Dame}, {Brunthaler}, \& {Menten}}]{Sato2014}
{Sato}, M., {Wu}, Y.~W., {Immer}, K., {et~al.} 2014, \apj, 793, 72

\bibitem[{{Schuller} {et~al.}(2009){Schuller}, {Menten}, {Contreras},
  {Wyrowski}, \& {Schilke}}]{schuller2009}
{Schuller}, F., {Menten}, K.~M., {Contreras}, Y., {Wyrowski}, F., \& {Schilke},
  e.~a. 2009, \aap, 504, 415

\bibitem[{{Siringo} {et~al.}(2009){Siringo}, {Kreysa}, {Kov{\'a}cs},
  {Schuller}, {Wei{\ss}}, {Esch}, {Gem{\"u}nd}, {Jethava}, {Lundershausen},
  {Colin}, {G{\"u}sten}, {Menten}, {Beelen}, {Bertoldi}, {Beeman}, \&
  {Haller}}]{Siringo2009}
{Siringo}, G., {Kreysa}, E., {Kov{\'a}cs}, A., {et~al.} 2009, \aap, 497, 945

\bibitem[{{Snell} {et~al.}(1990){Snell}, {Dickman}, \& {Huang}}]{snell1990}
{Snell}, R.~L., {Dickman}, R.~L., \& {Huang}, Y.-L. 1990, \apj, 352, 139

\bibitem[{{Sridharan} {et~al.}(2002){Sridharan}, {Beuther}, {Schilke},
  {Menten}, \& {Wyrowski}}]{sridharan2002}
{Sridharan}, T.~K., {Beuther}, H., {Schilke}, P., {Menten}, K.~M., \&
  {Wyrowski}, F. 2002, \apj, 566, 931

\bibitem[{Stephens(1974)}]{Stephens1974}
Stephens, M.~A. 1974, Journal of the American Statistical Association, 69, 730

\bibitem[{{Tan} {et~al.}(2016){Tan}, {Kong}, {Zhang}, {Fontani}, {Caselli}, \&
  {Butler}}]{Tan2016}
{Tan}, J.~C., {Kong}, S., {Zhang}, Y., {et~al.} 2016, \apjl, 821, L3

\bibitem[{{Thompson} {et~al.}(2004){Thompson}, {White}, {Morgan}, {Miao},
  {Fridlund}, \& {Huldtgren-White}}]{Thompson2004}
{Thompson}, M.~A., {White}, G.~J., {Morgan}, L.~K., {et~al.} 2004, \aap, 414,
  1017

\bibitem[{{Traficante} {et~al.}(2015){Traficante}, {Fuller}, {Peretto},
  {Pineda}, \& {Molinari}}]{Traficante2015}
{Traficante}, A., {Fuller}, G.~A., {Peretto}, N., {Pineda}, J.~E., \&
  {Molinari}, S. 2015, \mnras, 451, 3089

\bibitem[{{Urban} {et~al.}(2009){Urban}, {Evans}, \& {Doty}}]{Urban2009}
{Urban}, A., {Evans}, II, N.~J., \& {Doty}, S.~D. 2009, \apj, 698, 1341

\bibitem[{{Urquhart} {et~al.}(2007){Urquhart}, {Busfield}, {Hoare}, {Lumsden},
  {Clarke}, {Moore}, {Mottram}, \& {Oudmaijer}}]{urquhart2007}
{Urquhart}, J.~S., {Busfield}, A.~L., {Hoare}, M.~G., {et~al.} 2007, \aap, 461,
  11

\bibitem[{{Urquhart} {et~al.}(2014{\natexlab{a}}){Urquhart}, {Csengeri},
  {Wyrowski}, {Schuller}, {Bontemps}, {Bronfman}, {Menten}, {Walmsley},
  {Contreras}, {Beuther}, {Wienen}, \& {Linz}}]{urquhart2014c}
{Urquhart}, J.~S., {Csengeri}, T., {Wyrowski}, F., {et~al.} 2014{\natexlab{a}},
  \aap, 568, A41

\bibitem[{{Urquhart} {et~al.}(2014{\natexlab{b}}){Urquhart}, {Figura}, {Moore},
  {Hoare}, {Lumsden}, {Mottram}, {Thompson}, \& {Oudmaijer}}]{Urquhart2014a}
{Urquhart}, J.~S., {Figura}, C.~C., {Moore}, T.~J.~T., {et~al.}
  2014{\natexlab{b}}, \mnras, 437, 1791

\bibitem[{{Urquhart} {et~al.}(2012){Urquhart}, {Hoare}, {Lumsden}, {Oudmaijer},
  {Moore}, {Mottram}, {Cooper}, {Mottram}, \& {Rogers}}]{Urquhart2012}
{Urquhart}, J.~S., {Hoare}, M.~G., {Lumsden}, S.~L., {et~al.} 2012, \mnras,
  420, 1656

\bibitem[{{Urquhart} {et~al.}(2014{\natexlab{c}}){Urquhart}, {Moore},
  {Csengeri}, {Wyrowski}, {Schuller}, {Hoare}, {Lumsden}, {Mottram},
  {Thompson}, {Menten}, {Walmsley}, {Bronfman}, {Pfalzner}, {K{\"o}nig}, \&
  {Wienen}}]{Urquhart2014}
{Urquhart}, J.~S., {Moore}, T.~J.~T., {Csengeri}, T., {et~al.}
  2014{\natexlab{c}}, \mnras, 443, 1555

\bibitem[{{Urquhart} {et~al.}(2015){Urquhart}, {Moore}, {Menten}, {K{\"o}nig},
  {Wyrowski}, {Thompson}, {Csengeri}, {Leurini}, \& {Eden}}]{Urquhart2015}
{Urquhart}, J.~S., {Moore}, T.~J.~T., {Menten}, K.~M., {et~al.} 2015, \mnras,
  446, 3461

\bibitem[{{Urquhart} {et~al.}(2013{\natexlab{a}}){Urquhart}, {Moore},
  {Schuller}, {Wyrowski}, {Menten}, {Thompson}, {Csengeri}, {Walmsley},
  {Bronfman}, \& {K{\"o}nig}}]{Urquhart2013}
{Urquhart}, J.~S., {Moore}, T.~J.~T., {Schuller}, F., {et~al.}
  2013{\natexlab{a}}, \mnras, 431, 1752

\bibitem[{{Urquhart} {et~al.}(2009){Urquhart}, {Morgan}, \&
  {Thompson}}]{urquhart2009}
{Urquhart}, J.~S., {Morgan}, L.~K., \& {Thompson}, M.~A. 2009, \aap, 497, 789

\bibitem[{{Urquhart} {et~al.}(2013{\natexlab{b}}){Urquhart}, {Thompson},
  {Moore}, {Purcell}, {Hoare}, {Schuller}, {Wyrowski}, {Csengeri}, {Menten},
  {Lumsden}, {Kurtz}, {Walmsley}, {Bronfman}, {Morgan}, {Eden}, \&
  {Russeil}}]{urquhart2013b}
{Urquhart}, J.~S., {Thompson}, M.~A., {Moore}, T.~J.~T., {et~al.}
  2013{\natexlab{b}}, \mnras, 435, 400

\bibitem[{{Walsh} {et~al.}(2011){Walsh}, {Breen}, {Britton}, {Brooks},
  {Burton}, {Cunningham}, {Green}, {Harvey-Smith}, {Hindson}, {Hoare},
  {Indermuehle}, {Jones}, {Lo}, {Longmore}, {Lowe}, {Phillips}, {Purcell},
  {Thompson}, {Urquhart}, {Voronkov}, {White}, \& {Whiting}}]{walsh2011}
{Walsh}, A.~J., {Breen}, S.~L., {Britton}, T., {et~al.} 2011, in EAS
  Publications Series, Vol.~52, EAS Publications Series, ed. M.~{R{\"o}llig},
  R.~{Simon}, V.~{Ossenkopf}, \& J.~{Stutzki}, 135--138

\bibitem[{{Walsh} {et~al.}(1999){Walsh}, {Burton}, {Hyland}, \&
  {Robinson}}]{walsh1999}
{Walsh}, A.~J., {Burton}, M.~G., {Hyland}, A.~R., \& {Robinson}, G. 1999,
  \mnras, 309, 905

\bibitem[{{Walsh} {et~al.}(1997){Walsh}, {Hyland}, {Robinson}, \&
  {Burton}}]{walsh1997}
{Walsh}, A.~J., {Hyland}, A.~R., {Robinson}, G., \& {Burton}, M.~G. 1997,
  \mnras, 291, 261

\bibitem[{{Walsh} {et~al.}(2003){Walsh}, {Macdonald}, {Alvey}, {Burton}, \&
  {Lee}}]{walsh2003}
{Walsh}, A.~J., {Macdonald}, G.~H., {Alvey}, N.~D.~S., {Burton}, M.~G., \&
  {Lee}, J.-K. 2003, \aap, 410, 597

\bibitem[{{Walsh} {et~al.}(2014){Walsh}, {Purcell}, {Longmore}, {Breen},
  {Green}, {Harvey-Smith}, {Jordan}, \& {Macpherson}}]{walsh2014}
{Walsh}, A.~J., {Purcell}, C.~R., {Longmore}, S.~N., {et~al.} 2014, \mnras,
  442, 2240

\bibitem[{{Whitney} {et~al.}(2005){Whitney}, {Robitaille}, {Wood}, {Denzmore},
  \& {Bjorkman}}]{Whitney2005}
{Whitney}, B.~A., {Robitaille}, T.~P., {Wood}, K., {Denzmore}, P., \&
  {Bjorkman}, J.~E. 2005, in Protostars and Planets V Posters, Vol. 1286, 8460

\bibitem[{{Wienen} {et~al.}(2015){Wienen}, {Wyrowski}, {Menten}, {Urquhart},
  {Csengeri}, {Walmsley}, {Bontemps}, {Russeil}, {Bronfman}, {Koribalski}, \&
  {Schuller}}]{Wienen2015}
{Wienen}, M., {Wyrowski}, F., {Menten}, K.~M., {et~al.} 2015, \aap, 579, A91

\bibitem[{{Wood} \& {Churchwell}(1989)}]{wood1989}
{Wood}, D.~O.~S. \& {Churchwell}, E. 1989, \apjs, 69, 831

\bibitem[{{Wright} {et~al.}(2010){Wright}, {Eisenhardt}, {Mainzer}, {Ressler},
  {Cutri}, {Jarrett}, {Kirkpatrick}, {Padgett}, {McMillan}, {Skrutskie},
  {Stanford}, {Cohen}, {Walker}, {Mather}, {Leisawitz}, {Gautier}, {McLean},
  {Benford}, {Lonsdale}, {Blain}, {Mendez}, {Irace}, {Duval}, {Liu}, {Royer},
  {Heinrichsen}, {Howard}, {Shannon}, {Kendall}, {Walsh}, {Larsen}, {Cardon},
  {Schick}, {Schwalm}, {Abid}, {Fabinsky}, {Naes}, \& {Tsai}}]{Wright2010}
{Wright}, E.~L., {Eisenhardt}, P.~R.~M., {Mainzer}, A.~K., {et~al.} 2010, \aj,
  140, 1868

\bibitem[{{Wu} {et~al.}(2014){Wu}, {Sato}, {Reid}, {Moscadelli}, {Zhang}, {Xu},
  {Brunthaler}, {Menten}, {Dame}, \& {Zheng}}]{Wu2014}
{Wu}, Y.~W., {Sato}, M., {Reid}, M.~J., {et~al.} 2014, \aap, 566, A17

\bibitem[{{Wyrowski} {et~al.}(2016){Wyrowski}, {G{\"u}sten}, {Menten},
  {Wiesemeyer}, {Csengeri}, {Heyminck}, {Klein}, {K{\"o}nig}, \&
  {Urquhart}}]{Wyrowski2016}
{Wyrowski}, F., {G{\"u}sten}, R., {Menten}, K.~M., {et~al.} 2016, \aap, 585,
  A149

\bibitem[{{Xu} {et~al.}(2011){Xu}, {Moscadelli}, {Reid}, {Menten}, {Zhang},
  {Zheng}, \& {Brunthaler}}]{Xu2011}
{Xu}, Y., {Moscadelli}, L., {Reid}, M.~J., {et~al.} 2011, \apj, 733, 25

\bibitem[{{Xu} {et~al.}(2009){Xu}, {Reid}, {Menten}, {Brunthaler}, {Zheng}, \&
  {Moscadelli}}]{Xu2009}
{Xu}, Y., {Reid}, M.~J., {Menten}, K.~M., {et~al.} 2009, \apj, 693, 413

\bibitem[{{Zhang} {et~al.}(2014){Zhang}, {Moscadelli}, {Sato}, {Reid},
  {Menten}, {Zheng}, {Brunthaler}, {Dame}, {Xu}, \& {Immer}}]{Zhang2014}
{Zhang}, B., {Moscadelli}, L., {Sato}, M., {et~al.} 2014, \apj, 781, 89

\bibitem[{{Zhang} {et~al.}(2013){Zhang}, {Reid}, {Menten}, {Zheng},
  {Brunthaler}, {Dame}, \& {Xu}}]{Zhang2013}
{Zhang}, B., {Reid}, M.~J., {Menten}, K.~M., {et~al.} 2013, \apj, 775, 79

\bibitem[{{Zhang} {et~al.}(2009){Zhang}, {Zheng}, {Reid}, {Menten}, {Xu},
  {Moscadelli}, \& {Brunthaler}}]{Zhang2009}
{Zhang}, B., {Zheng}, X.~W., {Reid}, M.~J., {et~al.} 2009, \apj, 693, 419

\bibitem[{{Zinnecker} \& {Yorke}(2007)}]{Zinnecker2007}
{Zinnecker}, H. \& {Yorke}, H.~W. 2007, \araa, 45, 481

\end{thebibliography}

\Online

\appendix

\onecolumn{}
\begin{appendix}
\section{Full Data}\label{appendix.fulltable}

\begin{longtable}{ccccccccccccc}
\caption{\label{tbl.fulldata}Source parameters.}\\
\hline\hline
Name & $d$ & $d_\mathrm{ref}$ & V$_\mathrm{LSR}$ & $l_\mathrm{app}$ & $b_\mathrm{app}$ & $D_\mathrm{app}$ & Class & $T$ & $\Delta T$ & $L_\mathrm{bol}$ & $M_\mathrm{clump}$ & $\alpha_\mathrm{vir}$\\
 & (kpc) &  & (km/s) & ($\deg$) & ($\deg$) & ($''$) &  & (K) & (K) & $(\mathrm{L}_\odot)$ & $(\mathrm{M}_\odot)$ & \\
\hline
\endfirsthead
\caption{continued.}\\
\hline\hline
Name & $d$ & $d_\mathrm{ref}$ & V$_\mathrm{LSR}$ & $l_\mathrm{app}$ & $b_\mathrm{app}$ & $D_\mathrm{app}$ & Class & $T$ & $\Delta T$ & $L_\mathrm{bol}$ & $M_\mathrm{clump}$ & $\alpha$\\
 & (kpc) &  & (km/s) & ($\deg$) & ($\deg$) & ($''$) &  & (K) & (K) & $(\mathrm{L}_\odot)$ & $(\mathrm{M}_\odot)$ &  \\
\hline
\endhead
\hline
\endfoot
AGAL006.216$-$00.609 & $2.9$ & $(22)$ & 18.5 & 6.216 & $-$0.609 & 71.6 & IRw & 16.0 & 0.2 & $7.3\times10^{2}$ & $4.5\times10^{2}$ & $-$ \\
AGAL008.671$-$00.356 & $4.8$ & $(19)$ & 35.1 & 8.668 & $-$0.355 & 69.1 & H\textsc{ii} & 26.3 & 1.4 & $8.6\times10^{4}$ & $3.0\times10^{3}$ & $-$ \\
AGAL008.684$-$00.367 & $4.8$ & $(19)$ & 38.0 & 8.682 & $-$0.367 & 66.1 & IRw & 24.2 & 0.2 & $2.7\times10^{4}$ & $1.4\times10^{3}$ & $0.69$ \\
AGAL008.706$-$00.414 & $4.8$ & $(19)$ & 39.4 & 8.704 & $-$0.412 & 92.1 & IRw & 11.8 & 0.3 & $5.0\times10^{2}$ & $1.6\times10^{3}$ & $0.22$ \\
AGAL008.831$-$00.027 & $-$ & $-$ & 0.5 & 8.831 & $-$0.027 & 70.9 & IRw & 24.1 & 1.5 & $-$ & $-$ & $-$ \\
AGAL010.444$-$00.017 & $8.6$ & $(1)$ & 75.9 & 10.442 & $-$0.016 & 65.5 & IRw & 20.7 & 0.3 & $1.1\times10^{4}$ & $1.6\times10^{3}$ & $0.40$ \\
AGAL010.472+00.027 & $8.6$ & $(1)$ & 67.6 & 10.472 & +0.028 & 55.1 & H\textsc{ii} & 30.5 & 2.4 & $4.6\times10^{5}$ & $1.0\times10^{4}$ & $0.22$ \\
AGAL010.624$-$00.384 & $5.0$ & $(1)$ & $-$2.9 & 10.623 & $-$0.382 & 65.0 & H\textsc{ii} & 34.5 & 3.6 & $4.2\times10^{5}$ & $3.7\times10^{3}$ & $0.42$ \\
AGAL012.804$-$00.199 & $2.4$ & $(28)$ & 36.2 & 12.805 & $-$0.197 & 84.4 & H\textsc{ii} & 35.1 & 2.3 & $2.4\times10^{5}$ & $1.8\times10^{3}$ & $0.78$ \\
AGAL013.178+00.059 & $2.4$ & $(25)$ & 50.4 & 13.176 & +0.062 & 87.6 & 70w & 24.2 & 0.8 & $8.3\times10^{3}$ & $3.6\times10^{2}$ & $0.96$ \\
AGAL013.658$-$00.599 & $4.5$ & $(22)$ & 48.4 & 13.656 & $-$0.596 & 67.0 & IRb & 27.4 & 1.3 & $2.0\times10^{4}$ & $5.6\times10^{2}$ & $0.53$ \\
AGAL014.114$-$00.574 & $2.6$ & $(3)$ & 20.8 & 14.112 & $-$0.572 & 81.0 & IRw & 22.4 & 0.8 & $3.1\times10^{3}$ & $3.5\times10^{2}$ & $0.65$ \\
AGAL014.194$-$00.194 & $3.9$ & $(3)$ & 39.2 & 14.194 & $-$0.191 & 72.0 & IRw & 18.2 & 0.6 & $2.7\times10^{3}$ & $8.2\times10^{2}$ & $0.50$ \\
AGAL014.492$-$00.139 & $3.9$ & $(3)$ & 39.5 & 14.491 & $-$0.137 & 89.3 & 70w & 12.4 & 0.4 & $7.5\times10^{2}$ & $1.9\times10^{3}$ & $0.55$ \\
AGAL014.632$-$00.577 & $1.8$ & $(27)$ & 18.5 & 14.631 & $-$0.576 & 83.5 & IRw & 22.5 & 0.4 & $2.7\times10^{3}$ & $2.5\times10^{2}$ & $0.84$ \\
AGAL015.029$-$00.669 & $2.0$ & $(9)$ & 19.6 & 15.032 & $-$0.674 & 98.6 & IRb & 32.9 & 1.5 & $1.3\times10^{5}$ & $1.1\times10^{3}$ & $0.19$ \\
AGAL018.606$-$00.074 & $4.3$ & $(7)$ & 46.6 & 18.606 & $-$0.074 & 74.8 & IRw & 13.8 & 0.3 & $5.9\times10^{2}$ & $8.7\times10^{2}$ & $0.46$ \\
AGAL018.734$-$00.226 & $12.5$ & $(19)$ & 42.6 & 18.734 & $-$0.224 & 74.1 & IRw & 21.9 & 1.0 & $7.2\times10^{4}$ & $7.9\times10^{3}$ & $0.45$ \\
AGAL018.888$-$00.474 & $4.7$ & $(19)$ & 66.1 & 18.888 & $-$0.472 & 81.5 & IRw & 14.4 & 0.2 & $3.2\times10^{3}$ & $2.8\times10^{3}$ & $0.54$ \\
AGAL019.609$-$00.234 & $12.6$ & $(22)$ & 40.8 & 19.608 & $-$0.233 & 60.1 & H\textsc{ii} & 32.4 & 1.2 & $1.1\times10^{6}$ & $1.3\times10^{4}$ & $-$ \\
AGAL019.882$-$00.534 & $3.7$ & $(19)$ & 45.0 & 19.882 & $-$0.532 & 66.7 & IRb & 24.2 & 1.4 & $1.2\times10^{4}$ & $7.9\times10^{2}$ & $0.41$ \\
AGAL022.376+00.447 & $4.0$ & $(19)$ & 54.0 & 22.376 & +0.447 & 64.7 & IRw & 13.1 & 0.4 & $3.1\times10^{2}$ & $6.2\times10^{2}$ & $0.10$ \\
AGAL023.206$-$00.377 & $4.6$ & $(5)$ & 78.2 & 23.206 & $-$0.377 & 60.7 & IRw & 22.1 & 0.1 & $1.2\times10^{4}$ & $1.2\times10^{3}$ & $0.44$ \\
AGAL024.416+00.101 & $7.7$ & $(17)$ & 112.2 & 24.418 & +0.102 & 84.5 & IRw & 15.4 & 0.2 & $4.2\times10^{3}$ & $2.4\times10^{3}$ & $-$ \\
AGAL024.629+00.172 & $7.7$ & $(17)$ & 116.0 & 24.631 & +0.174 & 72.0 & IRw & 18.1 & 0.5 & $5.0\times10^{3}$ & $1.5\times10^{3}$ & $0.85$ \\
AGAL024.651$-$00.169 & $7.7$ & $(17)$ & 113.0 & 24.653 & $-$0.169 & 86.5 & 70w & 18.6 & 1.0 & $6.8\times10^{3}$ & $1.8\times10^{3}$ & $-$ \\
AGAL024.789+00.082 & $7.7$ & $(17)$ & 110.2 & 24.789 & +0.085 & 69.0 & H\textsc{ii} & 26.5 & 2.5 & $1.9\times10^{5}$ & $6.8\times10^{3}$ & $-$ \\
AGAL028.564$-$00.236 & $5.5$ & $(3)$ & 87.2 & 28.566 & $-$0.232 & 101.3 & IRw & 11.7 & 0.1 & $1.7\times10^{3}$ & $5.4\times10^{3}$ & $0.40$ \\
AGAL028.861+00.066 & $7.4$ & $(8)$ & 104.0 & 28.861 & +0.067 & 69.1 & IRb & 34.5 & 1.7 & $1.6\times10^{5}$ & $1.0\times10^{3}$ & $0.39$ \\
AGAL030.818$-$00.056 & $4.9$ & $(2)$ & 97.8 & 30.816 & $-$0.055 & 65.3 & IRb & 22.9 & 0.1 & $6.3\times10^{4}$ & $5.6\times10^{3}$ & $0.28$ \\
AGAL030.848$-$00.081 & $4.9$ & $(2)$ & 94.4 & 30.848 & $-$0.082 & 78.4 & 70w & 16.7 & 0.2 & $3.1\times10^{3}$ & $1.2\times10^{3}$ & $0.10$ \\
AGAL030.893+00.139 & $4.9$ & $(2)$ & 97.3 & 30.893 & +0.139 & 84.7 & 70w & 11.4 & 0.5 & $4.9\times10^{2}$ & $1.9\times10^{3}$ & $0.41$ \\
AGAL031.412+00.307 & $4.9$ & $(2)$ & 98.2 & 31.411 & +0.308 & 58.9 & H\textsc{ii} & 26.3 & 1.8 & $6.8\times10^{4}$ & $3.0\times10^{3}$ & $0.31$ \\
AGAL034.258+00.154 & $1.6$ & $(20)$ & 58.5 & 34.256 & +0.155 & 66.5 & H\textsc{ii} & 31.0 & 1.2 & $4.8\times10^{4}$ & $8.0\times10^{2}$ & $0.57$ \\
AGAL034.401+00.226 & $1.6$ & $(20)$ & 57.4 & 34.402 & +0.229 & 81.5 & H\textsc{ii} & 22.8 & 1.2 & $2.9\times10^{3}$ & $2.7\times10^{2}$ & $1.94$ \\
AGAL034.411+00.234 & $1.6$ & $(20)$ & 58.0 & 34.411 & +0.236 & 62.6 & IRb & 26.1 & 1.8 & $4.8\times10^{3}$ & $2.1\times10^{2}$ & $1.20$ \\
AGAL034.821+00.351 & $1.6$ & $(20)$ & 58.3 & 34.819 & +0.351 & 89.0 & IRb & 24.7 & 0.6 & $2.7\times10^{3}$ & $1.1\times10^{2}$ & $1.62$ \\
AGAL035.197$-$00.742 & $2.2$ & $(21)$ & 34.7 & 35.197 & $-$0.743 & 73.5 & IRb & 29.5 & 2.7 & $2.3\times10^{4}$ & $4.6\times10^{2}$ & $0.90$ \\
AGAL035.579+00.007 & $10.3$ & $(22)$ & 53.1 & 35.576 & +0.011 & 109.8 & 70w & 14.5 & 0.2 & $9.1\times10^{3}$ & $1.0\times10^{4}$ & $-$ \\
AGAL037.554+00.201 & $6.7$ & $(17)$ & 86.1 & 37.552 & +0.201 & 69.5 & IRb & 28.4 & 1.8 & $5.1\times10^{4}$ & $1.2\times10^{3}$ & $0.67$ \\
AGAL043.166+00.011 & $11.1$ & $(4)$ & 5.3 & 43.165 & +0.012 & 68.5 & H\textsc{ii} & 34.0 & 2.7 & $3.7\times10^{6}$ & $4.3\times10^{4}$ & $0.08$ \\
AGAL049.489$-$00.389 & $5.4$ & $(10)$ & 57.2 & 49.488 & $-$0.388 & 62.1 & H\textsc{ii} & 29.1 & 1.2 & $5.5\times10^{5}$ & $1.1\times10^{4}$ & $0.18$ \\
AGAL053.141+00.069 & $1.6$ & $(18)$ & 22.3 & 53.141 & +0.071 & 66.8 & IRb & 25.4 & 0.7 & $2.3\times10^{3}$ & $9.4\times10^{1}$ & $1.59$ \\
AGAL059.782+00.066 & $2.2$ & $(24)$ & 23.1 & 59.782 & +0.066 & 78.9 & IRb & 28.2 & 1.5 & $9.7\times10^{3}$ & $2.5\times10^{2}$ & $0.71$ \\
AGAL301.136$-$00.226 & $4.4$ & $(19)$ & $-$39.2 & 301.136 & $-$0.224 & 62.1 & H\textsc{ii} & 34.6 & 2.8 & $2.1\times10^{5}$ & $1.9\times10^{3}$ & $0.42$ \\
AGAL305.192$-$00.006 & $3.8$ & $(16)$ & $-$34.2 & 305.192 & $-$0.004 & 73.7 & IRw & 26.1 & 3.1 & $1.2\times10^{4}$ & $5.1\times10^{2}$ & $0.96$ \\
AGAL305.209+00.206 & $3.8$ & $(16)$ & $-$42.5 & 305.208 & +0.208 & 68.0 & IRb & 30.1 & 0.9 & $8.8\times10^{4}$ & $1.4\times10^{3}$ & $0.73$ \\
AGAL305.562+00.014 & $3.8$ & $(16)$ & $-$39.8 & 305.561 & +0.014 & 67.9 & IRb & 33.4 & 1.9 & $5.1\times10^{4}$ & $4.0\times10^{2}$ & $1.00$ \\
AGAL305.794$-$00.096 & $3.8$ & $(16)$ & $-$40.8 & 305.794 & $-$0.096 & 84.7 & 70w & 16.0 & 0.1 & $9.8\times10^{2}$ & $5.9\times10^{2}$ & $0.62$ \\
AGAL309.384$-$00.134 & $5.3$ & $(19)$ & $-$51.3 & 309.383 & $-$0.134 & 74.9 & IRb & 24.0 & 1.3 & $1.5\times10^{4}$ & $1.2\times10^{3}$ & $0.65$ \\
AGAL310.014+00.387 & $3.6$ & $(3)$ & $-$41.3 & 310.011 & +0.389 & 76.6 & IRb & 32.2 & 1.1 & $4.9\times10^{4}$ & $4.1\times10^{2}$ & $1.14$ \\
AGAL313.576+00.324 & $3.8$ & $(19)$ & $-$46.9 & 313.576 & +0.326 & 57.8 & IRb & 29.2 & 1.8 & $9.4\times10^{3}$ & $1.8\times10^{2}$ & $1.68$ \\
AGAL316.641$-$00.087 & $1.2$ & $(23)$ & $-$17.7 & 316.639 & $-$0.086 & 62.2 & IRb & 30.6 & 1.9 & $9.9\times10^{2}$ & $1.8\times10^{1}$ & $5.44$ \\
AGAL317.867$-$00.151 & $3.0$ & $(3)$ & $-$40.2 & 317.867 & $-$0.149 & 62.3 & IRw & 19.3 & 0.5 & $1.6\times10^{3}$ & $3.6\times10^{2}$ & $0.94$ \\
AGAL318.779$-$00.137 & $2.8$ & $(3)$ & $-$38.1 & 318.778 & $-$0.136 & 86.5 & IRw & 24.9 & 1.5 & $6.3\times10^{3}$ & $3.5\times10^{2}$ & $1.45$ \\
AGAL320.881$-$00.397 & $10.0$ & $(3)$ & $-$45.3 & 320.879 & $-$0.396 & 69.1 & 70w & 16.8 & 0.2 & $6.0\times10^{3}$ & $2.8\times10^{3}$ & $0.20$ \\
AGAL326.661+00.519 & $1.8$ & $(13)$ & $-$39.8 & 326.661 & +0.521 & 74.2 & IRb & 28.4 & 0.8 & $7.4\times10^{3}$ & $1.2\times10^{2}$ & $1.25$ \\
AGAL326.987$-$00.032 & $4.0$ & $(3)$ & $-$58.6 & 326.986 & $-$0.029 & 63.7 & IRw & 17.9 & 2.4 & $1.1\times10^{3}$ & $4.4\times10^{2}$ & $1.80$ \\
AGAL327.119+00.509 & $5.5$ & $(19)$ & $-$83.6 & 327.119 & +0.511 & 67.7 & IRb & 31.8 & 1.2 & $5.8\times10^{4}$ & $6.7\times10^{2}$ & $0.71$ \\
AGAL327.293$-$00.579 & $3.1$ & $(22)$ & $-$44.7 & 327.291 & $-$0.578 & 58.8 & IRb & 27.9 & 2.2 & $8.3\times10^{4}$ & $2.8\times10^{3}$ & $0.29$ \\
AGAL327.393+00.199 & $5.9$ & $(19)$ & $-$89.2 & 327.391 & +0.201 & 70.1 & IRb & 23.2 & 1.0 & $1.3\times10^{4}$ & $1.1\times10^{3}$ & $0.53$ \\
AGAL328.809+00.632 & $3.0$ & $(19)$ & $-$41.9 & 328.808 & +0.635 & 64.7 & H\textsc{ii} & 36.3 & 3.2 & $1.5\times10^{5}$ & $1.0\times10^{3}$ & $0.70$ \\
AGAL329.029$-$00.206 & $11.5$ & $(19)$ & $-$43.5 & 329.029 & $-$0.197 & 89.3 & IRw & 23.1 & 0.6 & $2.1\times10^{5}$ & $1.1\times10^{4}$ & $0.16$ \\
AGAL329.066$-$00.307 & $11.6$ & $(19)$ & $-$41.9 & 329.064 & $-$0.306 & 87.4 & IRb & 21.9 & 1.0 & $7.1\times10^{4}$ & $9.0\times10^{3}$ & $0.33$ \\
AGAL330.879$-$00.367 & $4.2$ & $(26)$ & $-$62.5 & 330.878 & $-$0.366 & 63.0 & H\textsc{ii} & 33.4 & 3.0 & $1.5\times10^{5}$ & $1.5\times10^{3}$ & $0.48$ \\
AGAL330.954$-$00.182 & $9.3$ & $(11)$ & $-$91.2 & 330.952 & $-$0.181 & 58.6 & H\textsc{ii} & 33.0 & 4.5 & $1.3\times10^{6}$ & $1.7\times10^{4}$ & $0.25$ \\
AGAL331.709+00.582 & $10.5$ & $(3)$ & $-$67.8 & 331.707 & +0.584 & 72.2 & IRw & 21.0 & 0.3 & $3.7\times10^{4}$ & $5.1\times10^{3}$ & $0.53$ \\
AGAL332.094$-$00.421 & $3.6$ & $(13)$ & $-$56.9 & 332.094 & $-$0.419 & 70.0 & IRb & 30.8 & 0.7 & $5.8\times10^{4}$ & $6.2\times10^{2}$ & $0.86$ \\
AGAL332.826$-$00.549 & $3.6$ & $(13)$ & $-$57.1 & 332.824 & $-$0.549 & 63.6 & H\textsc{ii} & 35.7 & 2.8 & $2.4\times10^{5}$ & $1.9\times10^{3}$ & $0.51$ \\
AGAL333.134$-$00.431 & $3.6$ & $(13)$ & $-$53.1 & 333.133 & $-$0.43 & 77.4 & H\textsc{ii} & 35.2 & 1.7 & $4.1\times10^{5}$ & $2.8\times10^{3}$ & $0.89$ \\
AGAL333.284$-$00.387 & $3.6$ & $(13)$ & $-$52.4 & 333.284 & $-$0.387 & 85.4 & H\textsc{ii} & 30.4 & 1.8 & $1.2\times10^{5}$ & $2.0\times10^{3}$ & $0.41$ \\
AGAL333.314+00.106 & $3.6$ & $(13)$ & $-$46.5 & 333.314 & +0.106 & 70.6 & IRb & 25.9 & 1.5 & $1.0\times10^{4}$ & $4.2\times10^{2}$ & $1.63$ \\
AGAL333.604$-$00.212 & $3.6$ & $(13)$ & $-$48.4 & 333.602 & $-$0.21 & 85.2 & H\textsc{ii} & 41.1 & 2.4 & $1.2\times10^{6}$ & $3.4\times10^{3}$ & $0.77$ \\
AGAL333.656+00.059 & $5.3$ & $(3)$ & $-$85.0 & 333.656 & +0.059 & 84.7 & 70w & 17.8 & 0.3 & $4.2\times10^{3}$ & $1.4\times10^{3}$ & $0.53$ \\
AGAL335.789+00.174 & $3.7$ & $(19)$ & $-$50.1 & 335.789 & +0.176 & 71.7 & IRw & 24.7 & 0.9 & $2.0\times10^{4}$ & $1.1\times10^{3}$ & $0.46$ \\
AGAL336.958$-$00.224 & $10.9$ & $(19)$ & $-$71.3 & 336.956 & $-$0.224 & 57.8 & IRw & 15.8 & 1.0 & $3.6\times10^{3}$ & $2.4\times10^{3}$ & $0.11$ \\
AGAL337.176$-$00.032 & $11.0$ & $(6)$ & $-$67.8 & 337.174 & $-$0.031 & 82.6 & IRw & 22.3 & 1.2 & $5.9\times10^{4}$ & $5.5\times10^{3}$ & $0.17$ \\
AGAL337.258$-$00.101 & $11.0$ & $(6)$ & $-$68.3 & 337.258 & $-$0.099 & 68.0 & IRw & 21.7 & 0.9 & $3.0\times10^{4}$ & $3.1\times10^{3}$ & $0.33$ \\
AGAL337.286+00.007 & $9.4$ & $(3)$ & $-$106.6 & 337.284 & +0.009 & 86.7 & 70w & 10.7 & 0.5 & $1.2\times10^{3}$ & $6.6\times10^{3}$ & $0.09$ \\
AGAL337.406$-$00.402 & $3.3$ & $(19)$ & $-$41.0 & 337.404 & $-$0.402 & 66.3 & H\textsc{ii} & 31.8 & 2.3 & $8.5\times10^{4}$ & $1.1\times10^{3}$ & $0.55$ \\
AGAL337.704$-$00.054 & $12.3$ & $(19)$ & $-$47.4 & 337.704 & $-$0.053 & 63.7 & H\textsc{ii} & 25.6 & 0.6 & $3.1\times10^{5}$ & $1.4\times10^{4}$ & $0.28$ \\
AGAL337.916$-$00.477 & $3.2$ & $(26)$ & $-$39.6 & 337.914 & $-$0.476 & 61.9 & IRb & 34.4 & 1.8 & $1.2\times10^{5}$ & $1.2\times10^{3}$ & $0.45$ \\
AGAL338.066+00.044 & $4.7$ & $(3)$ & $-$69.2 & 338.066 & +0.046 & 89.2 & 70w & 18.5 & 1.4 & $3.1\times10^{3}$ & $9.6\times10^{2}$ & $2.86$ \\
AGAL338.786+00.476 & $4.5$ & $(3)$ & $-$63.8 & 338.782 & +0.477 & 84.8 & 70w & 12.2 & 0.6 & $4.8\times10^{2}$ & $1.2\times10^{3}$ & $0.38$ \\
AGAL338.926+00.554 & $4.4$ & $(19)$ & $-$61.6 & 338.924 & +0.558 & 89.9 & IRb & 24.2 & 1.0 & $9.4\times10^{4}$ & $5.9\times10^{3}$ & $0.43$ \\
AGAL339.623$-$00.122 & $3.0$ & $(19)$ & $-$34.6 & 339.621 & $-$0.121 & 80.2 & IRb & 28.7 & 1.4 & $1.4\times10^{4}$ & $3.1\times10^{2}$ & $1.16$ \\
AGAL340.374$-$00.391 & $3.6$ & $(3)$ & $-$43.6 & 340.372 & $-$0.389 & 72.2 & IRw & 13.4 & 0.1 & $5.1\times10^{2}$ & $7.9\times10^{2}$ & $0.49$ \\
AGAL340.746$-$01.001 & $2.8$ & $(3)$ & $-$29.4 & 340.744 & $-$1.001 & 75.1 & IRb & 27.1 & 0.3 & $7.6\times10^{3}$ & $2.1\times10^{2}$ & $1.55$ \\
AGAL340.784$-$00.097 & $10.0$ & $(19)$ & $-$101.5 & 340.784 & $-$0.096 & 61.4 & IRw & 26.2 & 0.4 & $7.2\times10^{4}$ & $2.8\times10^{3}$ & $0.45$ \\
AGAL341.217$-$00.212 & $3.7$ & $(19)$ & $-$43.6 & 341.216 & $-$0.211 & 65.6 & IRb & 27.0 & 1.3 & $1.6\times10^{4}$ & $4.8\times10^{2}$ & $0.74$ \\
AGAL342.484+00.182 & $12.6$ & $(19)$ & $-$41.5 & 342.483 & +0.182 & 64.8 & IRw & 23.6 & 0.7 & $6.4\times10^{4}$ & $4.9\times10^{3}$ & $0.12$ \\
AGAL343.128$-$00.062 & $3.0$ & $(3)$ & $-$30.6 & 343.125 & $-$0.062 & 69.1 & H\textsc{ii} & 30.9 & 4.5 & $7.1\times10^{4}$ & $1.1\times10^{3}$ & $0.69$ \\
AGAL343.756$-$00.164 & $2.9$ & $(14)$ & $-$27.8 & 343.756 & $-$0.162 & 60.9 & IRw & 24.3 & 1.4 & $9.8\times10^{3}$ & $6.1\times10^{2}$ & $0.43$ \\
AGAL344.227$-$00.569 & $2.5$ & $(19)$ & $-$22.3 & 344.225 & $-$0.569 & 63.0 & IRw & 22.0 & 1.3 & $9.7\times10^{3}$ & $1.1\times10^{3}$ & $0.35$ \\
AGAL345.003$-$00.224 & $3.0$ & $(19)$ & $-$27.6 & 345.002 & $-$0.222 & 65.2 & H\textsc{ii} & 31.0 & 2.8 & $6.4\times10^{4}$ & $9.6\times10^{2}$ & $0.79$ \\
AGAL345.488+00.314 & $2.2$ & $(19)$ & $-$17.6 & 345.487 & +0.316 & 86.3 & H\textsc{ii} & 30.7 & 1.7 & $6.1\times10^{4}$ & $9.2\times10^{2}$ & $0.95$ \\
AGAL345.504+00.347 & $2.2$ & $(26)$ & $-$17.8 & 345.504 & +0.35 & 75.2 & IRb & 32.7 & 1.5 & $4.3\times10^{4}$ & $4.2\times10^{2}$ & $1.70$ \\
AGAL345.718+00.817 & $1.6$ & $(3)$ & $-$11.2 & 345.716 & +0.817 & 94.9 & IRb & 22.1 & 1.3 & $1.8\times10^{3}$ & $2.0\times10^{2}$ & $0.67$ \\
AGAL351.131+00.771 & $1.8$ & $(13)$ & $-$5.3 & 351.133 & +0.771 & 80.5 & 70w & 18.6 & 0.3 & $6.2\times10^{2}$ & $1.2\times10^{2}$ & $0.76$ \\
AGAL351.161+00.697 & $1.8$ & $(13)$ & $-$6.7 & 351.16 & +0.698 & 76.4 & IRb & 21.9 & 3.4 & $8.7\times10^{3}$ & $1.1\times10^{3}$ & $0.46$ \\
AGAL351.244+00.669 & $1.8$ & $(13)$ & $-$3.3 & 351.245 & +0.67 & 99.0 & IRb & 32.5 & 2.2 & $7.7\times10^{4}$ & $8.8\times10^{2}$ & $0.46$ \\
AGAL351.416+00.646 & $1.3$ & $(27)$ & $-$7.4 & 351.416 & +0.645 & 59.6 & H\textsc{ii} & 33.4 & 1.5 & $3.9\times10^{4}$ & $4.6\times10^{2}$ & $0.54$ \\
AGAL351.444+00.659 & $1.3$ & $(27)$ & $-$4.3 & 351.444 & +0.659 & 94.4 & IRw & 21.4 & 1.1 & $9.5\times10^{3}$ & $1.4\times10^{3}$ & $0.28$ \\
AGAL351.571+00.762 & $1.3$ & $(27)$ & $-$3.2 & 351.569 & +0.762 & 107.8 & 70w & 17.0 & 0.1 & $4.3\times10^{2}$ & $1.6\times10^{2}$ & $0.64$ \\
AGAL351.581$-$00.352 & $6.8$ & $(19)$ & $-$95.9 & 351.581 & $-$0.352 & 65.6 & IRb & 27.1 & 2.6 & $2.4\times10^{5}$ & $8.7\times10^{3}$ & $0.13$ \\
AGAL351.774$-$00.537 & $1.0$ & $(15)$ & $-$2.8 & 351.775 & $-$0.535 & 63.3 & IRb & 31.8 & 3.5 & $1.6\times10^{4}$ & $2.6\times10^{2}$ & $1.05$ \\
AGAL353.066+00.452 & $0.9$ & $(22)$ & 1.7 & 353.066 & +0.452 & 72.1 & IRw & 17.8 & 0.2 & $5.7\times10^{1}$ & $1.7\times10^{1}$ & $2.27$ \\
AGAL353.409$-$00.361 & $3.4$ & $(22)$ & $-$16.0 & 353.408 & $-$0.359 & 90.4 & IRb & 28.3 & 1.7 & $1.2\times10^{5}$ & $3.4\times10^{3}$ & $0.27$ \\
AGAL353.417$-$00.079 & $6.1$ & $(3)$ & $-$54.4 & 353.416 & $-$0.079 & 93.7 & 70w & 17.1 & 0.4 & $4.5\times10^{3}$ & $1.7\times10^{3}$ & $0.11$ \\
AGAL354.944$-$00.537 & $1.9$ & $(3)$ & $-$5.4 & 354.944 & $-$0.537 & 88.2 & 70w & 19.1 & 1.3 & $4.8\times10^{2}$ & $1.4\times10^{2}$ & $1.13$ \\
\end{longtable}
\tablefoot{The Columns are as follows: Name: the ATLASGAL catalog source name. $d$: distance. $d_\mathrm{ref}$ Distance reference (see list of references below table). $V_\mathrm{lsr}$: source velocity. $l_\mathrm{app}$: Galactic longitude of aperture center. $b_\mathrm{app}$: Galactic latitude of aperture center. $D_\mathrm{app}$: aperture diameter. Class: class of the source (H\textsc{ii}: H\textsc{ii} region, IRb: mid-infrared bright, 24d: mid-infrared weak (c: confused within the aperture), 70d: 70~$\mu$m weak). $T$: Dust temperature $\Delta T$: Error of the dust temperature. $L_\mathrm{bol}$: Bolometric luminosity. $M_\mathrm{clump}$: Clump mass. $\alpha_\mathrm{vir}$: virial parameter.\\}
\tablebib{Distance references: (1): \citet{Sanna2014}; (2): \citet{Zhang2014}; (3): \citet{Giannetti2014}; (4): \citet{Zhang2013}; (5): \citet{Brunthaler2009}; (6): \citet{Giannetti2015}; (7): \citet{Kurayama2011}; (8): \citet{Sato2014}; (9): \citet{Xu2011}; (10): \citet{Sato2010}; (11): \citet{Urquhart2012}; (13): \citet{moises2011}; (14): \citet{busfield2006}; (15): \citet{snell1990}; (16): \citet{Davies2012}; (17): Tangent Point; (18): \citet{roman2009}; (19): \citet{Green2011}; (20): \citet{Zhang2009}; (21): \citet{Wienen2015}; (22): \citet{Urquhart2014a}; (23): \citet{Xu2009}; (24): \citet{Immer2013}; (25): \citet{caswell1975}; (26): \citet{Wu2014}; (27): \citet{immer2012}}
\end{appendix}
\end{document}